   \documentstyle[12pt]{article}
\if@twoside\oddsidemargin25mm
\else\oddsidemargin25mm\evensidemargin25mm\marginparwidth25mm\fi%
\begin{document}
\newcommand{\ft}[2]{{\textstyle\frac{#1}{#2}}}
\newcommand{\QED}{{\hspace*{\fill}\rule{2mm}{2mm}\linebreak}}
\def\dop{{\rm d}\hskip -1pt}
\def\bfone{\relax{\rm 1\kern-.35em 1}}
\def\bfzero{\relax{\rm I\kern-.18em 0}}
\def\inbar{\vrule height1.5ex width.4pt depth0pt}
\def\IC{\relax\,\hbox{$\inbar\kern-.3em{\rm C}$}}
\def\ID{\relax{\rm I\kern-.18em D}}
\def\IF{\relax{\rm I\kern-.18em F}}
\def\IK{\relax{\rm I\kern-.18em K}}
\def\IH{\relax{\rm I\kern-.18em H}}
\def\II{\relax{\rm I\kern-.17em I}}
\def\IN{\relax{\rm I\kern-.18em N}}
\def\IP{\relax{\rm I\kern-.18em P}}
\def\IQ{\relax\,\hbox{$\inbar\kern-.3em{\rm Q}$}}
\def\IR{\relax{\rm I\kern-.18em R}}
\def\IG{\relax\,\hbox{$\inbar\kern-.3em{\rm G}$}}
\font\cmss=cmss10 \font\cmsss=cmss10 at 7pt
\def\ZZ{\relax\ifmmode\mathchoice
{\hbox{\cmss Z\kern-.4em Z}}{\hbox{\cmss Z\kern-.4em Z}}
{\lower.9pt\hbox{\cmsss Z\kern-.4em Z}}
{\lower1.2pt\hbox{\cmsss Z\kern
.4em Z}}\else{\cmss Z\kern-.4em
Z}\fi}
\def\a{\alpha} \def\b{\beta} \def\d{\delta}
\def\e{\epsilon} \def\c{\gamma}
\def\G{\Gamma} \def\l{\lambda}
\def\L{\Lambda} \def\s{\sigma}
+\def\cA{{\cal A}} \def\cB{{\cal B}}
\def\cC{{\cal C}} \def\cD{{\cal D}}
    \def\cF{{\cal F}} \def\cG{{\cal G}}
\def\cH{{\cal H}} \def\cI{{\cal I}}
\def\cJ{{\cal J}} \def\cK{{\cal K}}
\def\cL{{\cal L}} \def\cM{{\cal M}}
\def\cN{{\cal N}} \def\cO{{\cal O}}
\def\cP{{\cal P}} \def\cQ{{\cal Q}}
\def\cR{{\cal R}} \def\cV{{\cal V}}\def\cW{{\cal W}}
%
%
%
\def\crr{\crcr\noalign{\vskip {8.3333pt}}}
\def\tilde{\widetilde}
\def\bar{\overline}
\def\us#1{\underline{#1}}
\let\shat=\hat
\def\hat{\widehat}
\def\hyp{\vrule height 2.3pt width 2.5pt depth -1.5pt}
\def\square{\mbox{.08}{.08}}
\def\Coeff#1#2{{#1\over #2}}
\def\Coe#1.#2.{{#1\over #2}}
\def\coeff#1#2{\relax{\textstyle {#1 \over #2}}\displaystyle}
\def\coe#1.#2.{\relax{\textstyle {#1 \over #2}}\displaystyle}
\def\half{{1 \over 2}}
\def\shalf{\relax{\textstyle {1 \over 2}}\displaystyle}
\def\dag#1{#1\!\!\!/\,\,\,}
\def\to{\rightarrow}
\def\notin{\hbox{{$\in$}\kern-.51em\hbox{/}}}
\def\shdot{\!\cdot\!}
\def\ket#1{\,\big|\,#1\,\big>\,}
\def\bra#1{\,\big<\,#1\,\big|\,}
\def\equaltop#1{\mathrel{\mathop=^{#1}}}
\def\Trbel#1{\mathop{{\rm Tr}}_{#1}}
\def\inserteq#1{\noalign{\vskip-.2truecm\hbox{#1\hfil}
\vskip-.2cm}}
\def\attac#1{\Bigl\vert
{\phantom{X}\atop{{\rm\scriptstyle #1}}\phantom{X}}}
\def\exx#1{e^{{\displaystyle #1}}}
\def\del{\partial}
\def\delbar{\bar\partial}
\def\nex#1{$N\!=\!#1$}
\def\dex#1{$d\!=\!#1$}
\def\cex#1{$c\!=\!#1$}
\def\eg{{\it e.g.}} \def\ie{{\it i.e.}}
%
\def\cS{{\cal K}}
\def\IE{\relax{{\rm I\kern-.18em E}}}
\def\cE{{\cal E}}
\def\rt{{\cR^{(3)}}}
\def\IGam{\relax{{\rm I}\kern-.18em \Gamma}}
\def\IGa{\IA}
\def\LG{Lan\-dau-Ginz\-burg\ }
\def\cV{{\cal V}}
\def\Rt{{\cal R}^{(3)}}
\def\wabc{W_{abc}}
\def\WABC{W_{\a\b\c}}
\def\W{{\cal W}}
\def\tft#1{\langle\langle\,#1\,\rangle\rangle}
\def\IA{\relax{\hbox{{\rm A}\kern-.82em {\rm A}}}}
\let\picfuc=\fp
\def\hata{{\shat\a}}
\def\hatb{{\shat\b}}
\def\hatA{{\shat A}}
\def\hatB{{\shat B}}
\def\bv{{\bf V}}
\def\spg{special geometry}
\def\sc{SCFT}
\def\leel{low energy effective Lagrangian}
\def\pf{Picard--Fuchs}
\def\pfS{Picard--Fuchs system}
\def\el{effective Lagrangian}
\def\Fb{\overline{F}}
\def\nablab{\overline{\nabla}}
\def\Ub{\overline{U}}
\def\Db{\overline{D}}
\def\zb{\overline{z}}
\def\eb{\overline{e}}
\def\fb{\overline{f}}
\def\tb{\overline{t}}
\def\Xb{\overline{X}}
\def\Vb{\overline{V}}
\def\Cb{\overline{C}}
\def\Sb{\overline{S}}
\def\delb{\overline{\del}}
\def\Gammab{\overline{\Gamma}}
\def\Ab{\overline{A}}
\def\Anh{A^{\rm nh}}
\def\alphab{\bar{\alpha}}
\def\cy{Calabi--Yau}
\def\cabg{C_{\alpha\beta\gamma}}
\def\B{\Sigma}
\def\Bh{\hat \Sigma}
\def\Kh{\hat{K}}
\def\Knh{{\cal K}}
\def\A{\Lambda}
\def\Ah{\hat \Lambda}
\def\R{\hat{R}}
\def\V{{V}}
\def\T{T}
\def\Gammah{\hat{\Gamma}}
\def\twot{$(2,2)$}
\def\K{K\"ahler}
\def\rat{({\theta_2 \over \theta_1})}
\def\lv{{\bf \omega}}
\def\w{w}
\def\CP{C\!P}
\def\o#1#2{{{#1}\over{#2}}}
\newcommand{\be}{\begin{equation}}
\newcommand{\ee}{\end{equation}}
\newcommand{\ba}{\begin{eqnarray}}
\newcommand{\ea}{\end{eqnarray}}
\newtheorem{definizione}{Definition}[section]
\newcommand{\bd}{\begin{definizione}}
\newcommand{\ed}{\end{definizione}}
\newtheorem{teorema}{Theorem}[section]
\newcommand{\bth}{\begin{teorema}}
\newcommand{\eth}{\end{teorema}}
\newtheorem{lemma}{Lemma}[section]
\newcommand{\blem}{\begin{lemma}}
\newcommand{\elem}{\end{lemma}}
\newcommand{\brr}{\begin{array}}
\newcommand{\err}{\end{array}}
\newcommand{\nn}{\nonumber}
\newtheorem{corollario}{Corollary}[section]
\newcommand{\bcorol}{\begin{corollario}}
\newcommand{\ecorol}{\end{corollario}}
\def\twomat#1#2#3#4{\left(\begin{array}{cc}
 {#1}&{#2}\\ {#3}&{#4}\\
\end{array}
\right)}
\def\twovec#1#2{\left(\begin{array}{c}
{#1}\\ {#2}\\
\end{array}
\right)}
\begin{titlepage}
\hskip 5.5cm
\vbox{
\hbox{CERN-TH/96-348}
}
\hskip 1.5cm
\vbox{\hbox{hep-th/9612105}\hbox{December, 1996}}
\vfill
\vskip 3cm
\begin{center}
{\LARGE {  U--Duality and Central Charges in Various Dimensions Revisited }}\\
\vskip 1.5cm
  {\bf Laura Andrianopoli$^1$,
Riccardo D'Auria$^2$ and
Sergio Ferrara$^3$ } \\
\vskip 0.5cm
{\small
$^1$ Dipartimento di Fisica, Universit\'a di Genova, via Dodecaneso 33,
I-16146 Genova\\
and Istituto Nazionale di Fisica Nucleare (INFN) - Sezione di Genova, Italy\\
\vspace{6pt}
$^2$ Dipartimento di Fisica, Politecnico di Torino,\\
 Corso Duca degli Abruzzi 24, I-10129 Torino\\
and Istituto Nazionale di Fisica Nucleare (INFN) - Sezione di Torino, Italy\\
\vspace{6pt}
$^3$ CERN Theoretical Division, CH 1211 Geneva 23, Switzerland}
\end{center}
\vskip 3cm
\begin{center}
DEDICATED TO THE MEMORY OF ABDUS SALAM
\end{center}
\vfill
\begin{center} {\bf Abstract}
\end{center}
{\small
A geometric formulation which describes extended supergravities in any dimension in presence
of electric and magnetic sources is presented. In this framework the underlying duality
symmetries of the theories are  manifest. Particular emphasis is given to the construction of
central and matter charges and to the symplectic structure of all
 $D=4$, $N$--extended theories.
The latter may be traced back to the existence, for $N>2$, of a flat symplectic bundle
which is the $N>2$ generalization of $N=2$ Special Geometry.
}
\vspace{2mm} \vfill \hrule width 3.cm
{\footnotesize
\noindent
$^*$ Work supported in part by EEC under TMR contract ERBFMRX-CT96-0045
 (LNF Frascati, Politecnico di Torino and Univ. Genova) and by DOE grant
DE-FGO3-91ER40662.}
\end{titlepage}
 \section{Introduction}
Recent developments on duality symmetries \cite{dual} in supersymmetric quantum
theories of fields and strings seem to indicate that the known
different string theories are different manifestations, in different
regions of the coupling constant space, of a unique more fundamental
theory that, depending on the regime and on the particular
compactification, may itself reveal extra (11 or 12) dimensions \cite{witten}, \cite{vafa}.
A basic aspect that allows a comparison of  different theories is their
number of supersymmetries and their spectrum of massless and massive
BPS states.
Indeed, to explore a theory in the nonperturbative regime, the
power of supersymmetry allows one to compute to a large extent
all dynamical details encoded in the low energy effective action of
a given formulation of the theory and to study the moduli (coupling
constants) dependence of the BPS states.
This latter property is important in order to study more dynamical
questions such as phase transitions in the moduli space \cite{huto}, \cite{wi},
 \cite{phtrans},\cite{ssv}, \cite{mova}, \cite{wit3} or properties
of solitonic solutions of cosmological interest, such as extreme
black holes \cite{malda} and their entropy.
A major mathematical tool in these studies is the structure of
supergravity theories in diverse dimensions \cite{sase} and with different
numbers of supersymmetries.
These theories have a central extension that gives an apparent
violation of the Haag-Lopuszanski-Sohnius theorem \cite{hls}, since they include
``central charges'' that are not Lorentz-invariant \cite{tow}.
However, these charges are important because they are related to
$p$-extended objects (for charges with $p$ antisymmetrized indices)
whose dynamics is now believed to be as fundamental as that of points
and strings \cite{dual}.
In fact point-like and string-like BPS states can be obtained by
wrapping $p$ (or $p - 1$) of the dimensions of a $p$-extended object
living in $D$ dimensions when $d\geq p$ dimensions have been
compactified.
It is the aim of this paper to give a detailed analysis of central
extensions of different supergravities existing in arbitrary
dimensions $4\leq D < 10$ in a unified framework and to study the moduli dependence of the
BPS mass per unit of $p$-volume of generic BPS $p$-branes existing in
a given theory.
A basic tool in our investigation will be an exploitation of ``duality symmetries''\cite{cfs}, \cite{fsz} (rephrased nowadays
as U--duality) of the underlying supergravity theory which, for a theory with more than 8
supercharges, takes the form of a discrete subgroup of the continuous isometries of the scalar
field sigma model of the theory \cite{ht}.
Duality symmetries, which rotate electric and magnetic charges, correspond, in a string context,
 to certain perturbative or nonperturbatives symmetries of the BPS spectrum, playing a crucial role in the study
of string dynamics.
In many respects the present investigation can be considered as a completion
 of the list of theories reported in the collected volume
papers by A. Salam and E. Sezgin \cite{sase}.
This paper will cover, for self-consistency, material covered but
scattered in the literature, and then some new material, such as the
details of some theories in $D=6$ and $D=8,9$ dimensions.
The basic focus of our approach is that the central extension of the
supersymmetry algebra \cite{hls}\cite{fesaz} is encoded in the supergravity transformation
rules.
The latter can be derived from supersymmetric Bianchi identities,
even if the complete lagrangian has not yet been derived.
Of course a careful study of these identities also allows a complete
determination of the lagrangian, whenever it exists.
Among the novelties of this analysis is a new formulation of $D=4$, $N$--extended
theories with $N > 2$, in which a manifest symplectic formulation is used.
In particular all these theories have in common a flat symplectic bundle which encodes
the differential relations among the symplectic sections and therefore among the central
and matter charges.
In this respect the $N=2$ case, related to Special Geometry \cite{str}, \cite{voi}, simply differs from the
$N>2$ cases by the fact that the base space is not necessarily a coset space. This is related to
the physical fact that $N=2$ Special Geometry suffers quantum corrections.
For higher dimensional theories, relations between central and matter charges for different
$p$--extended objects are derived in analogy with previous known results in $D=4$ and $D=5$.
For $D>4$ the embedding in a duality group that rotates electric into magnetic charges is only possible for $D/2
= p + 2$, which covers the cases of $D=6,8$.
\par
A shorter version of this paper, with particular emphasis on the
existence of duality invariant entropy formulae in higher dimensions,  already
appeared in the literature \cite{ultimo}.
\par
The paper is organized as follows:
\\
In section 2 we recall how different supergravity theories are related to the
low-energy limit of different string theories and their M-theory or
F-theory extension.
\\
In section 3, which is the main body of the paper, a  general discussion of the
geometric framework for all $D$ and $N$ is presented and, in particular, the symplectic
embedding of all $D=4$ theories is formulated.
The subsequent sections, which may be skipped by a reader not interested in the details
of a particular theory, consider the higher $D$
 and higher $N$ theories in the formalism discussed in section $3$.
In section 4 the $D=4$, $N>2$ theories are presented.
In section 5 matter coupled $D>4$ supergravities are discussed.
In sections 6 and 7 the maximally extended theories in odd and even dimensions
respectively are reported.
\section{Extended supergravities and their relations with superstrings,
M-theory and F-theory}
It is worth while to recall the various compactifications of
superstrings in $4 \leq D < 10$ as well as of M-theory and their
relation to extended supergravities and their duality symmetries.
In the string context the latter symmetries are usually called S, T and U--dualities.
S--duality means exchange of small with large coupling constant, i.e. strong--weak coupling duality.
T--duality indicates the exchange of small with large volume of compactification while U--duality refers
to the exchange of NS with RR scalars.
The major virtue of space--time supersymmetry is that it links toghether these dualities; often
some of them are interchanged in comparing dual theories in the nonperturbative
regime.
In this paper we will only consider compactifications on smooth
manifolds since the analysis is otherwise more complicated (and
richer) due to additional states concentrated at the singular points
of the moduli space.
The key ingredient to compare different theories in a given
space-time dimension is Poincar\'e duality, which converts a theory
with a $(p+ 2)$-form into one with a $(D-p-2)$-form (and inverse
coupling constant).
For example, at $D=9$ Poincar\'e duality relates 4- and 5-forms, at
$D=7$ \cite{witten} 3- and 4-forms and at $D=5$ 2- and 3-forms  \cite{schw}.
These relations are closely related to the fact that type IIA and
type IIB are T-dual at $D=9$ \cite{dhs}, heterotic on $T_3$ is dual to M-theory
on $K3$ at $D=7$ \cite{witten} and heterotic on $K3 \times S_1$ is dual to M-theory
on $CY_3$ at $D=5$ \cite{cad}, \cite{wit3}.
Let us consider dualities by first comparing theories with maximal
supersymmetry (32 supersymmetries).
An example is the duality between M-theory on $S_1$ at large radius and
type IIA in $D=10$ at strong coupling.
For the sequel we will omit the regime where these theories should be
compared.
We will just identify their low-energy effective action
including BPS states.
Further compactifying type IIA on $S_1$, it becomes equivalent to IIB
on $S_1$ with inverse radius.
This is the T-duality alluded to before.
It merely comes by Poincar\'e duality, exchanging the five form of
one theory (IIB) with the 4-form of the other theory (IIA).
The interrelation between M-theory and type IIA and type IIB theories
at $D \leq 9$ explains most of the symmetries of all maximally
extended supergravities.
At $D=8$ we have a maximal theory with U-duality \cite{ht} group $Sl(3,\ZZ)
\times Sl(2,\ZZ)$.
The  $Sl(3,\ZZ)$ has a natural interpretation from an
M-theory point of view, since the $D=8$ theory is M-theory on $T_3$.
On the other hand, the additional $Sl(2,\ZZ) $ which acts on the
4-form and its dual has a natural interpretation from the type IIB
theory on $T_2$, in which the $Sl(2,\ZZ)$ is related to the complex
structure of the 2-torus \cite{kuva}.
At $D=7$ the $ Sl(5,\ZZ) $ U-duality has no obvious interpretation unless we move
to an F--theory setting \cite{kuva}
This is also the case for $D=6,5,4$, where the U-duality groups are
$O(5,5;\ZZ)$, $E_{6,6}(\ZZ)$ and  $E_{7,7}(\ZZ)$  respectively.
However, they share the property that the related continuous group has, as maximal compact
subgroup, the automorphism group of the supersymmetry algebra, i.e.
$Sp(4)$, $Sp(4) \times Sp(4) $, $Usp(8)$ and $SU(8)$ respectively for
$D=7,6,5$ and 4.
The U--duality group for any $D$ corresponds to the series of $E_{11-D}$ Lie algebras
whose quotient with the above automorphism group of the supersymmetry algebra
provides the local description of the scalar fields moduli space
 \cite{cj}.
Recently, a novel way to unravel the structure of the U--duality groups
 in terms of solvable Lie algebras has been proposed in   \cite{solv}.
Moving to theories with lower (16) supersymmetries, we start to have
dualities among heterotic, M-theory and type II theories on manifolds
preserving 16 supersymmetries.
For $D=7$, heterotic theory on $T_3$ is ``dual'' to M-theory on $K3$
in the same sense that M-theory is ``dual'' to type IIA at $D=10$.
Here the coset space  $O(1,1) \times {O(3,19) \over O(3)\times O(19)}$
identifies the dilaton and Narain lattice of the heterotic string
with the classical moduli space of $K3$, the dilaton in one theory
being related to the volume of $K3$ of the other theory \cite{witten}.
The heterotic string on $T_4$ is dual to type IIA on $K3$.
Here the coset space $O(1,1) \times {O(4,20) \over O(4)\times O(20)}$
identifies the Narain lattice with the ``quantum'' moduli space of $K3$
(including torsion).
The $O(1,1)$ factor again relates the dilaton to the $K3$ volume.
A similar situation occurs for the theories at $D=5$.
At $D=4$ a new phenomenon occurs since the classical moduli space
${SU(1,1) \over U(1)}\times {O(6,22) \over O(6)\times O(22)}$
interchanges S-duality of heterotic string with T-duality of type IIA
theory and U-duality of type IIB theory  \cite{dlr}.
If we compare theories with 8 supersymmetries, we may at most start
with $D=6$.
On the heterotic side this would correspond to $K3$ compactification.
However at $D=6$ no M-theory or type II correspondence is possible
because we have no smooth manifolds of dimension 5 or 4 which
reduce the original supersymmetry (32) by one quarter.
The least we can do is to compare theories at $D=5$, where heterotic
theory on $K3 \times S_1$ can be compared \cite{cad} and in fact is dual, to
M-theory on a Calabi-Yau threefold which is a $K3$ fibration \cite{klema}.
Finally, at $D=4$ the heterotic string on $K3 \times T_2$ is dual to type
IIA (or IIB) on a Calabi-Yau threefold (or its mirror) \cite{cdfv}, \cite{fehastva}, \cite{kava}.
It is worth noticing that these ``dualities'' predict new BPS states as
well as they identify perturbative BPS states of one theory with
non-perturbative ones in the dual theory.
A more striking correspondence is possible if we further assume the
existence of 12 dimensional F-theory such that its compactification
on $T_2$ gives type IIB at $D=10$  \cite{vafa}.
In this case we can relate the heterotic string on $T_2$ at $D=8$ to
F-theory on $K3$ and the heterotic string at $D=6$ on $K3$ with
F-theory on a Calabi-Yau threefold  \cite{mova}, \cite{femisa}.
To make these comparisons one has to further assume that the smooth
manifolds of F-theory are elliptically fibered \cite{mova}.
An even larger correspondence arises if we also include type I strings \cite{type1}
and D-branes \cite{dbrane} in the game.
However we will not further comment on the other correspondences
relating all string theories with M and F theory.

\section{Duality symmetries and central charges in diverse dimensions}
\subsection{The general framework}
 All supergravity
theories contain scalar fields whose kinetic Lagrangian is described
by $\sigma$--models of the form $G/H$, with the exception of $D=4$, $N=1,2$ and $D=5$, $ N=2$.\\
Here $G$ is a non compact group acting as an isometry group on the
scalar manifold while $H$, the isotropy subgroup, is of the form:
\begin{equation}
H=H_{Aut} \otimes H_{matter}
\end{equation}
 $H_{Aut}$ being the automorphism group of the supersymmetry algebra
 while $H_{matter}$ is related to the matter multiplets.
 (Of course $H_{matter}=\bfone$ in all cases where supersymmetric matter
 doesn't exist, namely $N>4$ in $D=4,5$ and in general in all
 maximally extended supergravities).
The coset manifolds $G/H$ and the automorphism groups for various
 supergravity theories for any $D$ and $N$  can be found in the literature
 (see for instance
\cite{sase}, \cite{bibbia}).
 As it is well known, the group G acts linearly on the ($n=p+2$)--forms
 field strengths $H^\Lambda _{a_1\cdots a_n}$ corresponding to the
 various ($p+1$)--forms appearing in the gravitational and matter
 multiplets. Here and in the following the index $\Lambda$ runs over
 the dimensions of some representation of the duality group $G$.
The true duality symmetry (U--duality), acting on integral  quantized electric
and magnetics charges,
 is  the restriction of  the continuous group $G$ to the integers
 \cite{ht}. The moduli space of these theories is $ G(\ZZ)\backslash G/H$.
 \par
 All the properties of the given supergravity theories for fixed $D$
 and $N$ are completely fixed in terms of the geometry of $G/H$,
 namely in terms of the coset representatives $L$ satisfying the
 relation:
 \begin{equation}
 L(\phi^\prime) =g L(\phi) h (g,\phi)
\end{equation}
where $g\in G$, $h\in H$  and $\phi ^\prime =   \phi ^\prime
 (\phi)$,
 $\phi$ being the coordinates of $G/H$.
Note that the  scalar fields in $G/H$ can be assigned, in the linearized theory,  to linear representations
 $R_H$ of the local isotropy group  $H$ so that dim $R_H$ = dim $G$ $-$ dim $H$ (in the full theory, $R_H $ is the representation
 which the vielbein of $G/H$ belongs to).
\par
 As explained in the following, the kinetic metric for
 the ($p+2$)--forms $H^\Lambda$ is fixed in terms of $L$ and the
 physical field strengths of the interacting theories are "dressed"
 with scalar fields in terms of the coset representatives.
 This allows us to write down the central charges associated to the
 ($p+1$)--forms in the gravitational multiplet in a neat way in terms
 of the geometrical structure of the moduli space.
In an analogous way also the matter ($p+1$)--forms of the matter
multiplets give rise to charges which, as we will see, are closely
related to the central charges. Note that when $p>1$ the central
charges do not appear in the usual supersymmetry algebra, but in the
extended version of it containing central generators $Z_{a_1 \cdots
a_p}$ associated to $p$--dimensional extended objects ($a_1 \cdots
a_p$ are a set of space--time antisymmetric
Lorentz indices) \cite{df, vpvh, tow,
ach, bars}
\par
Our main goal is to write down the explicit form of the dressed
charges and to find relations among them analogous to those worked
out in $D=4$, $N=2$  by means of the Special Geometry relations \cite{cdf}\cite{cdfv}.
\par
To any ($p+2$)--form $H^\Lambda$ we may associate a magnetic charge (($D-p-4$)--brane)
and  an
electric ($p$--brane) charge given respectively by:
\begin{equation}
g^\Lambda = \int _{S^{p+2}} H^\Lambda
\qquad \qquad
e_\Lambda = \int _ {S^{D-p-2}}  \cG _\Lambda
\end{equation}
where $\cG_{\Lambda}= {\partial \cL \over \partial H^\Lambda}$.
\par
These charges however are not the physical charges of the interacting
theory; the latter ones can be computed by looking at the
transformation laws of the fermion fields, where the physical
field--strengths appear dressed with the scalar fields \cite{ultimo}.
Let us first introduce the central charges:
they are associated to the dressed ($p+2$)--forms $T^i_{AB}$ appearing
in the supersymmetry transformation law of the gravitino 1-form.
Quite generally we have, for any $D$ and $N$:
\begin{equation}
\delta \psi_A = D\epsilon_A + \sum_{i} c_iT^i _ {AB\vert a_1\cdots a_{n_i}}\Delta^{a a_1\cdots a_{n_i}}
\epsilon^B V_a+ \cdots
\label{tragra}
\end{equation}
where:
 \begin{equation}
\Delta_{a a_1\cdots a_n}=\left( \Gamma _{a a_1 \cdots
a_{n}} - {n \over n-1} (D-n-1)\delta^a_{[a_1} \Gamma_{a_2\cdots
a_{n}]} \right).
\label{delta}
\end{equation}
Here $D$ is the covariant derivative in terms of the space--time spin connection
and the composite connection of the automorphism group $H_{Aut}$,
 $c_i$ are coefficients fixed by supersymmetry, $V^a$ is the
 space--time vielbein, $A=1,\cdots,N$ is the index acted on by the
 automorphism group, $\Gamma_{a_1\cdots a_n}$ are $\gamma$--matrices
 in the appropriate dimensions, and the sum runs over all the ($p+2$)--forms
 appearing in the gravitational multiplet. Here and in the following
 the dots denote trilinear fermion terms.
Each $n$-form field-strength $T^i_{AB}$ is constructed by dressing the bare
field-strengths  $H^\Lambda$ with the coset representative $L(\phi)$ of $G/H$,
$\phi$ denoting a set of coordinates of $G/H$.
In particular, for any $p$, except for $D/2 = p+2$, we have:
 \begin{equation}
T^i_{AB} = L_{AB \Lambda_i} (\phi)
H^{\Lambda_i}
\end{equation}
where we have used the following decomposition of $L$:
\begin{equation}
L=(L^\Lambda_{AB}, L^\Lambda_I) \quad\quad L^{-1}=(L^{AB}_{\ \Lambda}, L^I_{\ \Lambda})
\label{defl}
\end{equation}
Here $L^\Lambda_\Sigma$ belongs to the representation of $G$ under which the ($p+2$)--forms $H^\Lambda$
transform irreducibly and the couple of indices $AB$ and $I$ refer to the
transformation properties of $L$  under the right action of $H_{Aut} \times H_{matter}$.
More precisely, the couple of indices $AB$ transform in the twofold tensor representation of $H_{Aut}$, which
in general is a  $Usp(N)$
group (except in  $D=8$ and $D=9$ theories where $H_{Aut}$ is $SU(N)\times U(1)$
or $O(N)$ respectively).
and $I$ is an index in the fundamental representation of $H_{matter}$ which in general is an orthogonal group.
Note that in absence of matter multiplets
$L\equiv ( L^\Lambda_{\ AB})$.
In all these cases ($D/2 \neq p+2$) the kinetic matrix of the ($p+2$)--forms $H^\Lambda$
is given in terms of the coset representatives as follows:
\begin{equation}
{1 \over 2} L_{AB\Lambda}L^{AB}_{\ \ \Sigma} - L_{I\Lambda}L^{I}_{\ \Sigma}=
\cN_{\Lambda\Sigma}\label{ndil}
\end{equation}
 with the indices of $H_{Aut}$  raised
 and lowered with the appropriate
metric of $H_{Aut}$  in the given representation.
For maximally extended supergravities
 $\, \cN_{\Lambda\Sigma} = L_{AB\Lambda}L^{AB}_{\ \ \Sigma}$.
Note that both for matter coupled and maximally extended
supergravities we have:
\begin{equation}
L_{\Lambda AB} = \cN_{\Lambda\Sigma}L^{\Sigma}_{\ AB}
\label{invl}
\end{equation}
When $G$ contains an orthogonal factor $O(m,n)$, what happens
 for matter coupled supergravities in $D=5,7,8,9$, where $G=O(10-D,n)  \times O(1,1)$
 and in all the matter coupled $D=6$ theories,
 the coset
representatives of the orthogonal group satisfy:
\begin{eqnarray}
L^t\eta L = \eta &\to &  L_{r\Lambda}L_{r \Sigma} -  L_{I\Lambda}L_{I \Sigma}=
\eta_{\Lambda\Sigma} \label{etadil}\\
L^t L = \cN &\to & L_{r \Lambda}L_{r \Sigma} +  L_{I\Lambda}L_{I \Sigma}=
\cN_{\Lambda\Sigma}
\end{eqnarray}
 where $\eta=\pmatrix{\bfone_{m\times m} & 0 \cr 0 & -\bfone_{n\times n}\cr}$
 is the $O(m,n)$ invariant metric  and $A=1,\cdots,m$; $I = 1,\cdots,n$
(In particular,  setting the matter to zero, we have in these cases
$\cN_{\Lambda\Sigma}= \eta_{\Lambda\Sigma}$).\\
In these cases we have:
\begin{equation}
 L^\Lambda_{\ AB}= L^\Lambda_{\ r} (\gamma^r)_{AB},
\end{equation}
$(\gamma^r)_{AB}$ being the $\gamma$-matrices intertwining between orthogonal and
$USp(N)$ indices.
\par
When $D$ is even and $D/2 = p+2$ the previous formulae in general require modifications,
since in that case we have the
 complication  that the action of $G$ on the $p+2 =D/2$--forms ($D$ even) is
 realized through the embedding of $G$ in $Sp(2n,\IR)$ ($p$ even) or $O(n,n)$
($p$ odd)
 groups \cite{gz}, \cite{cfg}.\\
This happens for $D=4$, $N>1$, $D=6$ , $N=(2,2)$ and the maximally extended $D=6$ and $D=8$ supergravities.
\footnote{The 6 dimensional theories $N=(2,0)$ and $N=(4,0)$ do not require such embedding
since the 3--forms $H^\Lambda$ have definite self--duality and no lagrangian exists.
For these theories the formulae of the previous odd dimensional cases are valid.}
(The necessary modifications for the embedding are worked out in section $3.2$.)
\par
Coming  back to the case $D/2 \neq p+2$, it is now straightforward to compute  the central charges.
\par
Indeed, the magnetic central charges for BPS saturated ($D-p-4$)--branes
can be now defined
(modulo numerical factors to be fixed in each theory) by integration
of the dressed
field strengths as follows:
\begin{equation}
Z^{(i)}_{(m) AB} = \int _{S^{p+2}}T^i_{ AB}=\int _{S^{p+2}}L_{\Lambda_i AB}(\phi) H^{\Lambda_i}=
 L_{\Lambda_i AB}(\phi_0) g^{\Lambda_i}
 \label{carma}
\end{equation}
where $\phi_0$ denote the $v.e.v.$ of the scalar fields, namely
$\phi_0 = \phi(\infty)$ in a given background.
The corresponding electric central charges are:
 \begin{equation}
Z^{(i)}_{(e) AB} = \int _{S^{D-p-2}}L_{AB \Lambda_i}(\phi)
^{\ \star}H^{\Lambda_i}= \int _{S^{D-p-2}} \cN _{\Lambda_i\Sigma_i}
L^{\Lambda_i}_{\  AB}(\phi)
^{\ \star}H^{\Sigma_i}=
 L^{\Lambda_i}_{\ AB}(\phi_0) e_{\Lambda_i}
\end{equation}
 These formulae make it explicit that $L^\Lambda_{\ AB}$ and
 $L_{\Lambda AB}$ are related by electric--magnetic duality via the
 kinetic matrix.
 \par
 Note that the same field strengths $T^i_{AB}$ which appear in the gravitino
 transformation laws are also present in the dilatino transformation laws
  in the following way:
  \begin{equation}
\delta \chi_{ABC} = \cdots +
\sum_{i} b_i L_{\Lambda_i AB} (\phi)
H^{\Lambda_i} _ {a_1\cdots a_{n_i}}\Gamma^{ a_1\cdots a_{n_i}} \epsilon_C+\cdots
\label{tradil}
\end{equation}
\vskip5mm
\par
 In an analogous way, when vector multiplets are present,
 the matter vector field
 strengths are dressed with the columns $L_{\Lambda I}$
 of the coset element (\ref{defl})
 and they
 appear in the transformation laws of the gaugino fields:
  \begin{equation}
\delta \lambda^I_{A} = c_1 \Gamma^a P^I_{AB , i}
\partial_a \phi^i \epsilon^B +
 c_2 L_{\Lambda}^{\ I} (\phi)
F^{\Lambda} _ {ab}\Gamma^{ ab} \epsilon_A  + \cdots
\label{tragau}
\end{equation}
  where  $ P^I_{AB }=P^I_{AB,i }d\phi^i$ (see eq. (\ref{dllp}) in the following)
 is the vielbein of the coset manifold
  spanned by the scalar fields of the vector multiplets,
 $F^\Lambda_{ab}$ is the field--strength of the matter photons
 and $c_1, c_2$  are  constants fixed by supersymmetry (in $D=6$, $N=(2,0)$ and $N=(4,0)$ the 2--form
 $F^{\Lambda}_{ab} \Gamma^{ab}$ is replaced by the 3--form
  $H^{\Lambda}_{abc} \Gamma^{abc}$).
In the same way as for central charges, one finds the
magnetic matter charges:
\begin{equation}
  Z_{(m)} ^{\ I} = \int _{S^{p+2}} L_\Lambda^{\ I} F^\Lambda
   =   L_\Lambda^{\ I} (\phi_0) g^\Lambda
\end{equation}
while the electric matter charges are:
\begin{equation}
Z_{(e) I} = \int _{S^{D-p-2}}L_{\Lambda I}(\phi)
^{\ \star}F^{\Lambda}= \int _{S^{D-p-2}} \cN _{\Lambda\Sigma}
L^{\Lambda}_{\  I}(\phi)
^{\ \star}F^\Sigma =
 L^{\Lambda}_{\ I}(\phi_0) e_{\Lambda}
\end{equation}
 \par
 The important fact to note is that the central charges and matter
 charges satisfy relations and sum rules analogous to those derived
 in $D=4$, $N=2$ using Special Geometry techniques \cite{cdf}.
They are inherited from the
 properties of the coset manifolds $G/H$, namely from the
 differential and algebraic properties satisfied by the coset
 representatives $L^\Lambda_{\ \Sigma}$.
Indeed, for a general coset manifold we may introduce the
left--invariant 1--form $\Omega=L^{-1} d L$ satisfying the
relation (see for instance \cite{bibbia}):
\begin{equation}
d \Omega + \Omega \wedge \Omega =0
\label{mc}
\end{equation}
 where \begin{equation}
\Omega=\omega^i T_i + P^\alpha T_\alpha
\label{defomega}
\end{equation}
  $T_i, T_\alpha$ being the generators of $G$ belonging respectively to the Lie
  subalgebra $\IH$ and to the coset space algebra $\IK$ in the Cartan
  decomposition
  \begin{equation}
\IG = \IH + \IK
\label{hk}
\end{equation}
 $\IG$ being the Lie algebra of $G$. Here $\omega^i$ is the $\IH$
 connection and $P^\alpha$, in the representation $R_H$ of $H$,  is the vielbein of $G/H$.
Since in all the cases we will consider $G/H$ is a symmetric space
($[\IK , \IK ] \subset \IH$), $\omega^i C_i^{\ \alpha\beta}$
($C_i^{\ \alpha\beta}$ being the structure constants of $G$)
can be identified
with the Riemannian spin connection of $G/H$.
\par
Suppose now we have a  matter coupled theory. Then, using
the decomposition (\ref{hk}), from
(\ref{mc}) and (\ref{defomega}) we get:
\begin{eqnarray}
dL^\Lambda_{\ AB} &=& {1 \over 2}L^\Lambda_{\ CD} \omega^{CD}_{\ \ AB} +
L^\Lambda_{\ I} P^I_{AB} \nonumber\\
dL^\Lambda_{\ I}&=& L^\Lambda_{\ J}\omega^{J}_{\ I} +L^\Lambda_{\ AB} P_I^{AB}
\label{cosetmc}
\end{eqnarray}
where $P^I_{AB}$ is the vielbein on $G/H$ and
 $\omega^{CD}_{\ \ AB}$ and $\omega^J_{\ I}$ are the $\IH_{AUT}$ and $\IH_{matter}$ connections
 respectively in the given
 representation.
It follows:
\begin{equation}
\nabla^{(H)} L^\Lambda_{\ AB}= L^\Lambda_{\ I} P^I_{AB}
\label{dllp}
\end{equation}
where the derivative is covariant with  respect to the
$\IH$--connection  $\omega^{CD}_{\ \ AB}$.
Using the definition of the magnetic dressed charges given in
(\ref{carma}) we obtain:
\begin{equation}
\nabla^{(H)} Z_{AB}= Z_{I} P^I_{AB}
\label{dz}
\end{equation}
 This is a prototype of the formulae one can derive in the various
 cases for matter coupled supergravities \cite{ultimo}.
 To illustrate one possible application of this kind of formulae let
 us suppose that in a given background preserving some number of
 supersymmetries $Z_I=0$ as a consequence of $\delta\lambda^I_A=0$.
 Then we find:
 \begin{equation}
\nabla^{(H)} Z_{AB}=0 \to d(Z_{AB} \bar Z^{AB} )=0
\end{equation}
 that is the square of the
 central charge reaches an extremum with respect to the
 $v.e.v.$ of the moduli fields. Backgrounds
with such fixed scalars describe the horizon geometry of extremal black holes and behave
as attractor points for the scalar fields evolution in the black hole geometry
 \cite{fk1}.
\par
 For the maximally extended supergravities there are no matter
 field--strengths and the previous differential relations become
differential relations among central charges only.
As an example, let us consider $D=5$, $N=8$ theory. In this case the Maurer--Cartan equations become:
\begin{equation}
dL^{\Lambda}_{\ AB} ={1 \over 2} L^{\Lambda}_{\ CD} \Omega^{CD}_{\ \ AB}+
{1 \over 2}\bar L^{\Lambda CD}  P_{CD AB}
\end{equation}
 where the coset representative is taken in the
$27 \times 27$ fundamental  representation of $E_6$, $\Omega^{CD}_{\ \ AB}=
 2 Q^{[A}_{\ [C}\delta^{B]}_{D]}$, $AB$ is  a couple of antisymmetric symplectic--traceless
$USp(8)$ indices, $Q^A_{\ B}$ is the $USp(8)$ connection and the vielbein $P_{CDAB}$
is antisymmetric, $\IC_{AB}$--traceless and pseudo--real.
Note that $(L^\Lambda_{\ CD})^*= L^ {\Lambda CD}$.
 Therefore we get:
  \begin{equation}
\nabla^{(H)} L^\Lambda_{\ AB}={1 \over 2} \bar L^{\Lambda CD} P_{CDAB}
\label{dllp2}
\end{equation}
   that is:
  \begin{equation}
\nabla^{(H)} Z_{AB}= {1 \over 2}\bar Z^{CD} P_{CD AB}
\label{dz2}
\end{equation}
 This relation implies that the vanishing of a subset of central
 charges forces the vanishing of the covariant derivatives of some
 other subset.
 Typically, this happens in some
 supersymmetry preserving backgrounds where
 the requirement $\delta\chi_{ABC}=0$ corresponds to the vanishing of
 just a subset of central charges.
 Finally, from the coset representatives relations
 (\ref{ndil}) (\ref{etadil}) it
 is immediate to obtain sum rules for the central and matter charges
 which are
the counterpart of those found in $N=2$, $D=4$ case using Special
Geometry \cite{cdf}.
 Indeed, let us suppose e.g. that the group $G$ is
 $G=O(10-D,n)\times O(1,1)$, as it
 happens in general for all the minimally extended supergravities in
 $7 \leq D \leq 9$,  $D=6$ type $IIA$ and
 $D=5$, $N=2$.
  The coset representative is now a tensor product $L \to e^\sigma L$, where
  $e^\sigma$ parametrizes the $O(1,1)$ factor.\\
  We have, from   (\ref{etadil})
 \begin{equation}
L^t \eta L =\eta
\end{equation}
 where $\eta$ is the invariant metric of $O(10-D,n)$ and  from  (\ref{ndil})
 \begin{equation}
e^{-2\sigma}(L^{t} L)_{\Lambda\Sigma} =\cN_{\Lambda\Sigma}.
\end{equation}
 Using eq.s (3.13) and (3.17) one finds:
 \begin{equation}
{1 \over 2} Z_{AB} Z_{AB} - Z_I Z_I = g^\Lambda \eta _{\Lambda\Sigma} g^\Sigma
e^{-2\sigma}
\end{equation}
   \begin{equation}
{1 \over 2} Z_{AB} Z_{AB} + Z_I Z_I = g^\Lambda \cN _{\Lambda\Sigma} g^\Sigma
\end{equation}
 In more general cases analogous relations of the same kind can be
 derived.
 \subsection{The embedding procedure for $D/2 = p+2 $}
In this subsection we work out the modifications to the formalism developed in the previous subsection
in the case $D/2 = p+2$, which derive from the embedding procedure \cite{gz} of the group $G$
in $Sp(2n,\IR)$ ($p$ even) or in $O(n,n)$ ($p$ odd).
We mainly concentrate on $D=4$, while for $D=6,8$ we just outline the procedure, referring for more details to the
next sections.
Furthermore we show that the flat symplectic bundle formalism of the $D=4$, $N=2$ Special Geometry case \cite{str}, \cite{voi}
can be extended to $N>2$ theories.
The $N=2$ case differs from the other higher $N$ extensions by the fact that the base space
of the flat symplectic bundle is not in general a coset manifold.
\par
Let us analyze the structure of the four dimensional theories.
\par
 In $D=4$, $N>2$ we may decompose the vector field--strengths in self--dual and
 anti self--dual parts:
 \begin{equation}
F^{\mp} = {1\over 2}(F\mp {\rm i} ^{\ \star} F)
\end{equation}
  According to the Gaillard--Zumino construction, $G$ acts on the
  vector $(F^{- \Lambda},\cG^{-}_\Lambda)$
  (or its complex conjugate) as a subgroup of
  $Sp(2 n_v,\IR)$ ($n_v$ is the number of vector fields)
with duality transformations interchanging electric and magnetic
 field--strengths:
 \begin{equation}
{\cal S}
\left(\matrix {F^{-\Lambda} \cr
\cG^-_\Lambda\cr}\right)=
\left(\matrix {F^{-\Lambda} \cr
\cG^-_\Lambda\cr}\right)^\prime
\end{equation}
 where:
 \begin{eqnarray}
\cG^-_\Lambda&=&\bar \cN_{\Lambda\Sigma}F^{-\Sigma}\nonumber\\
 \cG^+_\Lambda&=& \cN_{\Lambda\Sigma}F^{+\Sigma}
 \label{defg}
\end{eqnarray}
\begin{equation}
 {\cal S}=\left( \matrix{A& B\cr C & D \cr}\right)\in G \subset Sp(2 n_v,\IR) \to \quad \left\{\matrix{
A^t C -C^t A &=&0 \cr B^t D -D^t B &=& 0 \cr A^t D -C^t B &=&1 } \right.
\label{defs}
\end{equation}
  and $\cN_{\Lambda\Sigma}$, is the symmetric matrix appearing in the kinetic
  part of the vector Lagrangian:
  \begin{equation}
  \cL_{kin}= {\rm i}\bar \cN_{\Lambda\Sigma}F^{-\Lambda} F^{-\Sigma} + h.
  c.
  \end{equation}
If $L(\phi)$ is the coset representative of $G$ in some representation, $S$ represents the embedded coset representative
belonging to $Sp(2n_v,\IR)$ and in each theory, $A,B,C,D$ can be constructed in terms of $L(\phi)$.
Using a complex basis in the vector space of $Sp(2 n_v)$, we may
rewrite the
symplectic matrix as an $Usp(n_v,n_v)$ element:
\begin{equation}
U = {1 \over \sqrt{2}}\pmatrix{f+{\rm i}h & \bar f+{\rm i}\bar h \cr
f-{\rm i}h &\bar f-{\rm i}\bar h \cr} =
  \cA^{-1} S \cA
  \label{defu}
\end{equation}
where:
\begin{eqnarray}
  f&=&{1 \over \sqrt{2}} (A-{\rm i} B) \nonumber\\
h &=&{1 \over \sqrt{2}} (C-{\rm i} D) \nonumber\\
 \cA &=& \pmatrix{1 & 1 \cr -{\rm i} & {\rm i}\cr}
\end{eqnarray}
The requirement $ {U} \in Usp(n_v, n_v)$ implies:
 \begin{equation}
\left\lbrace\matrix{{\rm i}(f^\dagger h - h^\dagger f) &=& \bfone \cr
(f^t  h - h^t f) &=& 0\cr} \right.
\label{specdef}
\end{equation}
The $n_v\times n_v$ subblocks of U are submatrices $f,h$ which can be decomposed with respect to the
isotropy group $H_{Aut} \times H_{matter}$ in the same way as $L$ in equation (\ref{defl}), namely:
\begin{eqnarray}
  f&=& (f^\Lambda_{AB} , f^\Lambda _I) \nonumber\\
h&=& (h_{\Lambda AB} , h_{\Lambda I})
\label{deffh}
\end{eqnarray}
where $AB$ are indices in the antisymmetric representation of $H_{Aut}= SU(N) \times U(1)$ and
$I$ is an index of the fundamental representation of $H_{matter}$.
Upper $SU(N)$ indices label objects in the complex conjugate representation of $SU(N)$:
$(f^\Lambda_{AB})^* = f^{\Lambda AB}$ etc.
\par
Note that we can consider $(f^\Lambda_{AB}, h_{\Lambda AB})$ and $(f^\Lambda_{I}, h_{\Lambda I})$
as symplectic sections of a $Sp(2n_v, \IR)$ bundle over $G/H$.
We will see in the following that this bundle is actually flat.
The real embedding given by $S$ is appropriate for duality transformations  of $F^\pm$
 and their duals $\cG^\pm$, according to equations (\ref{defs}), (\ref{defg}), while
the complex embedding in the matrix $U$ is appropriate in writing down the fermion transformation
laws and supercovariant field--strengths.
The kinetic matrix $\cN$, according to Gaillard--Zumino \cite{gz}, turns out to be:
\begin{equation}
\cN= hf^{-1}, \quad\quad \cN = \cN^t
\label{nfh-1}
\end{equation}
 and transforms projectively under $Sp(2n_v ,\IR)$ duality rotations:
 \begin{equation}
\cN^\prime = (C+ D \cN) (A+B\cN)^{-1}
\end{equation}
  By using (\ref{specdef})and (\ref{nfh-1}) we find that
   \begin{equation}
   (f^t)^{-1} = {\rm i} (\cN - \bar \cN)\bar f
\end{equation}
which is the analogous of equation (\ref{invl}), that is
 \begin{eqnarray}
  f_{AB \Lambda} &\equiv& (f^{-1})_{AB \Lambda} =
{\rm i} (\cN - \bar \cN)_{\Lambda\Sigma}\bar f^\Sigma_{AB}\\
f_{I \Lambda} &\equiv& (f^{-1})_{I \Lambda} =
{\rm i} (\cN - \bar \cN)_{\Lambda\Sigma}\bar f^\Sigma_{I}
\end{eqnarray}
It follows that the dressing factor $(L^\Lambda)^{-1}=(L_{\Lambda AB}, L_{\Lambda I})$
in equation (\ref{tragra}) which was given by the inverse coset representative
in the defining representation of $G$ has to be replaced by the analogous
inverse representative $(f_{\Lambda AB}, f_{\Lambda I})$ when, as in the
present $D=4$ case, we have to embed $G$ in $Sp(2n,\IR)$.
 As a consequence, in the transformation law of gravitino (\ref{tragra}), dilatino (\ref{tradil})
 and gaugino (\ref{tragau})
 we perform the following replacement:
 \begin{equation}
(L_{\Lambda AB}, L_{\Lambda I}) \to  (\bar f_{\Lambda AB}, \bar f_{\Lambda I})
\end{equation}
In particular, the dressed graviphotons and matter self--dual field--strengths take the
symplectic invariant form:
\begin{eqnarray}
T^-_{AB}&=& {\rm i} (\bar f^{-1})_{AB \Lambda}F^{- \Lambda} = f^\Lambda_{AB}(\cN - \bar \cN)_{\Lambda\Sigma}F^{-\Sigma}=
 h_{\Lambda AB} F^{-\Lambda} -f^\Lambda_{AB} \cG^-_\Lambda  \nonumber\\
  T^-_{I}&=&{\rm i} (\bar f^{-1})_{I\Lambda}F^{- \Lambda} =  f^\Lambda_{I}(\cN - \bar \cN)_{\Lambda\Sigma}F^{-\Sigma}=
 h_{\Lambda I} F^{-\Lambda} - f^\Lambda_{I} \cG^-_\Lambda \nonumber\\
\bar T^{+ AB} &=& (T^-_{AB})^* \nonumber\\
\bar T^{+ I} &=& (T^-_{I})^* \label{gravi}
\end{eqnarray}
(Obviously, for $N>4$, $L_{\Lambda I} = f _{\Lambda I}= T_I =0$).
 To construct the dressed charges one integrates
 $T_{AB} = T^+_{AB} + T^ -_{AB}  $ and  (for $N=3, 4$)
 $T_I = T^+_I + T^ -_I $ on a large 2-sphere.
 For this purpose we note that
 \begin{eqnarray}
 T^+_{AB} & = &  h_{\Lambda
 AB}F^{+\Lambda} - f^\Lambda_{AB} \cG_\Lambda^+  =0  \label{tiden0}\\
   T^+_I & = &  h_{\Lambda
 I}F^{+\Lambda}-f^\Lambda_{I} \cG_\Lambda^+   =0 \label{tiden}
\end{eqnarray}
as a consequence of eqs. (\ref{nfh-1}), (\ref{defg}).
Therefore we have:
\begin{eqnarray}
Z_{AB} & = & \int_{S^2} T_{AB} = \int_{S^2} (T^+_{AB} + T^ -_{AB}) = \int_{S^2} T^ -_{AB} =
  h_{\Lambda AB} g^\Lambda- f^\Lambda_{AB} e_\Lambda
\label{zab}\\
Z_I & = & \int_{S^2} T_I = \int_{S^2} (T^+_I + T^ -_I) = \int_{S^2} T^ -_I =
 h_{\Lambda I} g^\Lambda - f^\Lambda_I e_\Lambda   \quad (N\leq 4)
\label{zi}
\end{eqnarray}
where:
\begin{equation}
e_\Lambda = \int_{S^2} \cG_\Lambda , \quad
g^\Lambda = \int_{S^2} F^\Lambda \label{charges}
\end{equation}
and the sections $(f^\Lambda, h_\Lambda)$ on the right hand side now depend on the {\it v.e.v.}'s
of the scalar fields $\phi^i$.
We see that the  central and matter charges are given in this case by  symplectic invariants
and that the presence of dyons in $D=4$ is related to the
 symplectic embedding.
In the case $D/2\neq p+2$, $D$ even, or $D$ odd, we were able to derive the differential relations
(\ref{dz}), (\ref{dz2}) among
the central and matter charges using the Maurer--Cartan equations (\ref{dllp}), (\ref{dllp2}).
The same can be done in the present case using the embedded coset representative $U$.
Indeed, let $\Gamma = U^{-1} dU$ be the $Usp(n_v,n_v)$ Lie algebra left invariant one form
satisfying:
\begin{equation}
  d\Gamma +\Gamma \wedge \Gamma = 0
\label{int}
\end{equation}
In terms of $(f,h)$ $\Gamma$ has the following form:
\begin{equation}
  \label{defgamma}
  \Gamma \equiv U^{-1} dU =
\pmatrix{{\rm i} (f^\dagger dh - h^\dagger df) & {\rm i} (f^\dagger d\bar h - h^\dagger d\bar f) \cr
-{\rm i} (f^t dh - h^t df) & -{\rm i}(f^t d\bar h - h^t d\bar f) \cr}
\equiv
\pmatrix{\Omega^{(H)}& \bar \cP \cr
\cP & \bar \Omega^{(H)} \cr }
\end{equation}
where the $n_v \times n_v$ subblocks $ \Omega^{(H)}$ and  $\cP$
embed the $H$ connection and the vielbein of $G/H$ respectively .
This identification folllows from the Cartan decomposition of the $Usp(n_v,n_v)$ Lie algebra.
Explicitly, if we define the $H_{Aut} \times H_{matter}$--covariant derivative of a vector $V= (V_{AB},V_I)$ as:
\begin{equation}
\nabla V = dV - V\omega , \quad \omega = \pmatrix{\omega^{AB}_{\ \
CD} & 0 \cr 0 & \omega^I_{\ J} \cr }
\label{nablav}
\end{equation}
we have:
\begin{equation}
\Omega^{(H)} = {\rm i} [ f^\dagger (\nabla h + h \omega) - h^\dagger
(\nabla f + f \omega)] = \omega \bfone
\end{equation}
where we have used:
\begin{equation}
\nabla h = \bar \cN \nabla f ; \quad h = \cN f
\end{equation}
and  the fundamental identity (\ref{specdef}).
Furthermore, using the same relations, the embedded vielbein  $\cP$ can be written as follows:
\begin{equation}
\cP = - {\rm i} ( f^t \nabla h - h^t \nabla f) = {\rm i} f^t (\cN -
\bar \cN) \nabla f
\end{equation}
From  (\ref{defu}) and (\ref{defgamma}),  we obtain the $(n_v \times n_v)$ matrix equation:
\begin{eqnarray}
  \nabla(\omega) (f+{\rm i} h) &=& (\bar f + {\rm i} \bar h) \cP \nonumber \\
\nabla(\omega) (f-{\rm i} h) &=& (\bar f - {\rm i} \bar h) \cP
\end{eqnarray}
together with their complex conjugates. Using further the definition (\ref{deffh}) we have:
\begin{eqnarray}
  \nabla(\omega) f^\Lambda_{AB} &=&  \bar f^\Lambda_{I}  P^I_{AB} + {1 \over 2}\bar f^{\Lambda CD} P_{ABCD} \nonumber \\
\nabla(\omega) f^\Lambda_{I} &=&  {1 \over 2} \bar f^{\Lambda AB}  P_{AB I} +\bar f^{\Lambda J}  P_{JI}
 \label{df}
\end{eqnarray}
where we have decomposed the embedded vielbein $\cP$ as follows:
\begin{equation}
  \label{defp}
  \cP = \pmatrix{P_{ABCD} & P_{AB J} \cr P_{I CD} & P_{IJ}\cr }
\end{equation}
 the subblocks being related to the vielbein of $G/H$, $P = L^{-1} \nabla^{(H)} L$,
 written in terms of the indices of $H_{Aut} \times H_{matter}$.
Note that, since $f$ belongs to the unitary matrix $U$, we have:
$(  f^\Lambda _{AB},  f^\Lambda_I)^\star = (\bar f^{\Lambda AB},\bar f^{\Lambda I})$.
Obviously, the same differential relations that we wrote for $f$ hold true for the dual matrix $h$ as well.
\par
Using the definition of the charges (\ref{zab}), (\ref{zi}) we  then get the following differential
relations among charges:
\begin{eqnarray}
  \nabla(\omega) Z_{AB} &=&  \bar Z_{I}  P^I_{AB} +  {1 \over 2} \bar Z^{CD} P_{ABCD} \nonumber \\
\nabla(\omega) Z_{I} &=& {1 \over 2}  \bar Z^{AB}  P_{AB I} + \bar Z_{J}  P^J_{I}
 \label{dz1}
\end{eqnarray}
Depending on the coset manifold, some of the subblocks of (\ref{defp}) can be actually zero.
For example in $N=3$ the vielbein of $G/H = {SU(3,n) \over SU(3)
 \times SU(n)\times U(1)}$ \cite{ccdffm} is $P_{I AB}$ ($AB $ antisymmetric),
 $I=1,\cdots,n;
A,B=1,2,3$ and it turns out
that $P_{ABCD}= P_{IJ} = 0$.
\par
In $N=4$, $G/H= {SU(1,1) \over U(1)} \times { O(6,n)\over O(6) \times O(n)}$ \cite{bekose}, and we have
$P_{ABCD}= \epsilon_{ABCD} P$, $P_{IJ} = \bar P \delta_{IJ}$, where $P$ is the K\"ahlerian vielbein of ${SU(1,1)\over
U(1)}$,  ($A,\cdots , D $ $SU(4)$ indices and $I,J$ $O(n)$ indices)
and $P_{I AB}$ is the vielbein of ${ O(6,n)\over O(6) \times O(n)}$.
\par
For $N>4$ (no matter indices) we have that $\cP $ coincides with the vielbein $P_{ABCD}$ of the relevant $G/H$.
\par
For the purpose of comparison of the previous formalism with the $N=2$ Special Geometry case, it is interesting
to note that, if the connection $\Omega^{(H)}$ and the vielbein $\cP$  are regarded as data of $G/H$,
then the Maurer--Cartan equations (\ref{df}) can be interpreted as an integrable system of
differential equations for a section $V=(V_{AB}, V_I, \bar V^{AB} , \bar V^I)$ of the symplectic fiber bundle
constructed over $G/H$.
Namely the integrable system:
\begin{equation}
 \nabla \pmatrix{V_{AB} \cr V_I \cr \bar V^{AB} \cr \bar V^I } = \pmatrix{0 & 0 & {1 \over 2} P_{ABCD} & P_{AB J} \cr
0 & 0 & {1 \over 2} P_{I CD} & P_{IJ} \cr
{1 \over 2}P^{ABCD} & P^{ AB J} & 0 & 0 \cr {1 \over 2}P^{I CD} & P^{IJ} & 0 & 0 \cr}
\pmatrix{V_{CD} \cr V_J \cr \bar V^{CD} \cr \bar V^J }
 \label{intsys}
\end{equation}
 has $2n$ solutions given by $V = (f^\Lambda_{\ AB},f^\Lambda_{\ I}),( h_{\Lambda AB}, h_{\Lambda_I})$, $\Lambda = 1,\cdots,n$.
The integrability condition (\ref{int}) means that $\Gamma$ is a flat
connection of the symplectic bundle.
In terms of the geometry of $G/H$
this in turn implies that the $\IH$--curvature, and hence the Riemannian
curvature, is constant, being proportional to
the wedge product of two vielbein.
\par
Besides the differential relations (\ref{dz1}), the charges also satisfy sum rules quite analogous
to those found in \cite{cdf} for the $N=2$ Special Geometry case.
\par
The sum rule has the following form:
\begin{equation}
 {1 \over 2} Z_{AB} \bar Z^{AB} + Z_I \bar Z^I = -{1\over 2} P^t \cM (\cN) P
\label{sumrule}
\end{equation}
where $\cM(\cN)$ and $P$ are:
\begin{equation}
\cM = \left( \matrix{ \bfone & - Re \cN \cr 0 &\bfone\cr}\right)
\left( \matrix{ Im \cN & 0 \cr 0 &Im \cN^{-1}\cr}\right)
\left( \matrix{ \bfone & 0 \cr - Re \cN & \bfone \cr}\right)
\label{m+}
\end{equation}
\begin{equation}
P=\left(\matrix{g^\Lambda \cr e_ \Lambda \cr} \right)
\label{eg}
\end{equation}
In order to obtain this result we just need to observe that from the
 fundamental identities (3.40) and  from the definition of the
 kinetic matrix given in (\ref{nhf}) it  follows:
 \begin{eqnarray}
ff^\dagger &= &-{\rm i} \left( \cN - \bar \cN \right)^{-1} \\
hh^\dagger &= &-{\rm i} \left(\bar \cN^{-1} - \cN^{-1} \right)^{-1}\equiv
-{\rm i} \cN \left( \cN - \bar \cN \right) ^{-1}\bar \cN \\
hf^\dagger &= & \cN ff^\dagger  \\
fh^\dagger & = & ff^\dagger \bar \cN
\end{eqnarray}
We note  that the matrix $\cM$ is a symplectic tensor and in this
sense it is quite analogous to the matrix $\cN _{\Lambda\Sigma} $ of
the odd dimensional cases defined in equation (\ref{ndil}).
Indeed, one sees that the matrix $\cM$ can be written as
the symplectic analogous of (\ref{ndil}):
\begin{equation}
\cM(\cN) = \pmatrix{0&\bfone \cr -\bfone & 0} \pmatrix{f \cr h \cr}
\pmatrix{f & h \cr}^\dagger
 \pmatrix{0&\bfone \cr -\bfone & 0}
\end{equation}
where  $\pmatrix{0&\bfone \cr -\bfone & 0} \pmatrix{f \cr h \cr}$ is the embedded object corresponding
to the $L$ of equation (\ref{ndil})
\vskip 5mm
The formalism we have developed so far for the $D=4$, $N>2$ theories
is completely determined by the embedding of the coset representative of $G/H$ in $Sp(2n,\IR)$
and by the $Usp(n,n)$ embedded Maurer--Cartan equations (\ref{df}).
We want now to show that this formalism, and in particular the identities (\ref{specdef}), the differential relations
among charges (\ref{dz1}) and the sum rules (\ref{sumrule}),
are completely analogous to the Special Geometry relations of $N=2$ matter coupled
supergravity \cite{spegeo}, \cite{str}.
This follows essentially from the fact that, though that the scalar manifold $\cM_{N=2}$ of the $N=2$ theory is not in
general a coset manifold, nevertheless it has a symplectic structure identical to the
$N>2$ theories.
Furthermore we will show that the analogous of the Maurer--Cartan equations of $N>2$ theories
are given in the $N=2$ case by the Picard--Fuchs equations \cite{voi} for the symplectic sections
which enter in the definition of the Special Geometry flat symplectic bundle.
\par
Indeed, let us recall that Special Geometry can be defined in terms of the holomorphic flat
vector bundle of rank $2n$ with structure group $Sp(2n,\IR)$ over a K\"ahler--Hodge manifold \cite{str}.
\par
If we introduce the Special Geometry symplectic and covariantly holomorphic section
of $U(1)$-weight $p= -\bar p=1$:
\begin{equation}
V \equiv \left( f^\Lambda, h_\Lambda \right) =
\left( f^\Lambda(z^i,z^{\bar \imath}), h_\Lambda (z^i,z^{\bar \imath})\right) ;
\quad \Lambda =1,...,n
\label {defsez}
\end{equation}
and  its covariant derivatives with respect to the K\"ahler
connection:
\begin{eqnarray}
  \nabla_i V &=&
\left( \partial _i + {p \over 2} \partial _i K \right)V \equiv
\left( f^\Lambda_i, h_{\Lambda i}\right) \quad i=1,\cdots , n-1
\nonumber\\
 \nabla_{\bar\imath} V &=&
\left( \partial _{\bar\imath} + {\bar p \over 2} \partial _{\bar\imath} K \right)V =0 \quad
\mbox{covariantly holomorphic}
\label{dv}
\end{eqnarray}
then, defining the $n \times n$
matrices:
\begin{equation}
f^\Lambda_\Sigma \equiv \left( f^\Lambda,  \bar f^{\Lambda i} \right);
\quad
h_{\Lambda \Sigma} \equiv \left( h_ \Lambda,
 \bar h_{\Lambda}^{\ i} \right)
\label{matrici}
\end{equation}
where $\bar f^{\Lambda i} \equiv \bar f^\Lambda_{\
\bar\jmath}g^{\bar\jmath i}$, $\bar h_{\Lambda}^{\ i}\equiv \bar
h_{\Lambda \bar\jmath}g^{\bar\jmath i}$ ,
the set of algebraic relations of Special Geometry can  be written  in matrix form as:
\begin{equation}
\left\lbrace\matrix{{\rm i}(f^\dagger h - h^\dagger f) &=& \bfone \cr
(f^t h - h^t f) &=& 0\cr} \right.
\label{specdef1}
\end{equation}
Recalling equations (\ref{specdef}) we see that the previous relations imply that the matrix
 $U$:
\begin{equation}
U={1\over{\sqrt{2}}}\left ( \matrix{f+{\rm i} h & \bar f +
{\rm i} \bar h \cr
 f-{\rm i} h & \bar f - {\rm i} \bar h \cr} \right)
\end{equation}
belongs to $ Usp (n,n)$.
In fact if we set $f^\Lambda \to f^\Lambda \epsilon_{AB}\equiv f^\Lambda_{AB}$
and flatten the world-indices of $( f^\Lambda_i,\bar f^\Lambda_{\bar\imath})$ (or $(\bar h_i, h_{\bar\imath})$)
with the K\"ahlerian vielbein $P^I_i, \bar P^{\bar I}_{\bar \imath}$:
\begin{equation}
  ( f^\Lambda_I, \bar f^\Lambda_{\bar I}) = ( f^\Lambda_i P^i_I,
  \bar f^\Lambda_{\bar \imath}\bar P^{\bar \imath}_{\bar I} ),
\quad  \quad P^I_{i} \bar P^{\bar J}_{\bar\jmath}\eta_{I\bar J} = g_{i\bar\jmath}
\end{equation}
where $\eta_{I\bar J}$ is the flat K\"ahlerian metric and $P^i_{\ I} =
(P^{-1})^I_{\ i}$,
the relations (\ref{specdef1}) are just a particular case of equations (\ref{specdef}) since,
for $N=2$, $H_{Aut}= SU(2) \times U(1)$, so that $f^\Lambda_{AB}$ is actually an $SU(2)$ singlet.
\par
Let us now consider the analogous of the embedded Maurer--Cartan equations  of $G/H$.
Defining as before  the matrix one--form $\Gamma= U^{-1} dU$ valued in the $Usp(n,n)$ Lie algebra,
we see that the relation $d\Gamma+ \Gamma\wedge\Gamma =0$ again implies
a flat connection for the symplectic bundle over the K\"ahler--Hodge manifold.
However, this does not imply anymore that the base manifold is a coset or a constant curvature manifold.
Indeed, let us introduce the covariant derivative of the symplectic
section $(f^\Lambda,\bar f^\Lambda_{\bar I}, \bar f^\Lambda ,
f^\Lambda_I)$ with respect to the $U(1)$--K\"ahler connection $\cQ$
and the spin connection $\omega^{IJ}$ of $\cM_{N=2}$:
 \begin{eqnarray}
& \nabla (f^\Lambda,\bar  f^\Lambda_{\bar I}, \bar f^\Lambda,  f^\Lambda_I) = & \nonumber\\
& d (f^\Lambda,\bar f^\Lambda_{\bar I},  f^\Lambda,\bar f^\Lambda_I) +
(f^\Lambda,\bar f^\Lambda_{\bar J}, \bar f^\Lambda,  f^\Lambda_J)\pmatrix{-{\rm i} \cQ   & 0 & 0 & 0 \cr
0 &  {\rm i} \cQ \delta^{\bar I} _{\bar J} + \omega^{\bar I}_{\ \bar J}
& 0 & 0 \cr  0 & 0 &  {\rm i} \cQ & 0 \cr 0 & 0 & 0 &-{\rm i} \cQ \delta^{ I} _{ J} + \omega^{I}_{\ J} \cr} &
\end{eqnarray}
where:
 \begin{equation}
\cQ = -{{\rm i} \over 2} ( \partial _i \cK d z^i - \bar \partial_{\bar \imath} \cK  d \bar z^{\bar \imath})
\to d \cQ = {\rm i} g_{i\bar\jmath} dz^i \wedge d\bar z ^{\bar\jmath},
\end{equation}
($\cK$ is the K\"ahler potential)
the K\"ahler weight of $(f^\Lambda, \bar f^\Lambda_I)$ and $(\bar
f^\Lambda,\bar f^\Lambda_{\bar I})$ being $p=1$ and $p=-1$ respectively.
Using the same decomposition as in equation (\ref{defgamma}) and eq.s (\ref{nablav}), (\ref{defomega})
we have in the $N=2$ case:
\begin{eqnarray}
  \label{gammaspec}
  \Gamma &=& \pmatrix{\Omega & \bar \cP \cr \cP & \bar \Omega \cr}
  ,\nonumber\\
  \Omega &=& \omega = \pmatrix{-{\rm i}  \cQ  & 0 \cr 0 &
  {\rm i}   \cQ \delta^I_J+ \bar \omega^I_{\ J} \cr}
   \end{eqnarray}
For the subblocks $\cP$ we obtain:
\begin{equation}
\cP = - {\rm i} (f^t \nabla h - h^t \nabla f)  = {\rm i} f^t (\cN -
\bar \cN) \nabla f = \pmatrix{0 &  P_{\bar I} \cr  P^{ J} &
P^{ J}_{\ \bar I} \cr}  \label{viel2}
\end{equation}
where $\bar P^J \equiv \eta^{J\bar I} P_{\bar I} $ is the $(1,0)$--form K\"ahlerian vielbein
while $ P^{ J}_{\ \bar I} \equiv
{\rm i} \left( f^t (\cN - \bar \cN) \nabla f \right)^{  J}_{\ \bar I}$ is a one--form which in general cannot be expressed
in terms of the vielbein $P^I$ and therefore represents a new geometrical quantity on $\cM_{N=2}$.
Note that we get zero in the first entry of equation  (\ref{viel2})
by virtue of the fact that the identity (\ref{specdef1}) implies $f^\Lambda(\cN -
\bar \cN)_{\Lambda\Sigma}f^\Sigma_i =0 $ and that $f^\Lambda$ is covariantly holomorphic.
If $\Omega $ and $\cP $ are considered as data on $\cM_{N=2}$ then we may interpret
$\Gamma =U^{-1} dU$ as an integrable system of differential equations, namely:
\begin{equation}
  \label{picfuc}
  \nabla (V ,\bar V_{\bar I} , \bar V ,   V_I ) =  (V ,\bar V_{\bar J} , \bar V ,  V_J )
  \pmatrix{0& 0 & 0 & \bar P_{I} \cr
0 & 0 & \bar P^{\bar J} & \bar P^{\bar J}_{\ I} \cr 0 &  P_{\bar I} & 0 & 0 \cr  P^J & P^J_{\ \bar I} & 0 & 0 \cr}
\end{equation}
where the flat K\"ahler indices $I,\bar I, \cdots $ are raised and lowered with the flat K\"ahler metric
$\eta_{I\bar J}$.
As it is well known, this integrable system describes the Picard--Fuchs equations for the
periods $(V,\bar V_{\bar I}, \bar V,   V_{I})$ of the Calabi--Yau threefold
with solution given by the $2 n$ symplectic vectors
$V \equiv (f^\Lambda, h_\Lambda)$.
The integrability condition $d\Gamma + \Gamma\wedge \Gamma=0$ gives
three constraints on the K\"ahler base manifold:
\begin{eqnarray}
  d( {\rm i} \cQ) + \bar P_I \wedge  P^{I} &=& 0
  \to \partial_{\bar\jmath}\partial_i \cK = P^I{\ ,i} \bar P_{  I,\bar \jmath}=g_{i\bar\jmath}
  \label{spec1} \\
  (d\omega +\omega\wedge\omega)_{\ \bar I}^{\bar J} &=& P_{\bar I}
  \wedge\bar P^{\bar J} -{\rm i} d\cQ
  \delta_{\bar I}^{\bar J}
-\bar P^{\bar J}_{\ L} \wedge  P^L_{\ \bar I}
\label{spec2}\\
\nabla P_{\ \bar I}^{J} &=& 0
\label{spec3}\\
\bar P_J \wedge P^J_{\ \bar I}=0
\label{spec4}
\end{eqnarray}
Equation (\ref{spec1}) implies that $\cM_{N=2}$ is a K\"ahler--Hodge manifold.
Equation (\ref{spec2}),
written with holomorphic and antiholomorphic curved indices, gives:
\begin{equation}
  \label{curvspec}
  R_{\bar\imath j \bar k l} = g_{\bar\imath l} g_{j \bar k} + g_{\bar k l}
  g_{\bar\imath j} - P_{\bar\imath\bar k \bar m}
\bar P_{jln} g^{\bar m n}
\end{equation}
which is the usual constraint on the Riemann tensor of the special geometry.
The further Special Geometry constraints on the three tensor
$\bar P_{ijk}$  are then consequences of
 equations (\ref{spec3}), (\ref{spec4}), which imply:
\begin{eqnarray}
  \nabla_{[l} \bar P_{i]jk} &=& 0 \nonumber\\
\nabla_{\bar l} \bar P_{ijk}&=&0
\label{cijk}
\end{eqnarray}
In particular, the first of eq. (\ref{cijk}) also implies
that $ \bar P_{ijk} $ is a completely symmetric tensor.
\par
In summary, we have seen that the $N=2$ theory and the higher
$N$ theories have essentially the same symplectic structure,
the only difference being that since the scalar manifold of $N=2$
is not in general a coset manifold
the symplectic structure allows the presence of a new geometrical
quantity which physically corresponds to
the anomalous magnetic moments of the $N=2$ theory.
It goes without saying that, when $\cM_{N=2}$ is itself a coset manifold \cite{cp},
then the anomalous magnetic moments $ \bar P_{ijk}$
must be expressible in terms of the vielbein of $G/H$.
We give here two examples.
\begin{itemize}
\item{
Suppose $\cM_{N=2}= {SU(1,1) \over U(1)} \times {O(2,n)\over O(2) \times O(n)} $.
The symplectic sections entering the matrix $U$ can be written as follows:
\begin{eqnarray}
  f&=&{\rm i}e^{\cK\over 2}(L^ \Lambda , L^\Lambda_a)\nonumber\\
 h&=&{\rm i}e^{\cK\over 2}(SL^\Sigma\eta_{\Lambda\Sigma},\bar S L^{\Sigma }_a\eta_{\Lambda\Sigma})
\end{eqnarray}
where $\Lambda = 1,\cdots,2n$, $a= 3,\cdots,n$, $\eta_{\Lambda\Sigma}=(1,1,-1,\cdots,-1)$ and we have set
$L^\Lambda = {1\over \sqrt{2}} (L^\Lambda_1 + {\rm i} L^ \Lambda_2)$. In particular from the pseudoorthogonality of
$O(2,n)$ we have:
\begin{equation}
L^\Lambda L^\Sigma \eta_{\Lambda\Sigma} = 0
\end{equation}
Furthermore we have parametrized  ${SU(1,1)\over U(1)}$ as follows:
\begin{equation}
M(S)={1\over {\sqrt{{4 Im S \over 1+\vert S \vert ^2 + 2 Im S}}}}
\left(\matrix{\bfone & {{{\rm i} -S} \over {{\rm i} +S}} \cr
{{{\rm i} + \bar S} \over {{\rm i}- \bar S}} & \bfone \cr }\right)
\end{equation}
We can then compute the embedded connection and vielbein using (\ref{defgamma}).
In particular we find:
\begin{equation}
 \cP = \pmatrix{0 & P_I \cr P_I &- \bar P \delta_{ab} \cr}
\end{equation}
We see that the general $P_{I\bar J}$ matrix in this case can be expressed in terms
of the vielbein of $G/H$ and one finds that the only non vanishing anomalous magnetic moments are:
\begin{equation}
  \bar P_{ab S} = \delta_{ab}P_{,S}= e^{\cK} \delta_{ab}.
\end{equation}}
\item{As a second example we consider the special coset manifold ${SO^*(12) \over U(6)}$.
\par
Note that this manifold also appears as the scalar manifold of the $D=4$, $N=6$ theory,
and we refer the reader to section 4
for notations and parametrization of $G/H$.
\par
The integrable system in this case can be written as follows:
\begin{equation}
  \label{so12}
  \nabla \pmatrix{V \cr V_{AB} \cr \bar V \cr \bar V^{AB}\cr} = \pmatrix{0 & 0 & 0 &{1 \over 2} P_{CD}
  \cr 0 & 0 & P_{AB} & {1 \over 2} \bar P_{ABCD} \cr
0 & {1 \over 2}\bar P^{CD} & 0 & 0 \cr \bar P^{AB} & {1 \over 2} P^{ABCD} & 0 & 0 \cr}
\pmatrix{V \cr V_{CD} \cr \bar V \cr \bar V^{CD}\cr}
\end{equation}
where $P^{ABCD}$ is the K\"ahlerian vielbein $(1,0)$--form ($\bar P_{ABCD}=(P^{ABCD})^*$ is a $(0,1)$--form) and:
\begin{equation}
  P_{AB} = {1 \over 4!}\epsilon_{ABCDEF} P^{CDEF} ; \quad \bar P^{AB} = (P_{AB})^*.
\label{pab}
\end{equation}
Moreover, $f_{AB}$ transforms in the {\bf 15} of $SU(6)$, $f$ is an $SU(6)$ singlet and $ \bar f^{AB} = (f_{AB})^*$.
It follows:
\begin{equation}
  \label{so12ol}
  \nabla_i \pmatrix{V \cr V_{AB} \cr \bar V \cr \bar V^{AB}\cr} =
  \pmatrix{0 & 0 & 0 & {1 \over 2} P_{CD,i} \cr 0 & 0 & P_{AB,i} & 0 \cr
0 & 0 & 0 & 0 \cr 0 & {1 \over 2} P^{ABCD}_{,i} & 0 & 0 \cr}  \pmatrix{V \cr V_{CD} \cr \bar V \cr \bar V^{CD}\cr}
\end{equation}
Hence:
\begin{equation}
  \nabla_i \bar V^{AB} = {1 \over 2} P^{ABCD}_{,i} V_{CD}
\end{equation}
If we set:
\begin{equation}
  V_{CD} = P_{CD}^{\ \ \bar l} V_{\bar l} ; \quad \bar V^{AB} = \bar P^{AB,k}V_k
\end{equation}
where the curved indices $k,\bar l$ are raised and lowered with the K\"ahler metric, one easily obtains:
\begin{equation}
  \nabla_i V_j ={1 \over 2} P_{AB, i}P_{CD, j} P^{ABCD}_{,k}g^{k \bar l} \bar V_{\bar l} =
  {1 \over 4}\epsilon _{ABCDEF}
P_{AB, i}P_{CD, j} P_{EF,k}g^{k \bar l} \bar V_{\bar l}.
\end{equation}
Therefore the anomalous magnetic moment is given, in terms of the vielbein, as:
\begin{equation}
\bar P_{ijk} ={1 \over 4} \epsilon _{ABCDEF}P_{AB, i}P_{CD, j} P_{EF,k}
\end{equation}
The other two equations of the integral system give:
\begin{eqnarray}
  \nabla_i V = {1 \over 2} P_{CD,i} V^{CD} &\to & \nabla_i V = V_i \nonumber\\
\nabla_i V_{AB} = P_{AB,i} \bar V &\to & \nabla_i V_{\bar j} = g_{i \bar j} \bar V
\end{eqnarray}
which are the remaining equations defining $N=2$ Special Geometry.
}
\end{itemize}
To complete the analogy between the $N=2$ theory and the higher $N$ theories in $D=4$,
 we also give for completeness in the $N=2$ case  the central and matter charges, the differential
 relations among them and the sum rules.
\par
Let us  note that also in $N=2$ the kinetic matrix ${\cal N} _{\Lambda \Sigma}$ which appears in the
vector kinetic Lagrangian \footnote{ The same normalization for the vector kinetic lagrangian will be used
in section 4 when discussing $D=4$, $N>2$ theories}:
\begin{equation}
{\cal L}_{Kin}^{(vector)} = {\rm i} \bar {\cal N} _{\Lambda \Sigma}
{ F}^{-\Lambda}_{\mu\nu} { F}^{-\Sigma \mu\nu}+ h. c.
\end{equation}
\begin{equation}
 { F}^{\pm\Lambda} = {1\over 2} (\bfone \pm \rm i ^\star){ F}^{\Lambda} \label{defv}\\
\end{equation}
\begin{equation}
 {\cG}^-_{\Lambda}=\bar {\cN}_{\Lambda\Sigma}  F^{-\Sigma} \label{defgd}\\
\end{equation}
is  given in terms of $f, h$ defined in eq. (\ref{matrici}) by the
formula:
\begin{equation}
\cN _{\Lambda\Sigma}=h_{\Lambda \Gamma}(f^{-1})_ \Sigma ^\Gamma \label{nhf}
\end{equation}
The columns of the matrix $f$ appear in the supercovariant
electric field strength $\hat F^\Lambda$:
\begin{equation}
\hat F^\Lambda  = F^\Lambda + f^\Lambda \bar\psi^A \psi ^B
\epsilon_{AB} -{\rm i} \bar f^\Lambda_{\bar\imath}
\bar \lambda^{\bar\imath}_A\gamma_a\psi_B\epsilon^{AB}V^a + h. c.
\end{equation}
(The columns of $h^\Lambda_I$ would appear in the  dual theory
written in terms of the dual magnetic field strengths) .
\par
The transformation laws for the chiral gravitino $\psi_A$ and gaugino
$\lambda^{iA}$ fields are:
\begin{equation}
 \delta \psi_{A \mu}={ D}_{\mu}\,\epsilon _A\,+ \epsilon _{AB}
T_{\mu\nu} \gamma^\nu
\epsilon^B + \cdots
 \end{equation}
 \begin{equation}
 \delta\lambda^{iA} = {\rm i} \partial_\mu z^i \gamma^\mu\epsilon^A +
 {{\rm i} \over 2}
T_{\bar \jmath\mu\nu} \gamma^{\mu \nu}
g^{i\bar\jmath}\epsilon^{AB}\epsilon_B + \cdots
\end{equation}
where:
\begin{equation}
T \equiv  h_\Lambda F ^{\Lambda}  - f^\Lambda \cG_\Lambda
\end{equation}
\begin{equation}
 T_{\bar \imath} \equiv
 \bar h_{\Lambda_{\bar \imath}} F^{\Lambda}  - \bar f^\Lambda_{\bar \imath} \cG_\Lambda
\end{equation}
are respectively the graviphoton and the matter-vectors
 $z^i$ ($i=1,\cdots,n$) are the complex scalar fields and the position of
 the $SU(2)$ automorphism index A (A,B=1,2) is related to chirality
 (namely $(\psi_A, \lambda^{iA})$ are chiral, $(\psi^A,
 \lambda^{\bar\imath}_A)$ antichiral).
 In principle only the (anti) self dual part of $ F$ and $\cG$ should
 appear in the transformation laws of the (anti)chiral fermi fields; however,
 exactly as in eqs. (\ref{tiden0}),(\ref{tiden}) for $N>2$ theories, from equations (\ref{defgd}), (\ref{nhf}) it follows that :
 \begin{equation}
T^+ = h_\Lambda  F^{+\Lambda} - f^\Lambda \cG_\Lambda ^+   =0
\end{equation}
so that $T=T^-$ (and $\bar T = \bar T^+$).
 Note that both the graviphoton and the matter vectors are $Usp
 (n,n)$ invariant according to the fact that the fermions do not
 transform under the duality group (except for a possible R-symmetry
 phase).
To define the physical charges let us note that in presence of
electric and magnetic sources we can write:
\begin{equation}
\int_{S^2}  F^\Lambda = g^\Lambda , \quad
\int_{S^2} \cG _ \Lambda = e_ \Lambda
\end{equation}
The central charges and the matter charges are now defined as the integrals
over a $S^2$ of the physical graviphoton and matter vectors:
\begin{equation}
Z= \int_{S^2} T= \int_{S^2} ( h_\Lambda  F ^{\Lambda} - f^\Lambda \cG_\Lambda )
= ( h_\Lambda (z,\bar z) g^{\Lambda} - f^\Lambda(z,\bar z) e_\Lambda )
\end{equation}
where $z^i, \bar z^{\bar \imath}$ denote the v.e.v. of the moduli fields in a given
background.
Owing to eq (\ref{dv}) we get immediately:
\begin{equation}
Z_i= \nabla_i Z
\label{fundeq}
\end{equation}
We observe that if in a given background $Z_i =0$ the BPS states in
this configuration have a minimum mass. Indeed
\begin{equation}
\nabla_i Z =0 \rightarrow \partial _i \vert Z \vert ^2 =0.
\end{equation}
 As a consequence of the symplectic structure, one can derive
  two sum
 rules for  $Z$ and $Z_i$:
  \begin{equation}
  \vert Z \vert ^2 \pm  \vert Z_i \vert ^2 \equiv
  \vert Z \vert ^2 \pm   Z_i g^{i\bar \jmath} \bar Z_{\bar \jmath} =
  -{1\over 2} P^t \cM _\pm P
  \label{sumrules}
\end{equation}
  where:
  \begin{equation}
\cM_+ = \left( \matrix{ \bfone & - Re \cN \cr 0 &\bfone\cr}\right)
\left( \matrix{ Im \cN & 0 \cr 0 &Im \cN^{-1}\cr}\right)
\left( \matrix{ \bfone & 0 \cr - Re \cN & \bfone \cr}\right)
\label{m+2}
\end{equation}
\begin{equation}
\cM_- = \left( \matrix{ \bfone & - Re F \cr 0 &\bfone\cr}\right)
\left( \matrix{ Im F & 0 \cr 0 &Im F^{-1}\cr}\right)
\left( \matrix{ \bfone & 0 \cr - Re F & \bfone \cr}\right)
\label{m-2}
\end{equation}
and:
\begin{equation}
P=\left(g^\Lambda, e_ \Lambda \right)
\label{eg2}
\end{equation}
Equation (\ref{m+2}) is obtained by using exactly the same procedure as in (\ref{m+}).
 The sum rule (\ref{m-2}) involves a matrix $\cM_-$, which  has exactly the
same form as $\cM_+$ provided we perform the substitution $\cN_{\Lambda\Sigma}
\rightarrow F_{\Lambda\Sigma}={\partial^2 F \over {\partial X^\Lambda
\partial X^\Sigma}}$($X^\Lambda = e^{-{K\over 2}} f^\Lambda$).
It can be derived in an analogous way by observing
that, when a prepotential $F=F(X)$ exists, Special Geometry gives
the following
extra identity:
\begin{equation}
f^\Lambda_I(F-\bar F)_{\Lambda\Sigma}f^\Sigma_J
= - {\rm i}\eta_{IJ}
\quad \quad \eta =\left(\matrix{ -1 & 0 \cr 0 &  \bfone_{n\times
n}\cr}\right)
\end{equation}
from which it follows:
 \begin{eqnarray}
f \eta f^\dagger &= &{\rm i} \left( F - \bar F \right)^{-1} \\
h \eta h^\dagger &= &{\rm i} \left(\bar F^{-1} - F^{-1} \right)^{-1}\equiv
{\rm i}\bar F \left( F - \bar F \right) ^{-1} F
\end{eqnarray}
Note that while $Im \cN$ has a definite (negative) signature, $Im F$
is not positive definite.
\vskip 5mm
\par
To conclude this section, we outline the embedding
procedure in $D=6$ and in $D=8$ maximal supergravities.
More details are given in sections 5.2 and 7.
\par
  In $D=8$, $N=2$ the situation is exactly similar to the $D=4$ case,
  where instead of 2--forms field--strengths we now have  4--forms.
 In the case at hand the 4--form in the gravitational multiplet
and its dual are a doublet under the duality group $Sl(2,\IR)$.
The embedding procedure and the relevant relations are discussed in
section 7.
 \par
Finally in $D=6$
the  3--form field strengths $H^\Lambda$ which appear in the
gravitational and/or tensor multiplet have a definite self--duality
 \begin{equation}
H^{\pm \Lambda} = {1\over 2} (H^\Lambda \pm ^\star H^\Lambda)
\end{equation}
  In this case we have a T-duality group of the form $G=O(m,n)$.
  In the chiral $N= (2,0)$ and $N=(4,0)$ theories, the number of
  self--dual tensors $H^{+\Lambda_1}$ in the gravitational multiplet,
  $\Lambda_1 =1,\cdots m$
 and antiself--dual tensors $H^{-\Lambda_2} $ in the matter
 multiplet, $\Lambda_2 =1,\cdots n$
 are different in general and $G$
  acts in its fundamental representation on
  $(H^{+\Lambda_1}, H^{-\Lambda_2})$ so that no embedding is required.
The procedure to find the charges and their relations is thus
completely analogous to the odd dimensional case.
One finds e.g. for the magnetic charges:
$(Z, Z_{I}) = (L_{\Lambda} g^\Lambda,   L_{\Lambda I} g^\Lambda)$
for(2,0) theory ($I= 1,\cdots,n)$, $n$ being the number of vector multiplets, and
$(Z_r, Z_{I})= (L_{\Lambda r} g^\Lambda,   L_{\Lambda I} g^\Lambda)$
($ r= 1,\cdots,5$) for(4,0) theory, $r$ being the number of self--dual 2--forms in the gravitational multiplet
(see sect.5.2).
 However, due to the
relation:
  \begin{equation}
\cN_{\Lambda\Sigma} ^{\ \ \star}H^\Sigma = \eta_{\Lambda\Sigma} H^\Sigma,
\label{elmagn}
\end{equation}
where $\eta$ and $\cN$ are defined in terms of the coset representatives of ${O(n,m)\over O(m)\times O(n)}$
as in
(\ref{etadil}), (\ref{ndil}),
we have no distinction among electric and magnetic charges. Indeed
\begin{equation}
e_\Lambda= \int \cN_{\Lambda\Sigma} ^{\ \ \star}H^\Sigma =
\int  \eta_{\Lambda\Sigma} H^\Sigma =\eta_{\Lambda\Sigma} g^\Sigma
\end{equation}
The six dimensional $N= (2,0)$ and $N=(4,0)$ matter coupled theories are discussed in section
7.
  \par
 In $D=6$  maximally extended theory we have an equal number (five)
 of self--dual and anti self--dual field strengths
 and therefore a Lagrangian
exists.
 The group
 $G=O(5,5)$ rotates among themselves $H^+$ and $H^-$ in the
 representation $\underline {10}$.
 The analogous of the Gaillard--Zumino construction in this case would
 define an $O(5,5)$ embedding of $O(5)$ rotating among themselves
 $(H^+ ,\cG^+)$ or $(H^-, \cG^-)$ where
 \begin{equation}
\cG^\pm=\cN_\mp H^\pm
\end{equation}
  $\cN_+,\cN_-=-(\cN_+)^t$ is the kinetic metric of the 3-forms in the
  Lagrangian: $\cL_{kin} = \cN^+_{\Lambda \Sigma} H^ {+\Lambda}\wedge H^{-\Sigma} +
\cN^-_{\Lambda \Sigma} H^ {-\Lambda}\wedge H^{+\Sigma}$.
In this case we obtain a formula analogous to (\ref{zab}), (\ref{zi})
 which is however invariant under $O(5,5)$ instead of $Usp(n,n)$.
 Namely the central charges of $D=6$ , $N=(4,4)$ have the following
 dyonic form:
\begin{equation}
Z_{\pm AB} = f^\Lambda_\pm e_\Lambda + h_{\Lambda \pm}g^\Lambda
\label{z+_}
\end{equation}
 which is $O(5,5)$ invariant with respect to the off
 diagonal metric $\eta = \pmatrix{0 & \bfone \cr \bfone & 0 \cr}$.
 Even in this case the relation (\ref{elmagn}) holds so that only ten
 independent charges exist.
The maximally extended $D=6$ supergravity is discussed in detail in
 section 8.
     \par
 Finally, in the $N=(2,2)$ (Type IIA) six dimensional theory we
 have a single 2--form in the gravitational multiplet.
 Here the embedding group is $O(1,1)$ and acts in the two dimensional
 representation on the self--dual and antiself--dual parts of the
 3--form $H=dB$.
 More details on this case are given in section 7.
\section{$N>2$ four dimensional supergravities revisited}
\setcounter{equation}{0}
In this section and in the following ones we apply the general considerations of section 3
to the various ungauged supergravity theories with scalar manifold $G/H$ for any $D$ and $N$.
This excludes $D=4$ $N=2$, already discussed in section $3.2$, and $D=5$ $N=2$
for which we refer to the literature \cite{gusito}.
Our aim is to write down the group theoretical structure of each theory,
 their symplectic or orthogonal  embedding, the vector kinetic
matrix, the supersymmetric transformation laws, the structure
 of the central and matter charges, the differential relations
originating from the Maurer--Cartan equations and the sum rules the charges satisfy.
For each theory we give the group--theoretical assignments for the fields,
their supersymmetry transformation laws, the ($p+2$)--forms kinetic metrics
and the relations satisfied by central and matter charges.
As far as the boson transformation rules are concerned we prefer to write
 down the supercovariant definition of the
field strengths (denoted by a superscript hat), from which the susy--laws are immediately retrieved.
 As it has been mentioned in section 3 it is here that
the symplectic sections $( f^\Lambda _{AB},f^\Lambda_I,\bar f^\Lambda_{AB},
\bar f^\Lambda_I )$ appear as coefficients of the bilinear fermions
in the supercovariant field--strengths while the analogous symplectic
 section $(h_{\Lambda AB},h_{\Lambda I },\bar h_{\Lambda AB},
\bar h_{\Lambda I})$ would appear in the dual magnetic theory.
We include in the supercovariant field--strengths also the supercovariant
vielbein of the $G/H$ manifolds.
Again this is equivalent to giving the susy transformation laws of the scalar fields.
The dressed field strengths from which the central and matter charges
 are constructed appear instead in the susy transformation laws of the
 fermions for which we give the expression up to trilinear fermion terms.
It should be stressed that the numerical coefficients in the aforementioned
 susy transformations and supercovariant field strengths
are fixed by supersymmetry (or ,equivalently, by Bianchi identities in superspace ),
but we have not worked out the relevant computations
being interested in the general structure rather that
 in the precise numerical expressions. However the numerical factors could
also be retrieved by comparing our formulae with those written
 in the standard literature on supergravity and performing the necessary
redefinitions.
The same kind of considerations apply to the central and matter
 charges whose precise normalization has not been fixed.
In the  Tables of  the present  and of the following sections, we give the group assignements for
 the supergravity fields; in particular, we quote the representation
$R_H$ under which the scalar fields of the linearized theory (or the vielbein of $G/H$ of the full theory)
transform.
Furthermore, for $D=4,8$ only the left--handed fermions are quoted.
Right handed fermions transform in the complex conjugate representation of $H$.
In the present section we apply the cosiderations given in section 3.2
to the 4D--Supergravities for $N>2$.
Throughout the section we denote by $A,B,\cdots$ indices of $SU(N)$,
 $SU(N)\otimes U(1)$ being the automorphism group of the $N$--extended
supersymmetry algebra. Lower and upper $SU(N)$ indices on the fermion fields are related to
 their left or right chirality respectively.
If some fermion is a $SU(N)$ singlet chirality is denoted by the usual (L) or (R) suffixes.
Right--handed fermions of $D=4$ transform in the complex f representation of $SU(N)\times U(1) \times H_{matter}$.

Furthermore for any boson field $v$ carrying $SU(N)$ indices
 we have that lower and upper indices are related by complex conjugation,
namely:
\begin{equation}
(v_{AB\cdots})^\star \sim \bar v^{AB\cdots}
\end{equation}
\begin{itemize}
\item{Let us first consider the $N=3$ case \cite{ccdffm}.
The coset space is:
\begin{equation}
G/H= {SU(3,n)\over {SU(3)\otimes SU(n)\otimes U(1)}}
\end{equation}
and the field content is given by:
\begin{eqnarray}
&(V^a_\mu, \psi_{A \mu}, A^{AB}_\mu , \chi_{(L)}) \quad\quad  A=1,2,3 \quad\quad
\hbox{(gravitational multiplet)}& \\
&(A_\mu , \lambda_A, \lambda_{(R)} , 3z)^I  \quad \quad I=1,\cdots,n \quad \quad
\hbox{(vector multiplets)}&
\end{eqnarray}
The transformation properties of the fields are given in the
following Table \ref{4,3} \footnote{We recall that $R_H$  denotes
 the representation which  the vielbein of the scalar
manifold belongs to.}
\begin{table}[ht]
\caption{Transformation properties of fields in $D=4$, $N=3$}
\label{4,3}
\begin{center}
\begin{tabular}{|c||c|c|c|c|c|c|c|c|c|}
\hline
& $V^a_\mu$ & $\psi_{A\mu}$ & $A^\Lambda_\mu$ & $\chi_{(L)}$
& $\lambda^{I}_A$ &
$\lambda^I_{(L)}$ &$ L^\Lambda_{AB} $&$ L^\Lambda_I $& $R_H$ \\
\hline
\hline
$SU(3,n)$ & 1  & 1 & $3+n$ & 1 & 1 & 1 & $3+n$& $3+n$ & -  \\
\hline
$SU(3)$ & 1 & 3 & 1 & 1 & 3 & 1 & 3 & 1 & $3$  \\
\hline
$SU(n)$ & 1  & 1 & 1 & 1 & $n$ & $n$ & 1 & $n$ & $n$ \\
\hline
  $U(1)$ & 0  & ${n\over 2}$ & 0 & 3${n\over 2}$ & 3+${n\over 2}$
  &$-3(1  + {n\over 2})$ & $n$ & $-3$ &$3+n$ \\
\hline
\end{tabular}
\end{center}
\end{table}
The embedding of $SU(3,n)$ in $Usp(3+n,3+n)$ allows to express
the section $(f,h)$ in terms of $L$ as follows::
\begin{eqnarray}
f^\Lambda_{\ \Sigma}&\equiv& \frac{1}{\sqrt{2}} (L^\Lambda_{\ AB},\bar L^\Lambda_{\ I})
\label{n3f}  \\
h_{\Lambda \Sigma}&=&-{\rm i}(J fJ)_{\Lambda\Sigma} \quad \quad \quad \quad
J= \left(\matrix{\bfone _{3\times 3}&0\cr 0 &-\bfone _{n\times n} \cr} \right) \label{n3h}
\end{eqnarray}
where $AB$ are antisymmetric $SU(3) $ indices, $I$ is an index of
$SU(n)\otimes U(1)$ and $\bar L^\Lambda_{\ I}$ denotes the complex conjugate
of the coset representative.
We have:
\begin{equation}
 \cN_{\Lambda\Sigma} = (h f^{-1})_{\Lambda \Sigma}
 = -{\rm i}(J fJf^{-1})_{\Lambda\Sigma} \label{kin2}
\end{equation}
The supercovariant field-strengths and the supercovariant scalar vielbein
 $\hat P^A_I = (L^{-1} \nabla ^{(H)}L)_I^{\ A}$ are:
\begin{eqnarray}
\hat F^\Lambda &=& dA^\Lambda - {1\over 2} f^{\Lambda }_{AB}
 \bar \psi ^A \psi^B + {{\rm i}\over 2} f^{\Lambda }_I
\bar \lambda^I_A \gamma_a \psi^A V^a + {\rm i}f^\Lambda_{AB}\bar
\chi_{(R)}\gamma_a \psi_C \epsilon^{ABC} V^a \nonumber\\
&+& h.c.\\
\hat  P^{\ A}_I &=& P^{\ A}_{I} - \bar\lambda^I_B \psi_C
 \epsilon^{ABC} - \bar \lambda_{I(R)}\psi^A
\end{eqnarray}
where:
\begin{eqnarray}
  P^{ A}_I & = & {1 \over 2}  \epsilon^{ABC}P_{I BC} =    {1 \over 2}
  \epsilon^{ABC}(L^{-1} \nabla ^{(SU(3) \times U(1))}L)_{I BC}\nonumber\\
&=& P^{A}_{I,i}dz^i \\
\bar P^{IA}& =& P_{IA}
\end{eqnarray}
$z^i$ being the (complex) coordinates of $G/H$
 and $H=H_{Aut} = SU(3) \times U(1)$ .
The chiral fermions transformation laws are given by:
\begin{eqnarray}
\delta \psi_A &=& D \epsilon_A + 2{\rm i} T^{-}_{AB \vert
ab}\Delta^{abc} V_c \epsilon^B + \cdots \\
\delta \chi_{(L)} &=& 1/2 T^{-} _{AB \vert ab} \gamma^{ab} \epsilon_C
\epsilon^{ABC} + \cdots \\
\delta\lambda_{IA} &=& -{\rm i} P_{I\ \ ,i}^{\ B}\partial_a z^i\gamma^a \epsilon^C \epsilon_{ABC}
+ T_{I \vert ab} \gamma^{ab} \epsilon_A + \cdots
\\
 \delta\lambda^I_{(L)} &=& {\rm i} P_{I\ \ ,i}^{\ A} \partial_a z^i\gamma^a \epsilon_A
+ \cdots
\end{eqnarray}
 where $T_{AB} $ and $T_{I}$ have the general form given
 in equation (\ref{gravi}).
  Therefore, the general form of the dyonic charges $(Z_{AB}, Z_I)$ are given by eqns.
  (\ref{zab})-- (\ref{charges}).
From the general form of the Maurer-Cartan equations for the embedded
coset representatives $U \in Usp(n,n)$, we find:
\begin{equation}
\nabla^{(H)} \pmatrix{f^\Lambda_{AB} \cr
h_{\Lambda AB}} = \pmatrix{\bar f^\Lambda_I \cr \bar h_{\Lambda I}}
P_I^{\ C} \epsilon_{ABC}
\end{equation}
According to the discussion given in section 3,
using (\ref{zab}), (\ref{zi}) one finds:
 \begin{eqnarray}
\nabla^{(H)} Z_{AB} &=& \bar Z^I
P_I^{\ C} \epsilon_{ABC}   \\
\nabla^{(H)} Z_{I} &=&{1\over 2} \bar Z^{AB}
P_I^{\ C} \epsilon_{ABC}
\end{eqnarray}
and the sum rule:
\begin{equation}
      {1  \over 2}\bar Z ^{  AB} Z_{AB} + Z_{ I} \bar Z_{I} = -{1 \over 2}P^t\cM(\cN)P
\end{equation}
where the matrix $\cM(\cN)$ has the same form as in equation (\ref{m+}) in terms of the kinetic
matrix $ \cN$ of eq.(\ref{kin2}) and $P$ is the charge vector $P^t=(g,e)$.
\item{For $N=4$ \cite{bekose}, the coset space is a  product:
\begin{equation}
G/H= {SU(1,1)\over U(1)}\otimes {O(6,n)\over {O(6) \otimes O(n)}}
\end{equation}
The field content is given by:
Gravitational multiplet:
\begin{equation}
(V^a_\mu,\psi_{A\mu},A_\mu^{AB},\chi_{ABC},S) \quad\quad (A,B=1 ,\cdots,4)
\end{equation}
Vector multiplets:
\begin{equation}
(A_\mu,\lambda^{A},6 \phi)^I \quad \quad (I=1,\cdots,n)
\end{equation}
The coset representative can be written as:
\begin{equation}
L^\Lambda _{\ \Sigma}\to M(S) L^\Lambda_{\ \Sigma}
\end{equation}
where $L^\Lambda_{\ \Sigma}$ parametrizes the coset manifold
 ${O(6,n)\over {O(6)\otimes O(n)}}$
and
\begin{equation}
M(S)={1\over {\sqrt{{4 Im S \over 1+\vert S \vert ^2 + 2 Im S}}}}
\left(\matrix{\bfone & {{{\rm i} -S} \over {{\rm i} +S}} \cr
{{{\rm i} + \bar S} \over {{\rm i}- \bar S}} & \bfone \cr }\right)
\end{equation}
The group assignments of the fields are given in Table \ref{tab4,4}.
\begin{table}[ht]
\caption{$D=4$, $N=4$ transformation properties}
\label{tab4,4}
\begin{center}
\begin{tabular}{|c||c|c|c|c|c|c|c|c|}
\hline
& $V^a_\mu$  & $\psi_{A \vert \mu}$ & $A^\Lambda_{\mu}$ & $\chi_{ABC}$ &
$\lambda_{IA} $ & $ M(S)L^\Lambda_{AB} $ & $M(S) L^\Lambda_I$ & $R_H$ \\
\hline
\hline
$SU(1,1)$ & 1 & 1 & - & 1 & 1 & $2\times 1$ &  $2\times 1$ & - \\
\hline
$O(6,n)$ & 1 & 1 & $6+n$ & 1 & 1 & $1 \times (6+n)$ & $1 \times (6+n)$ & - \\
\hline
$O(6)$ & 1 & 4 & 1 & $\bar 4 $ & $ \bar 4$ & $ 1\times 6$ & 1 & 6 \\
\hline
$O(n)$ & 1 & 1 & 1 & 1 & $n$ & 1 & $n$ & $n$ \\
\hline
$U(1)$& 0 & ${1\over 2}$ & 0 & ${3\over 2}$ & $ -{1\over 2}$ & 1 & 1 & 0 \\
 \hline
\end{tabular}
\end{center}
\end{table}
With the given coset parametrizations the symplectic embedded
 section $(f^\Lambda_\Sigma, h_{\Lambda \Sigma})$ is
(apart from a unessential phase ${{\rm i}+S \over {\rm i} - \bar S} $):
\begin{eqnarray}
f^\Lambda_{\ \Sigma}&=&{\rm i} e^{{K\over 2}}(L^\Lambda_{\ AB},L^\Lambda_{\ I})\\
h_{\Lambda\Sigma}&=&{\rm i} e^{{K\over 2}}(S L^\Gamma_{\ AB}
\eta_{\Lambda\Gamma},\bar S
L^\Gamma_{\ I}\eta_{\Lambda\Gamma})
\end{eqnarray}
where $K=- \mbox{
log}[{\rm i}(S-\bar S)]$ is the K\" ahler potential of ${SU(1,1)\over U(1)}$, and the kinetic matrix
$\cN = h f^{-1}$ takes the form:
\begin{equation}
\cN_{\Lambda\Sigma} = {1\over 2}
(S - \bar S)\bar L_\Lambda^{\ AB} L_{\Sigma AB} +
\bar S \eta _{\Lambda\Sigma}
\label{n44}
\end{equation}
The supercovariant field strengths and the vielbein of the coset manifold are:
\begin{eqnarray}
\hat F^\Lambda &=& dA^\Lambda +\bigl [
f^\Lambda_{AB}(c_1 \bar \psi^A \psi ^B + c_2 \bar \psi_C \gamma_a
\chi^{ABC}V^a)\nonumber\\
&+& f ^\Lambda_I (c_3 \bar\psi^A
\gamma_a \lambda^I_A V^a + c_4 \bar \chi^{ABC} \gamma_{ab} \lambda^{ID}\epsilon_{ABCD}V^a V^b) + h.c. \bigr ]\\
\hat P &=&  P - \bar \psi_A \chi_{BCD} \epsilon^{ABCD}\\
\hat P^I_{AB}&=& P^I_{AB} - (\bar\psi_A \lambda^I_B +\epsilon_{ABCD} \bar\psi^C\lambda^{ID})\\
\end{eqnarray}
where $P= P_{,S} dS$ and $P^I_{AB} =P^I_{AB, i}d\phi^i $ are the vielbein of ${SU(1,1)\over U(1)}$ and ${O(6,n) \over O(6)
     \times O(n)}$
respectively.
The fermion transformation laws are:
\begin{eqnarray}
\delta\psi_A &=& D \epsilon_A +a_1 T^-_{AB \vert ab}\Delta^{abc} \epsilon^B V_c
 + \cdots\\
\delta\chi_{ABC} &=&a_2 P_{,S}\partial_a S \gamma^a \epsilon^D \epsilon_{ABCD} +
a_3 T^-_{[AB \vert ab}\gamma^{ab} \epsilon_{C]} + \cdots\\
\delta\lambda^I_A &=&a_4 P^I_{AB,i}\partial_a \phi^i \gamma^a \epsilon^B + a_5 T^{- I}_{ ab} \gamma^{ab}\epsilon_A + \cdots
\end{eqnarray}
where the 2--forms $T_{AB}$ and $T_{I}$ are defined in eq.(\ref{gravi})
By integration of these two-forms, using eq.(\ref{tiden})--(\ref{charges}) we find the  central and matter dyonic charges
given in eq.s (\ref{zab}), (\ref{zi}).
From the Maurer-Cartan equations for $f,h$ and the
definitions of the charges one easily finds:
\begin{eqnarray}
  \label{charge4,4}
  \nabla^{ SU(4)\otimes U(1)} Z_{AB} &=& \bar  Z^I P_{I AB} + {1 \over 2} \epsilon_{ABCD}\bar Z^{CD}P  \\
  \nabla^{ SO(n)} Z_{I} &=&{1 \over 2} \bar  Z^{AB} P_{I AB} + Z_I
  \bar P
\end{eqnarray}
In terms of the kinetic matrix (\ref{n44}) the sum rule for the charges is given by eqs.(\ref{sumrule})--(\ref{eg}):
 \begin{equation}
 {1 \over 2} Z_{AB} \bar Z^{AB} + Z_I \bar Z_I = -{1\over 2} P^t \cM (\cN) P
\end{equation}
}
}\end{itemize}
For $N>4$ the only available supermultiplet is the gravitational one,
so that $H_{matter}=\bfone$.
The embedding procedure is much simpler than in the matter coupled supergravities since for each $N>4$ there
exists a representation of the scalar manifold isometry group $G$ given in terms of $Usp(n_v,n_v)$ matrices.
\begin{itemize}
\item{ For the $N=5$ theory \cite{dwni}
 the coset manifold is:
\begin{equation}
G/H = {SU(1,5) \over U(5) }
\end{equation}
The field content and the group assignments are displayed in table
\ref{tab4,5}.
\begin{table}[ht]
\caption{Transformation properties of fields in $D=4$, $N=5$}
\label{tab4,5}
\begin{center}
\begin{tabular}{|c||c|c|c|c|c|c|}
\hline
& $V^a$ &$ \psi _A; $ &$\chi_{ABC},\chi_L $
&$A^{\Lambda\Sigma}$ &$L^{x}_{A  }$ & $R_H$  \\
\hline
\hline
$SU(1,5)$& 1 & 1 & 1 & -          &6   & -     \\
\hline
$SU(5)$ & 1 & 5 & $(10,1)$ & 1 & 5 &${\bar 5}$ \\
\hline
$U(1)$ & 0 & ${1\over 2}$ & $({3\over 2}, - {5\over 2}
)$ & 0 & 1 & 2 \\
\hline
\end{tabular}
\end{center}
\end{table}
\noindent
Here $x,y,\cdots = 1,\cdots,6$ and  $A,B,C\cdots
=1,\cdots,5$ are  indices of the fundamental representations of $SU(1,5)$ and
 $ SU(5)$,respectively.
$L^x_A$ denote as usual the coset representative in the
fundamental representation of $SU(1,5)$. The antisymmetric couple $\Lambda
\Sigma$, $\Lambda,
\Sigma = 1,\cdots,5$, enumerates the ten vectors.
 The embedding
of $SU(1,5)$ into  the Gaillard-Zumino  group $Usp(10,10)$ is given in terms
of the three-times antisymmetric representation of $SU(1,5)$, a generic element
$t^{xyz}$ satisfying:
\begin{equation}
  \label{txyz}
  t^{xyz}={1 \over 3!} \epsilon^{xyzuvw}t_{uvw}
\end{equation}
 We may decompose $t^{xyz}$ as follows:
\begin{equation}
  \label{txyz1}
  t^{xyz}=\pmatrix{t^{\Lambda\Sigma 6}\cr
t^{\Lambda\Sigma\Gamma}= \epsilon^{\Lambda\Sigma\Gamma \Delta\Pi 6}
t_{\Delta\Pi 6}} \qquad (\Lambda,\Sigma,\cdots= 1,\cdots,5)
\end{equation}
In the following we write $t^{\Lambda\Sigma 6}\equiv t^{\Lambda\Sigma}$.
The 20 dimensional vector $(F^{\mp\Lambda\Sigma}, \cG^\mp_{\Lambda\Sigma})$
transforms under $Sp(20,\IR)$, as well as, for fixed $AB$, each of the 20-- dimensional vectors
$(f^{\Lambda\Sigma}_{AB},h_{\Lambda\Sigma AB})$ of the
embedding matrix:
\begin{equation}
U = {1\over \sqrt{2}}\pmatrix{f + {\rm i } h & \bar f + {\rm i }\bar h \cr
f - {\rm i } h & \bar f - {\rm i }\bar h \cr}
\end{equation}

The supercovariant field-strengths and  vielbein are:
\begin{eqnarray}
\hat F^{\Lambda\Sigma} &=& d A ^{\Lambda\Sigma} +
\bigl (f^{\Lambda\Sigma}_{\ \ \ AB} (a_1 \bar\psi^A \psi^B + a_2
\bar \psi_C \gamma_a \chi^{ABC}V^a ) + h. c.\bigr ) \\
\hat P_{ABCD} &=& P_{ABCD}- \bar \chi_{[ABC}\psi_{D]}
- \epsilon_{ABCDE} \bar\chi^{(R)} \psi^E
\end{eqnarray}
where $P_{ABCD}= \epsilon_{ABCDF} P^F$ is the complex vielbein, completely antisymmetric
in $SU(5)$
indices and $(P_{ABCD})^{\star} = \bar P^{ABCD}$.\\
The fermion transformation laws are:
\begin{eqnarray}
\delta\psi_A &=& D \epsilon _A  + a_3 T^-_{
 AB \vert ab} \Delta^{abc} \epsilon^B V_c
+ \cdots\\
\delta\chi_{ABC} &=&a_4 P_{ABCD , i}\partial_a \phi^i \gamma^a \epsilon^D +
a_5 T^-_{
 [AB \vert ab}  \gamma^{ab}
\epsilon_{C]} + \cdots \\
\delta \chi_{(L)} &=& a_6 \bar P^{ABCD}_{,\bar \imath}\partial_a \bar\phi^{\bar\imath} \gamma^a \epsilon^E
\epsilon_{ABCDE} + \cdots
\end{eqnarray}
where:
\begin{eqnarray}
  T_{AB}& =&-{{\rm i} \over 2}(\bar f^{-1})_{\Lambda\Sigma AB}F^{\Lambda\Sigma} =
{1 \over 4}(\cN -\bar \cN)_{\Lambda\Sigma,\Gamma\Delta}f^{\Gamma\Delta}_{\ \ AB}
F^{\Lambda\Sigma}\nonumber\\& =&{1 \over 2} (h_{\Lambda\Sigma AB}F^{\Lambda\Sigma} -
f^{\Lambda\Sigma}_{\ \ AB} \cG_{\Lambda\Sigma})\\
\cN_{\Lambda\Sigma, \Delta\Pi}& =& {1 \over 2}
h_{\Lambda\Sigma\vert AB} (f^{-1})^{AB}_{\ \ \Delta\Pi}\\
\cG^\pm _{\Lambda\Sigma} &=& -{\rm i}/2{\partial \cL \over \partial F^{\pm\Lambda\Sigma}};
\quad  \cG_{\Lambda\Sigma}=\cG^+_{\Lambda\Sigma}+\cG^- _{\Lambda\Sigma}
\end{eqnarray}
With a by now familiar procedure one finds the following (complex)
central charges:
\begin{equation}
Z_{AB} = {1 \over 2}( h_{\Lambda\Sigma\vert AB} g^{\Lambda\Sigma}
- f^{\Lambda\Sigma}_{\ \ \ AB}e_{\Lambda\Sigma}
 )
\end{equation}
where:
\begin{eqnarray}
g^{\Lambda\Sigma} &=& \int_{S^2} F^{\Lambda\Sigma}\\
 e_{\Lambda\Sigma} &=& \int_{S^2} \cG_{\Lambda\Sigma}
\end{eqnarray}
From the Maurer--Cartan equation
\begin{equation}
\nabla^{(U(5))} f^{\Lambda\Sigma}{\ \  AB} = {1 \over 2}\bar
f^{\Lambda\Sigma \vert CD}P_{ABCD}
\end{equation}
and the analogous one for $h$ we find:
\begin{equation}
\nabla^{(U(5))} Z_{AB} = {1 \over 2}\bar
Z^{CD}P_{ABCD}
\end{equation}
Finally, the sum rule for the central charges is:
\begin{equation}
{1 \over 2} Z_{AB}\bar Z^{AB} = - {1 \over 2} (g^{\Lambda\Sigma},e_{\Lambda\Sigma})
\cM(\cN)_{\Lambda\Sigma , \Gamma\Delta} \pmatrix{g^{\Gamma\Delta}\cr
e_{\Gamma\Delta}}
\end{equation}
where the matrix $\cM (\cN)$ has exactly the same form as in eq
(\ref{m+}).
}
\item{The scalar manifold of the $N=6$ theory has the coset structure:
\begin{equation}
G/H = {SO^\star (12) \over U(6)}
\end{equation}
We recall that $SO^\star (2n)$ is defined as the subgroup of
$O(2n,\IC)$ that preserves the sesquilinear antisymmetric metric:
\begin{equation}
L^\dagger C L = C, \qquad C= \pmatrix{0 & \bfone \cr -\bfone & 0 \cr}
\end{equation}
The field content and transformation properties are given in
Table \ref{tab4,6},
\begin{table}[ht]
\caption{Transformation properties of fields in $D=4$, $N=6$}
\label{tab4,6}
\begin{center}
\begin{tabular}{|c||c|c|c|c|c|c|}
\hline
& $V^a$ &$ \psi _A$ &$\chi_{ABC},\chi_A$
&$A^\Lambda$ &$S^\alpha_r $  & $R_H$ \\
\hline
\hline
$SO^\star(12)$& 1 & 1 & 1 &- & 32 & - \\
\hline
$SU(6)$ & 1 & $6$ & $(20 + 6)$ & 1 &$( 15, 1)+(\bar{15},\bar 1)$ & $ {\bar 15}$  \\
\hline
$U(1)$ & 0 & ${1\over 2}$ & $({3\over 2}, -{5\over 2})$ & 0 &$(1,-3) + (-1,3)$ & 2 \\
\hline
\end{tabular}
\end{center}
\end{table}
where $A, B, C = 1,\cdots,6$ are $SU(6)$ indices in the fundamental
representation and $ \Lambda = 1,\cdots, 16 $.
 As it happens  in the $N=5$ theory,
 the $\underline{32}$ spinor
representation of $SO^\star (12)$  can be given in terms of
a  $Usp(16,16)$ matrix, which we denote by $S^\alpha_r$($\alpha,r = 1,\cdots,32 $), so that the embedding
is automatically realized in terms of the spinor representation.
Employing the usual notation we may set:
 \begin{equation}
S^\alpha_r = {1\over \sqrt{2}}\pmatrix{f^\Lambda _I + {\rm i} h_{\Lambda I} &
 \bar f^\Lambda _I + {\rm i} \bar h_{\Lambda I}    \cr
 f^\Lambda _I - {\rm i} h_{\Lambda I} &
 \bar f^\Lambda _I - {\rm i} \bar h_{\Lambda I}    \cr}
\end{equation}
where $\Lambda,I=1,\cdots,16$.
With respect to  $SU(6)$,the sixteen symplectic vectors
$(f^\Lambda_I,h_{\Lambda I})$,
($I = 1,\cdots,16$) are  reducible into the antisymmetric 15--
dimensional representation plus a singlet of $SU(6)$:
\begin{equation}
(f^\Lambda_I,h_{\Lambda I})\to (f^\Lambda_{AB},h_{\Lambda AB})+
(f^\Lambda,h_{\Lambda})
\end{equation}
It is precisely the existence of a $SU(6)$ singlet which allows
for the Special Geometry structure of ${SO^*(12) \over U(6)}$
as discussed in section $3.2$
Note that the coset element $S^\alpha_r$ has no definite $U(1)$
weight since the submatrices
$f^\Lambda_{AB},f^\Lambda$ have the weights 1 and -3 respectively.
The supercovariant field-strenghts and the coset manifold vielbein have the
following expression:
\begin{eqnarray}
\hat F^{\Lambda} &=& d A ^{\Lambda} +\bigl [
f^{\Lambda}_{\  AB} (a_1 \bar\psi^A \psi^B + a_2
\bar \psi_C \gamma_a \chi^{ABC}V^a ) \nonumber\\
&+& a_3
f^\Lambda \bar \psi_C \gamma_a \chi^{C}V^a + h. c. \bigr ] \\
\hat P_{ABCD} &=& P_{ABCD} -  \bar \chi_{[ABC}\psi_{D]}
- \epsilon_{ABCDEF} \bar\chi^E \psi^F
\end{eqnarray}
where $P_{ABCD}=P_{ABCD,i} dz^i$ is the K\"ahler vielbein of the coset.
The fermion transformation laws are:
\begin{eqnarray}
\delta\psi_A &=& D \epsilon _A  + b_1 T^-_{AB\vert ab}
 \Delta^{abc} \epsilon^B V_c
+ \cdots\\
\delta\chi_{ABC} &=&  b_2 P_{ABCD\vert i}\partial_a z^i \gamma^a \epsilon^D +
b_3T^-_{[AB\vert ab} \gamma^{ab}
\epsilon_{C]} + \cdots\\
\delta \chi_A &=& b_4 P^{BCDE}_{,i} \partial_a z^i \gamma^a \epsilon^F
\epsilon_{ABCDEF} +b_5T^-_{ab}\gamma^{ab} \epsilon_A + \cdots
\end{eqnarray}
where:
\begin{eqnarray}
T_{AB}&=&- {\rm i}  (\bar f^{-1})_{\Lambda AB} F^{- \Lambda}\\
T&=& - {\rm i}(\bar f^{-1}) _{\Lambda
} F^{-\Lambda}
\end{eqnarray}
With the usual procedure we have the following complex dyonic central
charges:
\begin{eqnarray}
Z_{AB} &=&  h_{\Lambda AB} g^\Lambda - f^\Lambda_{AB} e_\Lambda \\
 Z &=&   h_{\Lambda} g^\Lambda - f^\Lambda e_\Lambda
\end{eqnarray}
 in the $\underline{15}$ and singlet representation of $SU(6)$
 respectively.
Notice that although we have 16 graviphotons, only 15 central charges
are present in the supersymmetry algebra.
The singlet charge plays a role analogous to a ``matter'' charge.
From the Maurer--Cartan equations:
\begin{eqnarray}
\nabla f^\Lambda_{\ AB} &=&{1 \over 2} \bar f^{\Lambda \vert CD} P_{ABCD} +
{1 \over 4!}\bar f^{\Lambda } \epsilon_{ABCDEF} P^{CDEF} \\
\nabla f^\Lambda &=& {1 \over 2! 4!}f^{\Lambda \vert AB}\epsilon_{ABCDEF}
P^{CDEF}
\end{eqnarray}
and the relation (\ref{nfh-1}) one finds:
\begin{eqnarray}
\nabla ^{(U(6))} Z_{AB} &=& {1 \over 2} \bar Z^{CD} P_{ABCD} + {1 \over 4!}\bar Z
\epsilon_{ABCDEF}P^{CDEF} \\
\nabla^{(U(1))} Z &=& {1 \over 2! 4!}\bar Z^{AB} \epsilon_{ABCDEF}P^{CDEF}
\end{eqnarray}
and the sum-rule (\ref{sumrule}):
\begin{equation}
  \label{sumrule4,6}
{1 \over 2}Z_{AB}\bar Z^{AB} + Z\bar Z = -{1 \over 2} \pmatrix{g^\Lambda,& e_\Lambda \cr} \cM(\cN)_{\Lambda\Sigma}\pmatrix{
g^\Sigma \cr e_\Sigma \cr}
\end{equation}
with the usual meaning for $\cM(\cN)$ (see eq.(\ref{m+})).
}
\item{In  the $N=8$ case \cite{crju}  the coset manifold is:
\begin{equation}
G/H={E_{7(-7)}\over SU(8)}.
\end{equation}
The field content and group assignments are given in the following Table
\ref{tab4,8}:
\begin{table}[ht]
\begin{center}
  \caption{ Field content and group assignments in $D=4$, $N=8$ supergravity}
  \label{tab4,8}
  \begin{tabular}{|c||c|c|c|c|c|c|}
\hline
&$V^a_\mu $ &$ \psi_A $ & $A^{\Lambda\Sigma}_\mu$ &
 $ \chi_{ABC}$ & $S^\alpha_r$ & $R_H$ \\
\hline
\hline
$E_{7(-7)}$ & 1 & 1 & - & 1 & 56 & - \\
\hline
$SU(8)$ & 1 & 8 & 1 & 56 & $ 28 + \bar{28}$ & 70 \\
\hline
 \end{tabular}
\end{center}
\end{table}

As in $N=5,6$, the embedding is automatically realized in terms
 of the $\underline{56}$ defining representation for $E_7$
 which belongs to $Usp(28,28)$ and it is given by the usual coset element (\ref{defu})
where
\begin{eqnarray}
f+ {\rm i} h& \equiv &f^{\Lambda\Sigma}_{\ \ AB} + {\rm i}
 h_{\Lambda\Sigma AB}\\
\bar f - {\rm i} \bar h& \equiv & \bar f^{\Lambda\Sigma AB} - {\rm i}
 \bar h_{\Lambda\Sigma}^{\ \ AB}
\end{eqnarray}
$\Lambda\Sigma,AB$ are couples of antisymmetric  indices, with
 $\Lambda,\Sigma,A,B$
 running
 from 1 to 8 .
The supercovariant field-strengths and  coset manifold vielbein are:
\begin{eqnarray}
  \hat F^{\Lambda\Sigma} &=& dA^{\Lambda\Sigma} +[ f^{\Lambda\Sigma}_{\ \ AB}(
a_1\bar\psi^A \psi^B + a_2 \bar\chi^{ABC} \gamma_a \psi_C V^a) + h.c.]\\
\hat P_{ABCD} &=&  P_{ABCD} -   \bar\chi_{[ABC}\psi_{D]} + h.c.
\end{eqnarray}
where $ P_{ABCD}= {1 \over 4!} \epsilon_{ABCDEFGH}\bar P^{EFGH}\equiv (L^{-1} \nabla^{SU(8)} L)_{AB\vert CD}=
P_{ABCD,i}d\phi^i$
($\phi^i$ coordinates of $G/H$).
 The fermion transformation laws are given by:
\begin{eqnarray}
  \delta\psi_A &=& D \epsilon_A + a_3 T^-_{AB \vert ab} \Delta^{abc}
\epsilon^B V_c + \cdots\\
\delta\chi_{ABC} &=& a_4 P_{ABCD,i} \partial_a \phi^i \gamma^a\epsilon^D + a_5T^-_{[AB\vert ab}
\gamma^{ab} \epsilon_{C]}+ \cdots
\end{eqnarray}
where:
\begin{eqnarray}
  T_{AB}& =&-{{\rm i} \over 2}(\bar f^{-1})_{\Lambda\Sigma AB} F^{\Lambda\Sigma}=
{1 \over 4} (\cN - \bar \cN)_{\Lambda\Sigma, \Gamma\Delta} f^{\Lambda\Sigma} _{AB}
F^{\Gamma\Delta}\nonumber\\
&=& {1 \over 2}(h_{\Lambda\Sigma AB} F^{\Lambda\Sigma}- f^{\Lambda\Sigma}_{\ \ AB}
\cG_{\Lambda\Sigma} )
\end{eqnarray}
with:
\begin{eqnarray}
\cN_{\Lambda\Sigma, \Gamma\Delta}&=& {1 \over 2}h_{\Lambda\Sigma AB}
 (f^{-1})^{ AB}_{\ \ \Gamma\Delta}  \\
\cG_{\Lambda\Sigma} &=& -{\rm i/2}{\partial \cL \over \partial F^{\Lambda\Sigma}}
\end{eqnarray}
With the usual manipulations we obtain the central charges:
\begin{equation}
  Z_{AB}={1 \over 2}( h_{\Lambda\Sigma AB}
g^{\Lambda\Sigma} - f^{\Lambda\Sigma}_{\ \ AB} e_{\Lambda\Sigma}),
\end{equation}
the differential relations:
\begin{equation}
  \nabla^{SU(8)}Z_{\ AB}= {1 \over 2} \bar Z^{\ CD} P_{ABCD}
\end{equation}
and the sum rule:
\begin{equation}
 {1 \over 2} Z_{AB}\bar Z^{AB} = - {1 \over 8} (g^{\Lambda\Sigma}, e_{\Lambda\Sigma})
\cM(\cN)_{\Lambda\Sigma , \Gamma\Delta}\pmatrix{g^{\Gamma\Delta} \cr e_{\Gamma\Delta}}
\end{equation}
}
\end{itemize}
 \section{Matter coupled higher dimensional supergravities}
\setcounter{equation}{0}
 With the exception of $D=5$, $N=2$ supergravity in which
 the vector multiplets moduli space is described by Very Special Geometry
\cite{gusito}\cite{vpdw2}, all the higher
 dimensional supergravities exhibit a coset structure $G/H$ as in
 $D=4$, $N\geq 3$.
 As we are going to see, their structure is completely fixed in terms
 of the coset representative $L$ and in particular the ``dressed''
 central and matter charges satisfy relations quite analogous to
 those discussed in the four dimensional case for $N\geq 2$.
 \par
 In this section we discuss the matter coupled supergravities
 deferring the maximally extended cases to next section.
 \par
 We begin by considering all the matter coupled supergravities which
 have the coset structure:
 \begin{equation}
G/H = {O(10-D,n) \over O(10-D)\otimes O(n)} \times O(1,1)
\label{gsuh}
\end{equation}
Actually this class covers all the cases except $D=6$, $N=(2,0)$ and $D=6$,
type $ IIB$.
\par
It is convenient to treat separately the odd and the even dimensional
cases.
\subsection{Odd dimensional theories}
The three cases $D=5,7,9$ can be discussed at the same time \cite{cre},
 \cite{awto}, \cite{toni}, \cite{salsez}, \cite{bks}, \cite{gns}
 if one takes in $D=5$ one of the vectors dualized with a two form and
 disregard the presence of hypermultiplets.
 With these assumptions  the field content for all the three
 theories is given by:\\
Gravitational multiplet:
\begin{equation}
(V^a_\mu, \psi_{A\mu}, B_{\mu\nu},A^\alpha _ \mu, \chi _A ,\sigma)
\qquad\quad (\mu=1,\cdots,D)  \label{fcontent3}
\end{equation}
Vector multiplet:
\begin{equation}
(A_\mu,\lambda_A,(10 -D)\phi)^I \qquad \quad (I=1,\cdots,n)
\label{fcontent4}
\end{equation}
where $\alpha$ runs from 1 to $10 - D$, the coset manifold of the
scalar fields being given in equation (\ref{gsuh}).
\par
The transformation properties of the fields are given in Table \ref{tabodd},
\begin{table}[ht]
\begin{center}
\caption{Transformation properties of fields in
matter coupled $D=5$ ($N=4$), $D=7$ ($N=2$), $D=9$ ($N=1$) }
\label{tabodd}
\begin{tabular}{|c||c|c|c|c|c|c|c|c|c|}
\hline
&$ V^a_\mu $& $H_{\mu\nu\rho}$ & $F^\Lambda_{\mu\nu}$
&
$L^\Lambda_{\ \Sigma}$& $e^\sigma$ & $ \psi^A_\mu $ &
$ \chi^{ A}$& $\lambda^{IA}$ & $R_H$\\
\hline
\hline
 $O(10 -D,n)$ & 1 & 1 &$10 - D+n$  &$ 10 - D+n$  & 1 & 1 & 1 & 1 & - \\
\hline
$O(1,1)$ & 0 & 2 & 1 & 0 & 1 & 0 & 0 & 0 & - \\
\hline
$H_{Aut}$ & 1 &1 & 1  &
$10 - D$ & 0 &  $N$ &  $ N $ &  $ N $ & $10-D$ \\
\hline
$ H_{matter}$ & 1 & 1 & 1 & $n$ & 0 & 1 & 1 & $n$ & $n$ \\
\hline
\end{tabular}
\end{center}
\end{table}
where:
\begin{equation}
 H_{Aut}\times H_{matter}=O(10-D) \times O(n).
\end{equation}
  $H_{Aut}$ acts on the index $A=1,\cdots,N$ of the
  spinors as a unitary symplectic group  in
  $D=5,7$ according to:
  \begin{eqnarray}
   O(10-D) & \sim & Usp(4) \qquad \qquad D=5 \nonumber\\
& \sim & Usp(2) \qquad \qquad D=7.
\end{eqnarray}
In $D=9$ $H_{Aut} \equiv \bfone $. $H_{matter}$ always acts in the vector
 representation labelled by the index $I$.
 The coset representative of (\ref{gsuh}) is:
 \begin{equation}
e^\sigma L^\Lambda_{\ \Sigma}\equiv e^\sigma(L^\Lambda_{\
AB},L^\Lambda_{\ I})
\end{equation}
where $e^\sigma$ parametrizes $O(1,1)$, $\sigma$ being the real
 scalar field of the gravitational multiplet and $ L^\Lambda_{\ \Sigma}$ is the
representative of ${O(10-D,n) \over O(10-D)\otimes O(n)}$.
The indices $AB$ of $L^\Lambda_{\ AB} $ are $Usp(N)\equiv H_{(Aut)}$
antisymmetric
indices for $D=5$ ($N=4$) and symmetric indices for $D=7$ ($N=2$)
 intertwining between the vector representation of
$O(10- D)$ and  the
representation of  $Usp(N)$:
\begin{equation}
L^\Lambda_{\ AB} =L^\Lambda_r \gamma^r_{AB}
\end{equation}
where
$\gamma^r_{AB}$ are gamma--matrices of $O(10-D)$.
For $D=9$  $L^\Lambda_{\ AB} \to
L^\Lambda\delta_{AB} =L^\Lambda$ (A,B=1).
The index $I$ of $L^\Lambda_I$ is an index of $O(n)=H_{(matter)}$ in
the vector representation.
As usual, the set of $n_g$ gravitational and $n_v$ matter
field--strengths $F^\Lambda$ of the vectors $(A^\alpha, A^I)$
    $(\Lambda=1,\cdots, n_g+n_v)$
transform among themselves under $O(10-D,n)\times O(1,1)$ while
$H\equiv dB$ is charged under $O(1,1)$ only
(we have labelled the $O(1,1)$ representations by the
"charge" under the shift $e^\sigma \to e^{\sigma + c}$).
\par
The supercovariant field--strengths contain, as in 4 dimensions, the coset
representatives, but in the defining representation of $G$.
We have \footnote{In all higher dimensional theories the space--time $\gamma$--matrices
are denoted as $\Gamma^a$ ( $a= 0,\cdots,D-1$).}:
\begin{eqnarray}
\hat H &= & dB + a_1
\eta_{\Lambda\Sigma}dA^\Lambda \wedge A^\Sigma +
a_2 e^{2\sigma} \bar \psi^A \Gamma_a \psi^B \IC_{AB}V^a \nonumber \\
&+& a_3 e^{2\sigma} \bar \psi^A \Gamma_{ab} \chi^B \IC_{AB} V^a V^b
 \label{acca}\\
\hat F^\Lambda &= & dA^\Lambda + b_1
e^{\sigma} \bar \psi^A \psi^B L^\Lambda_{\ AB} +
 b_2 e^{\sigma} \bar \psi^A \Gamma_{a} \chi^B V^a L^\Lambda_{\ AB}  \nonumber \\
&+& b_3 e^{\sigma} \bar \psi^A \Gamma_{a} \lambda^I_A V^a L^\Lambda_{\ I}
 \label{effe}\\
\hat P^I_{r} &=& (\gamma_r)^{AB} \hat P^I_{AB}= (\gamma_r)^{AB} ( P^I_{AB} - \bar \psi_{A} \lambda^I_{B})\\
\hat d{ \sigma}& =& d \sigma - \bar \psi^A \chi_A\\
\end{eqnarray}
where $\IC_{AB}$ is the invariant metric of $Sp(N)$ in $D=5,7$ while
in $D=9$ we set
 $\IC_{AB} \equiv \delta_{AB}$, $A,B\equiv 1$.
$P^I_{AB}=P^I_{AB,i}d\phi^i $ and $ d\sigma$ are the vielbein 1--forms of
 ${O(10-D,n) \over O(10-D)\otimes O(n)}$ and of $O(1,1)$ respectively.
 \par
 The supercovariant field strengths appear in the transformation laws
 of the fermions dressed with coset representatives:
\begin{eqnarray}
\delta \psi_A &=& D\epsilon_A +  c_1
T_{AB \vert ab}\Delta^{abc}
 \epsilon^B V_c  +  c_2
T_{\vert abc}\Delta^{abcd}
 \epsilon_A V_d + \cdots  \label{gratr}\\
\delta \chi_A &=& d_1 \Gamma^a \partial_a \sigma \epsilon_A +d_2
T_{AB \vert ab}\Gamma^{ab}
 \epsilon^B   +  d_3
T_{\vert abc}\Gamma^{abc}
 \epsilon_A + \cdots \label{diltr}\\
\delta \lambda^I_A& =& f_1 \Gamma^a P^I_{AB,i}
\partial_a \phi^i \epsilon^B +f_2
T^I_{\vert ab}\Gamma^{ab}
 \epsilon_A + \cdots \label{gautr}
\end{eqnarray}
 where $\Delta_{a a_1 \cdots a_n}$ is given in equation (\ref{delta}),
  $\phi^i$ are
 the coordinates of ${O(10-D,n)\over O(10-D)
 \otimes O(n)}$ and
  \begin{equation}
\matrix{T_{AB}&=& e^\sigma \cN_{\Lambda\Sigma} L^\Sigma_{\ AB}
 F^\Lambda & \equiv & e^{-\sigma} L_{\Lambda AB}
 F^\Lambda \cr
  T_{I}&=& e^\sigma \cN_{\Lambda\Sigma} L^\Sigma_{\ I}
 F^\Lambda & \equiv & e^{-\sigma} L_{\Lambda I}
 F^\Lambda \cr
T&=& \cN^{(H)} e^{2\sigma} H & \equiv & e^{-2\sigma} H \cr
} \label{ti}
\end{equation}
 are the dressed graviphotons, matter vectors and 3--form.
Here the vector kinetic matrix $\cN_{\Lambda\Sigma}$ and the 3--form metric
$\cN^{(H)}$ are:
\begin{eqnarray}
\cN_{\Lambda\Sigma}&=&e^{-2\sigma} (L_{AB \Lambda } L^{ AB}_{\Sigma } +
L_{I\Lambda} L_{ I \Sigma}) \label{nv}\\
\cN ^{(H)}& =& e ^{-4\sigma} \label{nh}
\end{eqnarray}
 and we have used the pseudo--orthogonality relations of $O(10 - D,
 n)$ :
 \begin{equation}
\eta_{\Lambda\Sigma}= (L_{\Lambda AB} L_\Sigma ^{\ AB} -
 L_{\Lambda I} L_{\Sigma I}) \label{eta}
\end{equation}
 where $\eta$ is the $O(10-D,n)$ invariant metric.
At this point it is very easy to construct the dressed charges for
 the gravitational and matter multiplets.
 Defining the magnetic charge as:
 \begin{equation}
g^\Lambda = \int_{S^2} F^\Lambda
\end{equation}
\begin{equation}
g = \int_{S^3} H
\end{equation}
  we have the vector central charge:
\begin{equation}
Z_{AB} = \int_{S^2} T_{AB} = e^{-\sigma} L_{\Lambda AB}(\phi)
g^\Lambda
\end{equation}
the 3--form central charge:
  \begin{equation}
Z = \int_{S^3} T = e^{-2\sigma}g
\end{equation}
 and the gaugino matter charge:
\begin{equation}
Z_{I} = \int_{S^2} T_{I} = e^{-\sigma} L_{\Lambda I}(\phi)
g^\Lambda
\label{ziodd}
\end{equation}
where on the r.h.s. the scalar fields $\sigma,\phi^i$ are understood
as $v.e.v.$'s on a given background.\\
From the Maurer Cartan equation for the coset representative
$e^\sigma L^\Lambda_{\ \Sigma}$ we now have:
\begin{equation}
\nabla ( e^\sigma L^\Lambda_{\ AB})= e^\sigma L^\Lambda_{\ I}
P^I_{\ AB}
\end{equation}
where $\nabla$ denotes the derivative covariant under the composite
connection of $H_{Aut}\times H_{matter}\equiv O(10-D)\otimes
O(n)$.
It follows:
\begin{eqnarray}
\nabla _i Z_{AB} &=& Z_{I} P^I_{\ AB,i} \\
 \nabla Z_{I} &=& {1\over 4}(P_I^{AB} Z_{AB} + \bar P_{I AB} \bar Z^{AB})- Z_{I}d \sigma
\end{eqnarray}
 In particular, if $Z_I=0$, the invariant
 \begin{equation}
\cM(\phi)= {1 \over 2} Z_{AB} \bar Z^{AB}
\end{equation}
 reaches a minimum.\\
 Furthermore, using eq.s (\ref{nv}) and  (\ref{eta}) one also finds:
 \begin{eqnarray}
{1 \over 2}Z_{AB}\bar Z^{AB} + Z_IZ_I &=& g^\Lambda \cN_{\Lambda\Sigma} g^\Sigma\\
{1 \over 2}Z_{AB}\bar Z^{AB} - Z_IZ_I &=& g^\Lambda \eta_{\Lambda\Sigma} g^\Sigma e^{-2\sigma}\\
\end{eqnarray}
 These ``obvious'' sum rules are the counterpart of the sum rules found
 in $N=2$, $D=4$ for the central and matter charges.\\
 Finally we observe that the electric charges associated to
 $F^\Lambda$ and $H$ are defined as:
 \begin{equation}
e_\Lambda=\int \cN_{\Lambda\Sigma} ^{\ \ \star}F^\Sigma
\qquad \qquad  e =\int e^{-4 \sigma \ \star}H
\end{equation}
Let us note that $N=4$, $D=5$ is usually formulated dualizing the
  3--form $H^{(3)}$ in terms of an extra 2--form $F=dB$
  \begin{equation}
dB= ^\star H e^{-4\sigma}
\end{equation}
In this case, instead of (\ref{acca}), we have:
\begin{equation}
\hat F=dB + f_1 e^{-2\sigma}\bar\psi ^A \psi^B  \IC_{AB} +
f_2  e^{-2\sigma}\bar\psi ^A\gamma_a \chi^B V^a \IC_{AB}
\end{equation}
 and the transformation rules of $\delta\psi_A,\delta\chi_A$ are
 changed accordingly:
 \begin{eqnarray}
\delta \psi_A &=& D\epsilon_A +  a \bigl (
T_{AB \vert ab}\Delta^{abc}
 \epsilon^B V_c  +  b
T_{\vert ab}\Delta^{abc} \IC_{AB}
 \epsilon^B V_c \bigr ) + \cdots \\
\delta \chi_A &=& c \Gamma^a \partial_a \sigma \epsilon_A +
d (T_{AB \vert ab}\Gamma^{ab}
 \epsilon^B - 2b T_{\vert ab}\Gamma^{ab}
 \epsilon_A) + \cdots \\
\end{eqnarray}
where $T_{ab}$ =$ e^{2 \sigma} F_{ab}$.
Note that in this case the kinetic matrix for the singlet is:
\begin{equation}
\cN= e^{4\sigma}
\end{equation}

We now have that the central charges are:
\begin{eqnarray}
Z_{AB} &=& e^{-\sigma} g^\Lambda L_{\Lambda AB} -
b e^{2\sigma} m\, \IC_{AB}\\
Z &=& e^{2\sigma} m
\end{eqnarray}
where $m=\int_{S^2} F$ ,
while the matter charges are the same as in (\ref{ziodd}).
\par
Note that the 2--forms $T_{AB}$ and $T$ appear in $\delta \chi_A$
 with a relative weight, $-2b$, which is fixed by supersymmetry.
By  $Usp(4)$--covariant differentiation  one obtains:
\begin{eqnarray}
\nabla Z_{AB} &=& Z^I P_{IAB} - 2 Z^{(\chi)}_{AB} d \sigma  \\
\nabla Z &=& 2 Z d \sigma \\
\nabla Z_{I} &=&  { 1\over 4} (Z_{AB} P^{AB}_{\ I} + \bar Z^{AB} P_{AB I})
- Z_I d\sigma
\end{eqnarray}
where:
\begin{equation}
Z^{(\chi)}_{AB} = {1 \over 2} (Z_{AB} + 3 b Z \IC _{AB})
\label{zeta}
\end{equation}

The sum rules are:
 \begin{eqnarray}
{1 \over 2}Z_{AB} \bar Z^{AB}-2 Z^2  -Z_I Z^I &=&
g^\Lambda \cN_{\Lambda\Sigma} g^\Sigma \\
Z^2 &=& m^2 \cN
\end{eqnarray}

\subsection{Even dimensional matter coupled theories}
\begin{itemize}
\item{In $D=8$ , $N=1$ \cite{sasez2}, \cite{awto2} we have again a coset manifold of the class
(\ref{gsuh}), namely:
\begin{equation}
G/H= {O(2,n) \over O(2) \times O(n)} \times O(1,1)
\end{equation}
The field content is: \\
 Gravitational multiplet:
\begin{equation}
(V^a_\mu, \psi_{\mu}, B_{\mu\nu},A^\alpha _ \mu, \chi,\sigma)
\qquad\quad (\mu=1,\cdots,8)  \label{fco1}
\end{equation}
 Vector multiplets:
\begin{equation}
(A_\mu,\lambda,2\phi)^M \qquad \quad (M=1,\cdots,n_v)
\label{fco2}
\end{equation}
where $\psi, \chi, \lambda$ are complex  Weyl spinors.
The group assignments can be read from Table \ref{tab8},
 \begin{table}[ht]
\caption{Transformation properties of fields in
matter coupled $D=8$ ($N=1$) }
\label{tab8}
\begin{center}
\begin{tabular}{|c||c|c|c|c|c|c|c|c|c|}
\hline
&$ V^a_\mu $& $H_{\mu\nu\rho}$ & $F^\Lambda_{\mu\nu}$
&
$L^\Lambda_{\ \Sigma}$& $e^\sigma$ &
$ \psi_{\mu L} $ & $ \chi_L$& $\lambda^{I}_L$ & $R_H$ \\
\hline
\hline
 $O(2,n)$ & 1 & 1 &$2+n$  &$ 2+n$  & 1 & 1 & 1 & 1 & -  \\
\hline
$O(1,1)$ & 0 & 2 & 1 & 0 & 1 & 0 & 0 & 0 & - \\
\hline
$U(1)\times O(n)$ & $(0,1)$ &  $(0,1)$ & $(0,1)$  &
$(2,n)$ &  $(0,0)$ &  $(1,1)$ &  $(-1,1)$ &  $(1,n)$ & $(2,n)$ \\
\hline
\end{tabular}
\end{center}
\end{table}
 \noindent
where the first entries in the last row are the weight under
$H_{Aut} = U(1) $.
From the coset representative decomposition:
\begin{equation}
e^\sigma L^{\Lambda}_{\Sigma} = e^\sigma ( L^\Lambda_A , L^\Lambda _I)
\qquad \quad ( A = 1,2)
\end{equation}
setting $L^{\Lambda}=L_1^{\Lambda} + {\rm i} L_2^{\Lambda}$ we have the following supercovariant
field--strenghts:
 \begin{eqnarray}
\hat H &= & dB + a_1
\eta_{\Lambda\Sigma}dA^\Lambda \wedge A^\Sigma +
(a_2 e^{2\sigma} \bar \psi_R \Gamma_a \psi_L  + h.c.)
 V^a\nonumber\\
&+&( a_3 e^{2\sigma} \bar \psi_L \Gamma_{ab} \chi_L + h. c.) V^a V^b
 \label{acca8}\\
\hat F^\Lambda &= & dA^\Lambda + ( b_1
e^{\sigma} \bar \psi_L \psi_L \bar L^\Lambda + h. c.) \nonumber\\
&+&
 ( b_2 e^{\sigma} \bar \psi_L \Gamma_{a} \chi_R L^\Lambda+ h.c.)
 V^a +
 ( b_3 e^{\sigma} \bar \psi_L \Gamma_{a} \lambda^I_R L^\Lambda_{\ I} + h.c.) V^a\\
\hat P^I& =& P^I - \bar \psi_L \lambda^I_L - \bar \psi_R \lambda^I_R\\
\hat {d \sigma} &=& d\sigma - \bar \psi_L \chi_L - \bar \psi_R \chi_R\\
 \label{effe8}
\end{eqnarray}
where $P^I \equiv P^I_{i} dz^i $ is the K\"ahlerian vielbein of
$ {O(2,n) \over O(2) \times O(n)}$ and $d\sigma$ the einbein of $ O(1,1)$.
We have now a complex graviphoton:
\begin{equation}
T^{(2)} = e^{-\sigma} L_\Lambda  F^\Lambda
\end{equation}
while the  dressed vector matter and tensor charges are:
\begin{eqnarray}
T^I &=&  L_\Lambda^{\ I} F^\Lambda  \\
T^{(3)} &=& e^{-2\sigma} H
\end{eqnarray}
$T^{(2)}$,$T^{(3)}$ and $T^I$ appear in the susy transformation laws of the
fermions:
\begin{eqnarray}
\delta \psi_L &=& D \epsilon_L +c_1 T_{ab}^{(2)} \Delta^{abc}\epsilon_R V_c
+ c_2 T_{abc}^{(3)} \Delta^{abcd} \epsilon_L V_d + \cdots \\
\delta \chi_L &=& d_1\Gamma^a \partial_a \sigma \epsilon_R +
 d_2 T_{ab}^{(2)} \Gamma^{ab}\epsilon_R
+d_3 T_{abc}^{(3)}\Gamma^{abc} \epsilon_L + \cdots  \\
\delta \lambda^I_L&=& f_1 \Gamma^a P^I_{ i}\partial_a z^i \epsilon_R
+ f_2 T_{ab}^I \Gamma^{ab} \epsilon_L  + \cdots
\end{eqnarray}
Taking into account:
\begin{eqnarray}
\eta_{\Lambda\Sigma} &=& L_{(\Lambda} \bar L_{ \Sigma)} - L_{\Lambda
I} L_{ \Sigma I} \\
\cN_{\Lambda\Sigma} &=& \left(L_{(\Lambda} \bar L_{ \Sigma)} + L_{\Lambda
I} L_{ \Sigma I} \right) e^{-2 \sigma}
\end{eqnarray}
 we find:
 \begin{eqnarray}
Z^{(2)} &=& e^{-\sigma} L_\Lambda (z) g^\Lambda \\
 Z^{(3)} &=& e^{-2 \sigma} g \\
 Z_I  &=&  L_{\Lambda I}(z) g^\Lambda
\end{eqnarray}
where
 \begin{equation}
  g = \int_{S^3}H \quad \quad  g^{\Lambda} = \int_{S^2}F^{\Lambda}
  \label{yyy}
 \end{equation}
The differential relations and sum rules among the charges are:
\begin{eqnarray}
\nabla Z^{(2)} &=& Z_I P^I - Z^{(2)} d \sigma \\
 \nabla Z^{(3)} &=& -2 Z^{(3)} d \sigma \\
 \nabla Z_I &=& {1 \over 2} (\bar Z P_I + Z \bar P_I)\\
 Z^{(2)}\bar Z^{(2)} - Z_I Z_I &=& g^\Lambda \eta _{\Lambda\Sigma}
 g^\Sigma\\
 Z^{(2)}\bar Z^{(2)} + Z_I Z_I &=& g^\Lambda \cN _{\Lambda\Sigma}
 g^\Sigma
\end{eqnarray} }
\vskip 5mm
Let us now consider the $D=6$ matter coupled theories.
\item{
For type IIA \cite{rom}, \cite{gpn} ($(N_+,N_-) =( 2,2)$) the coset manifold is
\begin{equation}
G/H= {O(4,n) \over O(4) \times O(n)} \times O(1,1)
\end{equation}
 and the field content is:\\
 Gravitational multiplet:
\begin{equation}
(V^a_\mu, \psi_{A\mu},\psi_{\dot A\mu}, B_{\mu\nu},A^\alpha _ \mu, \chi _A , \chi _{\dot A},\sigma)
\qquad\quad (\alpha = 1,\cdots,4)  \label{fcontent1}
\end{equation}}
Vector multiplets:
\begin{equation}
(A_\mu,\lambda_A,\lambda_{\dot A}, 4 \phi)^i \qquad \quad (i=1,\cdots,n)
\end{equation}
The group theoretical assignments under the duality group
$O(4,n)\times O(1,1)$ can be read from Table \ref{tab2a}.
\begin{table}[ht]
\caption{Transformation properties of fields in
matter coupled $D=6$ $N=(2,2)$}
\label{tab2a}
\begin{center}
\begin{tabular}{|c||c|c|c|c|c|c|c|c|c|}
\hline
&$ V^a_\mu $& $H_{\mu\nu\rho}$ & $F^\Lambda_{\mu\nu}$
&
$L^\Lambda_{\ \Sigma}$& $e^\sigma$ & $ \psi_{A \mu} $ &
$ \chi_{ A} $& $\lambda^{I}_A$ & $R_H$ \\
\hline
\hline
 $O(4,n)$ & 1 & 1 & $4+n$ & $ 4+n$  & 1 & 1 & 1 & 1 & -  \\
\hline
$O(1,1)$ & 0 & 2 & 1 & 0 & 1 & 0 & 0 & 0 & - \\
\hline
$SU_L(2) \times SU_R(2) $ & $(1,1)$ &  $(1,1)$ & $(1,1)$  &
$4$ &  $(1,1)$ &  $(2,1)$ &  $(2,1)$ &  $(2,1)$  & $(2,2)$ \\
\hline
$O(n)$ & 1 & 1 & 1 & $n$ & 1 & 1 & 1 & $n$  & $n$\\
\hline
\end{tabular}
\end{center}
\end{table}
In the present case
$H_{Aut}= O(4)$  acts as $ SU_L(2)\times SU_R(2)$ on the two
 left--handed and two right--handed fermions. In particular, left--handed spinors
transform as doublets under $SU_L(2)$ and as singlets $SU_R(2)$ (the opposite happens
for right--handed ones, which are not quoted in the Table).
The two chiralities  are distinguished by indices $A, \dot A$ respectively ($A,\dot A =1,2$).
$SU_{L}(2),SU_{R}(2)$ indices are raised and lowered with the metric $\epsilon_{AB}$, $\epsilon_{\dot A \dot B}$
respectively.
The coset representative is decomposed as follows:
 \begin{equation}
e^\sigma L \equiv e^\sigma (L^\Lambda _r,
L^\Lambda _{\ I} )
\end{equation}
where $r=1,\cdots,4$ is a vector index of $O(4)$, and we also define
\begin{equation}
L^\Lambda_{\ A\dot A} \equiv L^\Lambda_r {\rm e}^r_{A\dot A}
\end{equation}
where  $ {\rm e}^r_{A\dot A}$ is a quaternion converting $O(4)$
into $SU(2)\times SU(2)$ indices.
The supercovariant field--strengths have the following
form:
\begin{eqnarray}
\hat H^+ & = & \left[ dB + a_1
\eta_{\Lambda\Sigma}dA^\Lambda \wedge A^\Sigma \right]^+ \nonumber \\
&+& e^{2\sigma}\left(a_2 \bar \psi^A \Gamma_a \psi^B \epsilon_{AB}V^a
+  a_3 \bar\psi^{\dot A} \Gamma _{ab} \chi^{\dot B} \epsilon_{\dot A \dot B} V^a V^b\right)
\nonumber\\
\hat H^- & = & \left[ dB + a_1
\eta_{\Lambda\Sigma}dA^\Lambda \wedge A^\Sigma \right]^- \nonumber \\
&-& e^{2\sigma} \left(
 a_2 \bar\psi^{\dot A} \Gamma_a \psi^{\dot B} \epsilon_{\dot A \dot B}V^a  
+ a_3\bar \psi^A \Gamma_{ab} \chi^B \epsilon_{AB}V^a V^b
\right)
 \label{h62a}\\
\hat F^\Lambda &= & dA^\Lambda +
e^{\sigma} \left ( b_1 \bar \psi^A \psi^{\dot B}L^\Lambda_{\ A \dot B} +
\bar b_1 \bar \psi^{\dot A} \psi^B L^\Lambda_{\ \dot A B} \right)  \nonumber \\
&+& e^{\sigma} \left(b_2 \bar \psi^A \Gamma_{a} \chi^{\dot B}L^\Lambda_{\ A \dot B} +
\bar b_2 \bar \psi^{\dot A} \Gamma_{a} \chi^{ B} L^\Lambda_{\ \dot A B} \right) \nonumber\\
&+&  e^{\sigma} \left( b_3 \bar \psi^A \Gamma_{a} \lambda^I_A
+\bar b_3 \bar \psi^{\dot A} \Gamma_{a} \lambda^I_{\dot A} \right)
V^a L^\Lambda_{\ I}\\
\hat P^{Ir}&= & P^{Ir} - \left(\bar \psi^A \lambda^{I\dot B} (e_r)_{A \dot B} + h.c. \right)\\
\hat {d \sigma} & =&d \sigma - \left ( \bar\psi^A \chi_A + h.c. \right )
 \label{f62a}
\end{eqnarray}
where $P^{Ir} \equiv (L^{-1} \nabla^{(H)} L)^{Ir}= P^{Ir}_{,i}d\phi^i$.
The supersymmetry
 transformation laws for the chiral fermions are as follows:
 \begin{eqnarray}
\delta \psi_A &=& D\epsilon_A +  c_1
T_{A\dot B \vert ab}\Delta^{abc}
 \epsilon^{\dot B} V_c  +  c_2
T^+_{\vert abc}\Delta^{abcd}
 \epsilon_A V_d + \cdots \label{gratr6a}\\
\delta \chi_{ A} &=& d_1 \Gamma^a \partial_a \sigma \epsilon_{ A}
+d_2
T_{ A \dot B \vert ab}\Gamma_{ab}
 \epsilon^{\dot B}   +  d_3
T_{\vert abc}\Gamma^{abc}
 \epsilon_{ A} + \cdots \label{diltr6a}\\
\delta \lambda^I_A& =& f_1 \Gamma^a P^I_{A\dot B,i}
\partial_a \phi^i \epsilon^{\dot B} +f_2
T^I_{\vert ab}\Gamma^{ab}
 \epsilon_A + \cdots \label{gautr6a}
\end{eqnarray}
with analogous expressions for the antichiral ones.
Here
\begin{equation}
\matrix{T_{A\dot B}&=& e^{-\sigma} L_{\Lambda A \dot B}
 F^\Lambda \cr
  T_{I}&=& e^{-\sigma} L_{\Lambda I}
 F^\Lambda \cr
T^\pm &=& \mp e^{-2\sigma}  H^\pm  \cr
} \label{ti6a}
\end{equation}
By integration of the above 2-- and 3--forms on  $S^2$ and $S^3$ respectively
we find the central charges.
Turning back to the $O(4)$ vector indices we have:
\begin{eqnarray}
Z_a &=& e^{-\sigma} L_{\Lambda  a}
g^\Lambda\\ 
Z_\pm &=& \half (e^{-2\sigma} g \pm e^{+2\sigma} e ) \\
Z_I &=& e^{-\sigma} L_{\Lambda I} g^\Lambda
\end{eqnarray}
Furthermore, from Maurer--Cartan equations and:
\begin{eqnarray}
\cN_{\Lambda\Sigma} &=&
e^{-2\sigma} (
L_{r \Lambda }L_{\ \Sigma}^{r}  + L_{I \Lambda }
L_{I\Sigma})\\
\eta_{\Lambda\Sigma} &=&
L_{r \Lambda }L_{\ \Sigma}^{r}  - L_{I \Lambda }
L_{\ \Sigma}^{I}
\end{eqnarray}
we obtain:
\begin{eqnarray}
 \nabla Z_r &=& Z_I P^I_{r} - Z_r d \sigma\\
 \nabla Z_\pm &=& \mp 2 Z_\mp  d \sigma  \\
 \nabla Z_I &=& Z^r P_{Ir} - Z_I d \sigma  
\end{eqnarray}
and:
\begin{eqnarray}
  Z_r Z^r + Z_I Z^I
 &=& g^\Lambda \cN_{\Lambda\Sigma} g^\Sigma \\
 Z_r  Z^r  - Z_I Z^I
 &=& g^\Lambda \eta_{\Lambda\Sigma} g^\Sigma \\
Z_+ Z_+ + Z_- Z_-
 &=& \pmatrix{g&e\cr}\pmatrix{\half e^{-4\sigma} & 0 \cr 0 &\half e^{+4\sigma}\cr}\pmatrix{g\cr e \cr}\\
Z_+ Z_+ - Z_- Z_-
 &=& eg
\end{eqnarray}
\vskip 5mm
  The last two cases in $D=6$ are $N=(2,0)$ and Type $IIB$ ($N=(4,0)$) theories  \cite{rom2}.
\item{
For Type $IIB$ the field content is:\\
Gravitational multiplet:
\begin{equation}
(V^a_\mu, \psi_{A\mu}, 5B^+_{\mu\nu})
\qquad\quad (\mu=1,\cdots,6 \,; \,A=1,\cdots,4)  \label{fcon1}
\end{equation}
Tensor multiplets:
\begin{equation}
(B^-_{\mu\nu},\lambda_A,5\phi)^I \qquad \quad (I=1,\cdots,n)
\label{fcon2}
\end{equation}
 and the scalar coset manifold is:
 \begin{equation}
G/H = {O(5,n)\over {O(5) \times O(n)}}
\end{equation}
The transformation properties of the fields are encoded in the
following Table \ref{tab2b}
\begin{table}[ht]
\caption{ Transformation properties of the fields in $D=6$, $N=(4,0)$}
\label{tab2b}
\begin{center}
\begin{tabular}{|c||c|c|c|c|c|c|}
\hline
 $D=6$, $N=(4,0)$&$ V^a_\mu $&$H^{+\Lambda}_{\mu\nu\rho},H^{-\Lambda}_{\mu\nu\rho}$&
$L^\Lambda_{\ \Sigma}$&$ \psi_{A\mu} $ & $ \lambda^{IA}$ & $R_H$ \\
\hline
\hline
 $O(5,n)$ & 1 & $5+n$ &$5+n$ &1 &1 & -  \\
\hline
$O(5)\times O(n)$  &$(1,1)$  &$(1,1)$  &$(5,n)$  & $(4,1)$ & $(4,n)$ & $(5,n)$\\
\hline
\end{tabular}
\end{center}
\end{table}
\noindent
where $H_{Aut}= O(5) $ acts as $Usp(4)$ on the indices $A,B$ of the
left--handed gravitino and right--handed spin $1/2$ fermions $\lambda^{IA}$ of the
tensor multiplet.
$H_{matter}\equiv O(n)$ acts on the indices $I$.
As usual we decompose the coset representative of $G/H$ as follows:
\begin{equation}
L^\Lambda _{\ \Sigma}= (L^\Lambda_r, L^\Lambda_I)  \qquad \quad
r=1,\cdots,5
\end{equation}
and, setting $ L^\Lambda_{AB} \equiv L^\Lambda_r \gamma^r_{AB}$
($\gamma^r_{AB}$are gamma--matrices of $O(5)$), the supercovariant
three--form and the vielbein  can be written as follows:
\begin{eqnarray}
\hat H^\Lambda &=& dB^\Lambda + c_1L^\Lambda_{AB} \bar \psi^A
\Gamma_a\psi^B V^a +c_2 L^\Lambda_I \bar \lambda^I_A \Gamma_{ab} \psi_B
\IC^{AB} V^aV^b\\
\hat P^{Ir}& =& P^{Ir} - \bar \psi^A \lambda^{IB} (\gamma^r)_{AB}
\end{eqnarray}
where $P^{Ir} = P^{Ir}_{, i} d\phi^i$ is the vielbein of $G/H$.
 The transformation laws of the fermions are:
\begin{eqnarray}
 \delta \psi_A &=& D \epsilon_a + a_1 T_{AB\vert abc} \Delta^{abcd} \IC^{BC}\epsilon_C V_d + \cdots\\
\delta \lambda^{IA}& =& P^I_{r,i}\partial_a \phi^i \Gamma^a
(\gamma^r)^{AB} \epsilon_B  \label{xxx} + b_2 T^I_{abc} \Gamma^{abc} \IC^{AB} \epsilon_B + \cdots
\end{eqnarray}
where the 3-- form dressed field--strenghts appearing in the gravitino
and dilatino transformation laws are:
\begin{eqnarray}
T_{AB}&=& L_{\Lambda AB} \ H^\Lambda \\
T_I &=& L_{\Lambda I} \ H^\Lambda.
\end{eqnarray}
We get the following central and matter charges
($Z_r \equiv {1\over 8} Z_{AB}\gamma_r^{AB}$):
\begin{eqnarray}
 Z_r&=& L_{\Lambda r} g^\Lambda \\
Z_I &=& L_{\Lambda I} g^\Lambda
\end{eqnarray}
We note that in this case there is no distinction between magnetic
and electric charges.
Indeed from the previous definitions it follows:
\begin{equation}
\cN_{\Lambda\Sigma} ^{\ \ \ \star} H^\Sigma = \eta_{\Lambda\Sigma}H^\Sigma
\end{equation}
Integrating both sides on a three--sphere we get:
\begin{equation}
e_\Lambda = \eta_{\Lambda\Sigma}g^\Sigma
\end{equation}
The differential relation derived from the Maurer--Cartan equations and the sum rules are:
\begin{eqnarray}
\nabla Z_r &=& Z_I P^I _r \\
 \nabla Z_I &=& Z^r P_{r I} \\
Z_r Z_r +Z_I Z_I &=& g^\Lambda \cN_{\Lambda\Sigma} g^\Sigma\\
 Z_r Z_r - Z_I Z_I &=& g^\Lambda \eta_{\Lambda\Sigma} g^\Sigma
\end{eqnarray}}
\item{
Finally, we consider shortly the $N=(2,0)$, $D=6$ theory \cite{dafrr}, \cite{sagn}, \cite{femisa}. \\
The field content is: \\
Gravitational multiplet:
\begin{equation}
(V^a_\mu, \psi_{A\mu}, B^+_{\mu\nu})
\qquad\quad (\mu=1,\cdots,6;\, A=1,2 )  \label{fcont1}
\end{equation}
 Tensor multiplets:
\begin{equation}
(B^-_{\mu\nu},\chi^A,5\phi)^I \qquad \quad (I=1,\cdots,n)
\label{fcont2}
\end{equation}
Vector multiplets:
\begin{equation}
(A_{\mu},\lambda_A)^\alpha \qquad \quad (\alpha=1,\cdots,m)
\label{fcont3}
\end{equation}
Hypermultiplets:
\begin{equation}
(\xi^A,4q)^l \qquad \quad (l=1,\cdots,p)
\label{fcont4}
\end{equation}
The coset manifold is in this case:
\begin{equation}
G/H = {O(1,n) \over O(n)} \times \cQ
\end{equation}
 where $\cQ$ is a quaternionic manifold  parametrized by the
 hypermultiplet scalars $q$.
 Notice that in this theory the automorphism group coincides
 with the holonomy factor
 $SU(2)$ of $\cQ$. However,
 since the hypermultiplets do not enter in the definition of the
 supersymmetry charges, we forget about them in the following.
Table \ref{tab61} shows the transformation properties of the fields
where $\psi_A$ and $\lambda^{\alpha }_A$ are chiral while $\chi^{IA}$ are antichiral.
  \begin{table}[ht]
\caption{Transformation properties of fields in D=6, N=(2,0)}.
\label{tab61}
\begin{center}
\begin{tabular}{|c||c|c|c|c|c|c|c|c|}
\hline
D=6, N=(2,0)&$ V^a_\mu $&
$H^{+\Lambda}_{\mu\nu\rho},H^{-\Lambda}_{\mu\nu\rho}$
&$F^{\alpha }_{\mu\nu}$&
$L^\Lambda_{\ \Sigma}$&$ \psi_{A\mu}$ & $ \chi^{IA}$& $\lambda^{\alpha}_{ A}$ & $R_H$ \\
\hline
\hline
 $O(1,n)$ & 1 & $n+1$ & 1  &$ n+1$  & 1 &1 & 1 & - \\
\hline
$O(n)$ & 1 & 1  & 1  &$(n,1)$  & 1 & $n$&1 & $n$  \\
\hline
$Usp(2)$ & 1 & 1 &1 & 1 & 2  & 2 & 2 & 1 \\
\hline
\end{tabular}
\end{center}
\end{table}
We set
\begin{equation}
L=(L^\Lambda, L^\Lambda_I)
\end{equation}
and write down the supercovariant field--strenghts as:
 \begin{eqnarray}
\hat H^\Lambda &= & dB^\Lambda + a_1
C^\Lambda _{\alpha\beta}dA^\alpha \wedge A^\beta +
a_2  L^\Lambda \bar \psi^A \Gamma_a \psi_A  V^a  \nonumber\\
&+&
 a_3  L^\Lambda_I \bar \psi^A \Gamma_{ab} \chi _A V^a V^b
 \label{acca1}\\
\hat F^\alpha &= & dA^\alpha + b_1
\bar \psi^A \Gamma_{a} \lambda^\alpha _A V^a
 \label{effe1}\\
\hat P^I &=& P^I - \bar \psi_A \chi^{IA}
\end{eqnarray}
 where $P^I=P^I_{,i} d\phi^i$ is the vielbein of $G/H$, $C^\Lambda_{\alpha\beta} $ are constants and the $SU(2)$
 indices $A,B,\cdots=1,2$ are contracted with $\epsilon_{AB}$.
The fermions transformation laws are:
\begin{eqnarray}
  \delta\psi_A &=& D \epsilon_A + b_1 T_{abc} \Gamma^{ab} \epsilon_A V^c + \cdots \\
\delta \chi^{IA} &=& b_2 P^I_{,i} \partial_a\phi^i \Gamma^a \epsilon^{AB}\epsilon_B
+b_3 T^I_{abc} \Gamma^{abc} \epsilon^{AB}\epsilon_B +\cdots \\
\delta\lambda^\alpha_A &=& b_4 F^\alpha_{ab} \Gamma^{ab}\epsilon_A + \cdots
\end{eqnarray}
with:
\begin{equation}
T = L_\Lambda H^{+\Lambda} , \quad T^I = L^I_{\ \Lambda} H^{-\Lambda}
\end{equation}
Using a by now familiar procedure we may construct the charges
associated to  three--form $ H^\Lambda \equiv ( L^\Lambda  H^{+
} + L ^\Lambda _I H^{- I})$:
\begin{eqnarray}
Z&=& L_\Lambda g^\Lambda \\
Z_I & = & L_{\Lambda I } g^\Lambda
\end{eqnarray}
satisfying:
\begin{eqnarray}
\nabla Z &=& Z_I P^I \\
 \nabla Z_I &=& Z P_I \\
Z Z + Z_I Z_I & = & g^\Lambda \cN _{\Lambda\Sigma} g^\Sigma\\
Z Z - Z_I Z_I & = & g^\Lambda \eta _{\Lambda\Sigma} g^\Sigma
\end{eqnarray}
As in Type $IIB$ theory there is no distinction between  electric and
magnetic charges.
We have of course also the charges associated to the vector two--form
$F^\alpha$, namely the magnetic charge:
\begin{equation}
 m^\alpha = \int _{S^2} F^\alpha
\end{equation}
and the electric charge:
 \begin{equation}
 e_\alpha = \int _{S^4} L_ \Lambda C _{\alpha\beta}^{\Lambda
 \ \ \star} F^\alpha
\end{equation}
since in this case the kinetic matrix  for the vector is:
\begin{equation}
\cN_{\alpha\beta} = L_\Lambda C^\Lambda _{\alpha\beta}.
\end{equation}}
\end{itemize}
 \section{Maximally extended supergravities in odd dimensions}
\setcounter{equation}{0}
  The common feature of all maximally extended supergravities is that
  the relations between central and matter charges are now
  substituted by relations among central charges only, in the same
  way as it happens in $D=4$, $N=5,6,8$. By abuse of language we also
include in this section the $D=5$, $N=6$ theory which, though
 not maximally extended, does not admit matter coupling.
  \begin{itemize}
\item{
$D=5$, $N=6$ \cite{awto}
Coset manifold:
\begin{equation}
G/H = {SU^\star (6) \over Usp(6)}
\end{equation}
Field group assignments:
\begin{table}[ht]
\caption{Transformation properties of fields in $D=5$, $N=6$}
\label{tab5,6}
\begin{center}
\begin{tabular}{|c||c|c|c|c|c|c|}
\hline
& $V^a$ & $\psi_A$ & $F^{\Lambda\Sigma} $ & $\chi_{ABC}$ &
$L^{\Lambda\Sigma}_{\ \ AB} $ & $R_H$
\\
\hline
\hline
$SU^\star(6)$ & 1 & 1 & 15 & 1 & 15 & -\\
\hline
$Usp(6)$ & 1 & 6 & 1 & $14^\prime$ + 6 & 14 + 1 & 14
\\
\hline
\end{tabular}
\end{center}
\end{table}

 where $A,B  ;
  \Lambda ,\Sigma =1,\cdots,6$, the spinors are pseudo-Majorana and the dilatino field $\chi_{ABC}$
 can be decomposed as follows:
 \begin{equation}
\chi_{ABC} = \buildrel \circ \over \chi _{ABC} + \IC_{[AB} \chi_{C]}
\end{equation}
$\buildrel \circ \over \chi _{ABC} $ being the antisymmetric
$\IC_{AB} $-traceless representation of $Usp(6)$.
In $D=5$, $N=6$, we have 15 vectors
 in the antisymmetric
irrep of $SU^*(6)$.  We take the coset representatives in the same
representation, namely $ L^{\Lambda\Sigma}_{\ \ AB}
=L^{[\Lambda}_{[A}L^{\Sigma ]}_{B]} $, where $L^\Lambda_A$ is in the fundamental representation of $SU^*(6)$ and
$\Lambda,\Sigma $ and $A,B$ are $SU^*(6)$ and $Usp(6) $
indices
respectively.
Note that with respect to $Usp(6)$
we have:
\begin{equation}
L^{\Lambda\Sigma}_{\ \ AB} =
{\buildrel \circ \over L}^{\Lambda\Sigma}_{\ \ AB} + \IC_{AB}
L^{\Lambda\Sigma}
\end{equation}
where ${ \buildrel \circ \over L}^{\Lambda\Sigma}_{\ \ AB} $ is
$\IC_{AB}$-traceless.
The coset representative $L^\Lambda_\Sigma$ satisfies:
\begin{equation}
\left \{\matrix{L^\dagger &=& L \cr
\IC L^t \IC^{-1} &=& L} \right.
\end{equation}
 The supercovariant vector field-strengths and vielbein are:
 \begin{eqnarray}
\hat F^{\Lambda\Sigma} &=& dA^{\Lambda\Sigma} +
\left [L^{\Lambda\Sigma} _{\ \ AB} ( a_1 \bar \psi^A \psi^B + a_2 \bar
\psi_C \gamma_a \chi^{ABC} V^a ) + h.c. \right ]\\
\hat P_{AB} &=& P_{AB} -\bar \psi^C \buildrel \circ \over \chi _{ABC}
\end{eqnarray}
where $P_{AB} = P_{AB,i} d\phi^i$ belongs to the {\bf 14} irrep. of $USp(6)$
 (that is it is antisymmetric and $\IC_{AB} $ traceless).
The fermion transformation laws define the physical graviphotons $T_{AB}$:
\begin{eqnarray}
\delta\psi_A &=& D \epsilon_A + a_3 T_{AB \vert ab}
\Delta^{abc} \epsilon^B V_c + \cdots \\
\delta\chi_{ABC} &=&a_4 P_{[AB} \gamma_a \epsilon_{C]} +
a_5 \gamma^{ab} T_{ ab\vert  [AB} \epsilon_{C]} + \cdots
\end{eqnarray}
where the two-form $T_{AB}$ is given by:
\begin{equation}
  T_{AB} = {1\over 2} L_{\Lambda\Sigma \vert AB}
F^{ \Lambda\Sigma}
\end{equation}
The magnetic central charges are:
\begin{equation}
  Z_{AB} = \int_{S^2} T_{AB} ={1\over 2} L_{\Lambda\Sigma AB}g^{\Lambda\Sigma}
\end{equation}
and with the usual procedure, setting
\begin{equation}
Z_{AB}={\buildrel \circ \over Z}_{ AB} + \IC_{AB}
Z
\end{equation}
we get \footnote{We have adopted the convention, for lowering
and raising indices, that:
\begin{equation}
\IC_{AB} V^B =V_A ; \qquad  \IC^{AB} V_B = -V^A.
\end{equation}}:
\begin{eqnarray}
   \nabla Z &=& {1 \over 4} { \buildrel \circ \over Z}_{ AB} P^{AB}   \\
  \nabla { \buildrel \circ \over Z}_{AB} &=& { \buildrel \circ \over Z}_{C[A} \IC^{CD}P_{B]D}
  + {1 \over 6} \IC_{AB}{ \buildrel \circ \over Z}_{ LM} P^{LM} +{2 \over 3} Z
  P_{AB}
\end{eqnarray}
and the sum rule:
\begin{equation}
 {1\over 4} g^{\Lambda\Sigma}\cN_{\Lambda\Sigma , \Gamma\Delta} g^{\Gamma\Delta}
 = {1 \over 2} Z_{AB}\bar Z^{AB}
\end{equation}
where:
\begin{equation}
\cN_{\Lambda\Sigma , \Gamma\Delta}
={1 \over 2} L_{\Lambda\Sigma AB}\bar L^{AB}_{\ \ \Gamma\Delta}
\end{equation}
  }
\item{$\underline {D=5, N=8}$ \cite{gurowa} \\
  The coset manifold is ${E_{6(6)} / Usp(8)}$ and
  the coset representative $L^{\Lambda \Sigma}_{AB}$, antisymmetric and
$\IC_{AB}$ traceless in the $USp(8)$ indices $AB$, is taken in the $27\times 27$ defining
  representation of $E_{6(6)}$. The $Usp(8)$ indices $A,B$ are raised and lowered with the
symplectic metric $\IC_{AB}$ ($i.e.$ $ Z_{AB} = \IC_{AL} \IC_{BM} \bar Z^{LM}$).
Spinors are symplectic--Majorana.
 \par
 The field content and the transformation properties are given in
 Table \ref{tab5,8}.
   \begin{table} [ht]
\caption{  Transformation properties of fields in $D=5$, $N=8$}
\label{tab5,8}
\begin{center}
\begin{tabular}{|c||c|c|c|c|c|c|}
\hline
&$ V^a $&$\psi_A$&$F^{\Lambda\Sigma}$&$\chi_{ABC}$&
$L^{\Lambda\Sigma}_ {\ \ AB}$ & $R_H$ \\
\hline
\hline
$E_{6(6)}$ &1 &1& $27 $& 1 & 27 & -  \\
\hline
$Usp(8)$ & 1 & 8 & 1 & 48 & $\bar{27}$ & 42 \\
\hline
\end{tabular}
\end{center}
\end{table}
 As usual the coset representatives appear in the supercovariant
 field strengths as follows:
 \begin{equation}
\hat F^{\Lambda\Sigma} = F^{\Lambda\Sigma}
 +a_1 \bar \psi^A \psi ^B L^{\Lambda\Sigma}_{\ \ AB}
+ a_2 \bar \psi^A\Gamma_a \chi_{ABC}
L^{\Lambda\Sigma BC} V^a
\end{equation}
while the vielbein $P_{ABCD}=P_{ABCD,i}d\phi^i$, completely
 antisymmetric and pseudoreal, is related to its supercovariant part $\hat P_{ABCD}$ as:
\begin{equation}
\hat  P_{ABCD} = P_{ABCD} - \bar \psi_{[A} \chi_{BCD]}    \label{zzz}
    \end{equation}
The physical graviphotons appear in the transformation laws for the
fermions:
\begin{eqnarray}
\delta\psi_A &=& D \epsilon_A + b_1 T_{AB\vert ab} \Delta^{abc}
\epsilon^B V_c + \cdots  \\
 \delta \chi _{ABC} &=& b_2 P_{ABCD, i}\partial _ a \phi^i \Gamma^a \epsilon ^D +
 b_3( T_{[AB \vert ab}\Gamma^{ab} \epsilon_{C]} - {1 \over 3}
 \IC_{[AB} T_{C]D \vert ab}\Gamma^{ab} \epsilon^D)
\end{eqnarray}
where
\begin{equation}
 T_{AB} = {1 \over 2}L_{AB \Lambda \Sigma} F^{\Lambda \Sigma} \label{ccc}
\end{equation}
 and the vectors kinetic matrix is:
 \begin{equation}
\cN_{\Lambda\Sigma \vert \Gamma\Delta} = {1 \over 2} L_{AB \Lambda\Sigma }
L^{AB}_{\ \ \Gamma\Delta}  \equiv
{1 \over 2} L_{AB \Lambda\Sigma }
L_{CD \Gamma\Delta} \IC^{AC} \IC^{BD}
\label{n58}
\end{equation}
where $L_{AB \Lambda \Sigma}
=(L^{\Lambda\Sigma}_{\ \ AB})^{-1}$.
\\
Defining the magnetic and electric charges
\begin{equation}
g^{\Lambda\Sigma} = \int _{S^2} F^{\Lambda\Sigma} \quad;\quad
e_{\Lambda\Sigma} = \int _{S^3}{1 \over 2}\cN_{\Lambda\Sigma\Gamma\Delta} F^{\Gamma\Delta}
\end{equation}
the magnetic central charges are:
\begin{equation}
Z_{AB} = {1 \over 2}L_{\Lambda\Sigma AB} (\phi)
g^{\Lambda\Sigma}
\end{equation}
The Maurer Cartan equations for the $L$'s are:
\begin{equation}
d L^{\Lambda\Sigma}_{\ \ AB} = {1 \over 2}L^{\Lambda\Sigma}_{\ \ CD}
\Omega^{CD}_{\ \ AB} + {1 \over 2}L^{\Lambda\Sigma CD} P_{CD AB}
\end{equation}
where $ \Omega^{CD}_{\ \ AB} = 2 Q^{[C}_{\ [A} \delta^{D]}_{B]}$,
$Q^C_{\ A}$ belongs to the Lie algebra of $Usp(8)$.
It follows:
\begin{equation}
\nabla^{(Usp(8))} Z_{AB}={1 \over 2} \bar Z^{CD}P_{CDAB}
\end{equation}
 Furthermore, from (\ref{n58}) we find
 \begin{equation}
{1 \over 2}Z_{AB} \bar Z^{AB} ={1 \over 4} g^{\Lambda\Sigma}
\cN_{\Lambda\Sigma \vert \Gamma\Delta} g^{\Gamma\Delta}
\end{equation}
The electric central charges are given by:
 \begin{equation}
Z^{(e)}_{AB} = {1 \over 2}L^{\Lambda\Sigma}_{\ \  AB} (\phi)
e_{\Lambda\Sigma}
\end{equation}
}
 \item{
 $\underline {D=7, N=4}$ \cite{ppn}\\
 The coset space is:
 \begin{equation}
G/H =
{Sl(5,\IR) \over O(5)}
\end{equation}
 and the field content with the transformation properties under $G$ and $H$
 is given in Table \ref{tab3},
   \begin{table}[ht]
\caption{ Transformation properties of fields in $D=7$, $N=4$}
\label{tab3}
\begin{center}
\begin{tabular}{|c||c|c|c|c|c|c|c|}
\hline
&$ V^a_\mu $&$B_{\Lambda\mu\nu}$&$A^{\Lambda\Sigma}_\mu$&
$L^\Lambda_{\ I}$&$ \psi^A_\mu $ & $ \chi^{IA}$ & $R_H$\\
\hline
\hline
 $Sl(5)$ & 1 & 5 & 10 & 5 & 1 & 1 & -\\
\hline
$O(5)$ & 1 & 1 & 1 & 5 & 4 & 16 & 14 \\
\hline
\end{tabular}
\end{center}
\end{table}
 where $\psi ^A$ and $\chi ^{IA}$  are symplectic--Majorana
 spinors and $\chi ^{IA}$ satisfies the irreducibility constraint
 \begin{equation}
(\gamma_I)_{AB} \chi ^{IB} =0
\end{equation}
 $\gamma_I$ being the $O(5)$ gamma matrices.
$USp(4)$ symplectic indices $A,B,\cdots$ are raised and
lowered with $\IC_{AB}$.
Besides $L^\Lambda_{\ I}$ in the fundamental representation of
$Sl(5)$ we also introduce the coset representatives
$ L^{\Lambda\Sigma}_{\ \ IJ}=L^{[\Lambda}_ {\ [I} L^{\Sigma]}_ {\ J]}$
in the
representation $\underline{10}$  of $O(5)$.
\\
Then the supercovariant field strengths are:
\begin{eqnarray}
\hat H^\Lambda&\equiv& dB_\Lambda + a_1 L_{\Lambda
I}(\gamma^I)_{AB} \bar \psi^A \Gamma_a\psi^B V^a +
a_2 \epsilon_{\Lambda\Sigma_1\cdots \Sigma_4}F^{\Sigma_1 \Sigma_2}
\wedge A^{\Sigma_3 \Sigma_4} \nonumber\\
&+&
a_3 L_{\Lambda I} \psi_A \Gamma_{ab}\chi^{IA} V^aV^b\\
\hat F^{\Lambda\Sigma}&\equiv& dA^{\Lambda\Sigma} + b_1
  L^{\Lambda\Sigma}_{\ \ IJ}
(\gamma^{IJ})_{AB} \bar \psi^A \psi^B
 \nonumber\\
&+& b_2
  L^{\Lambda\Sigma}_{\ \ IJ}
(\gamma^{J})_{AB} \bar \psi^A \Gamma_a\chi^{IB} V^a\\
\hat P_{IJ} &\equiv &  P_{IJ} - \bar \psi ^A \chi_{(I}^{\ B}(\gamma_{J)})_{AB}
\end{eqnarray}
where $P_{IJ}=P_{IJ,i}d\phi^i $  ( $I,J$ symmetric and traceless) is the coset vielbein.
\par
The fermion transformation laws are:
 \begin{eqnarray}
\delta\psi_A &=& D\epsilon_A + c_1 T^{(3)}_{AB\vert abc}
\Delta^{abcd}\epsilon^B V_d + c_2 T^{(2)}_{AB\vert ab}
\Delta^{abc}\epsilon^B V_c + \cdots \\
\delta\chi^I_A&=& P^{IJ}_ {,i}\partial_a\phi^i  \Gamma^a
(\gamma_{J})_{AB}\epsilon^B + c_3 T^{(2)I}_{AB\vert ab}\Gamma^{ab}
\epsilon^B+ c_4 T^{(3)I}_{AB\vert abc}\epsilon^B + \cdots
\end{eqnarray}
 where ($F^{\Lambda\Sigma} \equiv dA^{\Lambda\Sigma}$, $ H_{\Lambda} \equiv dB_\Lambda$)
 \begin{eqnarray}
 T^{(3)}_{[AB]} &=& L^\Lambda_{ I}
(\gamma^I)_{AB} H_{\Lambda} \\
T^{(3)I}_{(AB)} &=& L^\Lambda_{ J}
(\gamma^{IJ} -4 \delta^{IJ}\bfone)_{AB} H_{\Lambda} \\
T^{(2)}_{(AB)} &=& L_{\Lambda\Sigma}^{\ \ IJ}
(\gamma_{IJ})_{AB} F^{\Lambda\Sigma} \\
T^{(2)}_{IAB} &=& L_{\Lambda\Sigma}^{\ \ KJ}
(\gamma_{IKJ} -3 \delta_{IK} \gamma_J)_{AB} F^{\Lambda\Sigma}
\end{eqnarray}
By integrating the 3--form $T^{(3)}_{AB}, T^{I(3)}_{AB}$ on $S^3$ and
the 2--form $T^{(2)}_{AB}, T^{I(2)}_{AB}$ on $S^2$
we get the following set of central and matter charges:
 \begin{eqnarray}
Z^{(2)}_{(AB)} &=&{1\over 2} L_{\Lambda\Sigma}^{\ \ IJ}(\gamma_{IJ})_{AB}
g^{\Lambda\Sigma}\\
  Z^{(3)}_{[AB]} &=& L^{\Lambda }_I(\gamma^{I})_{AB}
g^{\Lambda}\\
Z^{(3)\ I}_{\ (AB)} &=& L^{\Lambda }_J(\gamma^{IJ} -4 \delta^{IJ}\bfone)_{AB}
g^{\Lambda}\\
Z^{(2)}_{AB\vert K} &=&{1\over 2} L_{\Lambda\Sigma}^{\ \ IJ}
(\gamma_{KIJ}-3\delta_{KI} \gamma_J)_{AB}
g^{\Lambda\Sigma}
\end{eqnarray}
   where
   \begin{equation}
g^{\Lambda\Sigma} =  \int _{S^2}F^{\Lambda\Sigma} \qquad ; \qquad
 g_{\Lambda} =  \int _{S^3} H_\Lambda
\end{equation}
and $Z^{I (3)}_{(AB)}$ and $Z^{(2)}_{K\vert AB}$ satisfy the
constraint:
\begin{equation}
(\gamma_I Z^{I(3)})_{AB} =  (\gamma_I Z^{I(2)})_{AB} =0
\end{equation}
From the Maurer Cartan equations
\begin{equation}
\nabla^{(O(5))} L_{\Lambda I} = L_\Lambda^{\ J} P_{JI}
\end{equation}
and from the definition of the kinetic matrix
\begin{equation}
\cN_{\Lambda\Sigma} = L_\Lambda^{\ I} L_{\Sigma I}
\end{equation}
we find:
\begin{eqnarray}
\nabla^{(O(5))} Z^{I (3)}& =& Z^{(3)}_J P^{IJ}\\
 \nabla^{(O(5))} Z^{IJ (2)} &=&  Z^{(2)K[J}P^{I]}_K\\
Z^{(3)I}Z^{(3)}_I & = & g^\Lambda \cN_{\Lambda\Sigma} g^\Sigma
\end{eqnarray}
where we have traded the $Usp(4)$ indices of the dressed charges with $O(5)$
indices according to:
\begin{eqnarray}
Z^{I(3)} &=& Z^{(3)}_{AB} (\gamma^I)^{AB}\\
Z^{IJ(2)} &=& Z^{(2)}_{AB} (\gamma^{IJ})^{AB}
\end{eqnarray}
}
\item{
$ \underline{D=9, N=2}$
 The coset manifold is:
 \begin{equation}
G/H = {  Sl(2, \IR) \over O(2)} \times O(1,1)
\end{equation}
and the field content and  group assignements are given in Table \ref{tab9,2}.
 \begin{table}[ht]
\caption{Transformation properties of fields in $D=9$, $N=2$}
\label{tab9,2}
\begin{center}
\begin{tabular}{|c||c|c|c|c|c|c|c|c|c|c|}
\hline
&$ V^a_\mu $& $C_{\mu\nu\rho}$ & $B^\Lambda_{\mu\nu}$
&$A^{\Lambda}_\mu$& $A_\mu $&
$L^\Lambda_{\ AB}$& $e^\sigma$& $ \psi^A_\mu $ & $ \chi^{ABC}$ & $R_H$\\
\hline
\hline
 $Sl(2, \IR)$ & 1 & 1 &2 & 2 & 1 & 2 & 1 & 1 & 1 & - \\
\hline
$O(1,1)$ & 0 & 1 & 1 & 0 & 1 & 0 & 1 & 0 & 0 & -\\
\hline
$O(2)$ & 1 & 1 & 1 & 1 & 1 & 2 & 1 & 2 & 2 + 2 & 2 \\
\hline
\end{tabular}
\end{center}
\end{table}
Here $A,B,C$ are $O(2)$ vector indices, $L^\Lambda_{\ AB}$ is the
coset representative of ${Sl(2, \IR) \over O(2)}$ symmetric and
traceless in $A,B$, $e^\sigma$ parametrizes $O(1,1)$, $\Lambda =1,2$
are indices of $Sl(2, \IR)$ in the defining representation.
$\psi_A$, $\chi
_{ABC}$ are Majorana spinors.
 $\chi
_{ABC}$ is completely symmetric and can be decomposed as
\begin{equation}
\chi_{ABC} = \buildrel \circ \over \chi _{ABC} + \delta_{(AB} \chi _{C)}
\end{equation}
The supercovariant field strengths are defined as follows:
\begin{eqnarray}
\hat H^{\Lambda (3)} & \equiv & dB^\Lambda + a_1
\epsilon_{\Lambda\Sigma}
(A^\Lambda \wedge dA^\Sigma + A^\Sigma \wedge dA^\Lambda)+
a_2 e^{\sigma} \bar \psi_A \Gamma_a \psi_B V^a L^\Lambda_{\ AB}
\nonumber\\
&+&
 a_3 e^{\sigma} \bar \psi_C \Gamma_{ab} \chi^{ABC} V^a V^b \IC_{AB}\\
\hat H^{(4)}&\equiv & dC + b_1 (B^\Lambda \wedge dA_\Lambda + A
\wedge A^\Lambda \wedge dA^\Sigma \epsilon_{\Lambda\Sigma} ) + b_2
e^{\sigma} \bar \psi^A \Gamma_{ab}\psi^B V^a \wedge V^b \epsilon_{AB}
\nonumber\\
&+&
 b_3 e^{\sigma} \bar \chi_C \Gamma_{abc} \psi^C V^a V^b V^c \\
 \hat F^\Lambda &\equiv & d A^\Lambda + c_1 L^\Lambda_{\ AB} \bar \psi ^A \psi ^B+
 c_2 L^\Lambda_{\ AB} \bar \chi^{ABC} \Gamma_a \psi_C V^a \\
 \hat F &\equiv & d A + d_1 e^\sigma \bar \psi ^A \psi _A +
 d_2 e^\sigma \bar \chi_{A} \Gamma_a \psi^A V^a \\
\hat P_{AB} &\equiv & P_{AB} -\bar \psi^C  \buildrel \circ \over \chi _{ABC}\\
\hat {d\sigma} &=& d\sigma - \bar \psi^A \chi_A
\end{eqnarray}
where $P_{AB} = P_{AB,i} d\phi^i$ is the real vielbein of ${SL(2,\IR)\over O(2)}$.
The fermion transformation laws are:
\begin{eqnarray}
\delta\psi_A &=& D\epsilon_A + l_1 T^{(4)}_{abcd}
\Delta^{abcdf}\epsilon_A V_f + l_2 T^{(3)}_{AB\vert abc}
\Delta^{abcd}\epsilon^B V_d \nonumber\\
&+& l_3 T^{(2)}_{AB\vert ab}
\Delta^{abc}\epsilon^B V_c +l_4 T^{(2)}_{ab}
\Delta^{abc}\epsilon_A V_c +\cdots  \\
\delta\chi_{ABC}&=& h_1 P_{(AB,i}\partial_a\phi^i  \Gamma^a
\epsilon_{C)} + h_2 T_{abcd}
\Gamma^{abcd}\delta_{(AB}\epsilon_{C)} \nonumber\\
&+& h_3 T_{(AB \vert abc}
\Gamma^{abc}\epsilon_{C)} + h_4 T_{(AB\vert ab}
\Gamma^{ab}\epsilon_{C)} +h_5 T_{ab}
\Gamma^{ab}\delta_{(AB}\epsilon _{C)} +\cdots
\end{eqnarray}
 where
 \begin{eqnarray}
 T^{(4)} &=& e^{-\sigma} H^{(4)} \\
 T^{(3)}_{AB} &=&e^{-\sigma} L_{\Lambda AB}
 H^{{(3)}\Lambda} \\
T^{(2)}_{(AB)} &=&  L_{\Lambda AB}
F^{(2)\Lambda} \\
T^{(2)} &=& e^{-\sigma} F^{(2)}
\end{eqnarray}
By integration of the previous ($p+2$)--forms on $S^{p+2}$ we get the magnetic
central charges:
\begin{eqnarray}
 Z^{(4)} &=& e^{-\sigma} g  ; \qquad g= \int H^{(4)} \\
 Z^{(3)}_{AB} &=& e^{-\sigma}L_{\Lambda AB}
 g^{\Lambda} ; \qquad
 g^\Lambda= \int H^{(3)\Lambda} \\
Z^{(2)}_{(AB)} &=&  L_{\Lambda AB}
m^{\Lambda}     \qquad
 m^\Lambda= \int F^{(2)\Lambda}\\
Z^{(2)} &=& e^{-\sigma} m ;  \qquad
 m= \int F^{(2)}
\end{eqnarray}
Using now the Maurer Cartan equations for the coset representative
$e^{-\sigma} L_{\Lambda AB}$ we find:
\begin{equation}
\nabla^{O(2)} (e^{-\sigma}L_{\Lambda AB}) = e^{-\sigma}
( L_{\Lambda C(A} P_{B)C} - d\sigma L_{\Lambda AB} )
\end{equation}
where the indices between brackets are symmetric and traceless.
   Therefore:
   \begin{eqnarray}
\partial _\sigma \left( \matrix{ Z^{(4)} \cr Z^{(3)}_{AB} \cr Z^{(2)} }
\right) &=& - \left( \matrix{ Z^{(4)} \cr Z^{(3)}_{AB} \cr Z^{(2)} }
\right) \\
\nabla_i \left( \matrix{ Z^{(3)}_{AB} \cr Z^{(2)}_{AB} }
\right) &=&  \left( \matrix{ Z^{(3)}_{C(A} \cr Z^{(2)}_{C(A} }
\right)P_{B)C,i}
\end{eqnarray}
  Finally the kinetic matrices for the $p$--forms are:
  \begin{eqnarray}
\cN^{(4)} & = & e^{-2\sigma} \\
 \cN^{(3)}_{\Lambda\Sigma} & = & {1\over 2} e^{-2\sigma}
 L_{\Lambda AB} L_\Sigma^{\ AB}\\
 \cN^{(2)}_{\Lambda\Sigma} & = & {1\over 2}L_{\Lambda AB} L_\Sigma^{\ AB} \\
 \cN^{(2)} & = & e^{-2\sigma} \\
\end{eqnarray}
It follows:
\begin{eqnarray}
(Z^{(4)})^2 & = & e^{-2\sigma} g^2\\
{1\over 2} Z_{AB}^{(3)}Z^{(3)AB} & =& g^\Lambda \cN^{(3)}_{\Lambda\Sigma} g^\Sigma\\
{1\over 2} Z^{(2)}_{AB} Z^{(2) AB} & = & m^\Lambda\cN^{(2)}_{\Lambda\Sigma}m^\Sigma \\
 (Z^{(2)})^2 & = & e^{-2\sigma} m^2 \\
\end{eqnarray}
}
\end{itemize}
\section{$D=6$ and $D=8$ maximally extended supergravities}
\setcounter{equation}{0}
\begin{itemize}
\item{The  field content of $D=6$ $N = (4,4)$ supergravity \cite{tanii} is given by the
following
gravitational multiplet:
\begin{equation}
(V^a_\mu, \psi_{A\mu}, \psi_{\dot A\mu},
B^{+I}_{\mu\nu}, B^{- \dot I}_{\mu\nu},
A^{\alpha \dot \alpha} _ \mu,\chi _{AI} ,
\chi _{\dot A \dot I} ,L^x_{\ y})
\label{fcontent}
\end{equation}
  $(I,\dot I=1,\cdots,5; \, \alpha,
\dot\alpha=1,\cdots,4; \,A, \dot A = 1,\cdots, 4
; \, x,y = 1,\cdots,10)$ where $L^x_{\ y}$ is the coset representative of:
 \begin{equation}
G/H= {O(5,5) \over O(5) \times O(5)},
\end{equation}
and
 \begin{equation}
H_{Aut}\equiv Usp(4)\times Usp(4) \sim O(5) \times O(5)
\end{equation}
The group theoretical assignments for the fields are defined in Table
\ref{tabpari}.
\begin{table}[ht]
\caption{ Transformation properties of fields in
maximally extended $D=6$, $N=(4,4)$ supergravity}
\label{tabpari}
\begin{center}
\begin{tabular}{|c||c|c|c|c|c|c|c|}
\hline
 $D=6$, $N=(4,4)$&$ V^a_\mu $&$H^{+\Lambda}_{\mu\nu\rho},
 H^{-\Lambda}_{\mu\nu\rho}$&$F^{\alpha \dot\alpha}_{\mu\nu}$&
$L^\Lambda_{\ I}, L^\Lambda_{\ \dot I},$
&$ \psi^A_\mu ,\psi^{\dot A}_\mu $ & $ \chi^{I\dot A},\chi^{\dot I A}$ & $R_H$ \\
\hline
\hline
 $O(5,5)$ & 1 & $\underline{10}$ &$\underline{ 16}$
 &$\underline{ 10}$ & $1;1$ &$ 1;1$ & - \\
\hline
$O(5)\times O(5)$ & 1 &$(1,1)$  &$(1,1)$  &$(5,1);(1,5)$
& $(4,1);(1,4)$ & $(5,4);(4,5)$ & $(5,5)$ \\
\hline
\end{tabular}
\end{center}
\end{table}
Here, $\psi_A, \psi_{\dot A}$ are chiral  and antichiral
gravitinos and transform under $H_{Aut}$ in the fundamental
representation  of the two $Usp(4)$ factors respectively.
The dilatinos $\chi _{I \dot A},\chi _{\dot I A} $ are instead spinor--vectors
under each $O(5)$ with the couple of
indices $I \dot A$ ($\dot I A$) transforming in the $ \underline {\bf 5}$
and $\underline {\bf 4}$ of the two $O(5)$ factors respectively.
The dilatinos
chiralities are related to dotted and undotted indices in the
opposite way as for the gravitinos.
The conversion between $O(5)$ and $USp(4)$ indices is performed via the
$O(5)$ $\gamma$--matrices $(\gamma^I)_{AB}$, $(\gamma^{\dot I})_{\dot A\dot B}$.
Raising and lowering of the $Usp(4)$ indices are performed with the
symplectic metric $\IC_{AB}$ according to the rule:
$\psi_A = \IC_{AB} \psi^B$ (and the same for dotted indices).\\
Note that we have labelled the index of the spinorial representation ${\underline{16}}$
of $O(5,5)$ with a couple of indices $\alpha, \dot \alpha$.
The coset representative in the ${\underline 16}$ irrep will be denoted
 in the following $U^{\alpha\dot \alpha}_{\ \ A \dot A}$
Let us recall that in $D=6$ we can decompose $H^\Lambda$ into real
self--dual and antiself--dual parts which transform irreducibly
under the Lorentz group.
Defining
\begin{equation}
H^x = (H^{\Lambda +}, H^{\Lambda -}  )  \quad ( x=1,\cdots,10;
\Lambda= 1,\cdots,5)
\end{equation}
$O(5,5)$ acts as a T--duality group on $H^x$ in the fundamental
representation.
However, in $D/2 = p+2 $, $p$ odd, the $p+2$--forms are also acted on by
the Gaillard--Zumino duality group $O(n,n)$ ($n $ being the number of
$p+2$ tensors).
In $D=6$, $p=1$ and $n=5$, so that the Gaillard--Zumino duality group is again $O(5,5)$.
Quite generally, when $D/2 = p+2$, $p$ odd, the kinetic lagrangian for the
$p+2$--forms $H^{\pm\Lambda}$ has the following form:
\begin{equation}
\cL _{kin} = H^{+ \Lambda} \cN^+ _{\Lambda\Sigma} H^{-\Sigma} + H^{- \Lambda} \cN^- _{\Lambda\Sigma} H^{+\Sigma}
\end{equation}
where  the kinetic matrices $\cN^\pm$ satisfy:
\begin{equation}
\cN^\pm = - (\cN ^{\mp })^t
\end{equation}
and transform projectively under $O(n,n)$:
\begin{equation}
\cN^{\pm \prime} = (C+ D \cN ^\pm) \times (A+B \cN^\pm)^{-1}.
\end{equation}
Here we have used a notation where  a generic element of $O(n,n)$ is decomposed in $n\times n$ blocks
as follows:
\begin{equation}
S= \pmatrix{A & B \cr C & D \cr} \in O(n,n)
\end{equation}
 satisfying:
\begin{eqnarray}
C^t A + A^t C &=&0 \\
C^t B + A^t D &=& 1 \\
D^t A + B^t C & =& 1 \\
D^t B + B^t D &=& 0
\end{eqnarray}
where the $O(n,n)$ invariant metric has the off diagonal form:
\begin{equation}
 \eta = \pmatrix{{\bf 0}& \bfone \cr \bfone & {\bf 0} \cr}.
\end{equation}
Identifying $S$ with the coset representative of $O(n,n)$ it is convenient,
in analogy with the 4 dimensional case, to express the $n\times n$ subblocks
of $L$ in the following way:
\begin{equation}
  \label{d6coset}
  S=\sqrt{2}\pmatrix{f_+ + f_- & f_+ - f_- \cr h_+ + h_- & h_+ - h_- \cr}
\end{equation}
and the orthogonality relations become:
\begin{eqnarray}
  \label{orto6}
  h_+^t f_- + f^t_+ h_- &=&0 \\
h_\pm^t f_\pm + f^t_\pm h_\pm &=& \pm 1
\end{eqnarray}
The kinetic matrices $\cN_+, \cN_-$ can be expressed in terms of
$f_\pm$ and $h_\pm$ as follows:
\begin{eqnarray}
  \label{hnf6}
  h_- &=& \cN_+ f_- \\
h_+ &=& \cN_- f_+
\end{eqnarray}
so that (\ref{orto6}) can be also rewritten as:
\begin{eqnarray}
  f_\pm^t (\cN_- -\cN_+ ) f_\pm = 1
\label{orto'6}
\end{eqnarray}
As in the four dimensional case, the matrices $\cN_\pm$ transform projectively
under the $S$-duality group $O(5,5)$:
\begin{equation}
  \cN^\prime_\pm = (C+ D\cN_\pm) (A+B\cN_\pm)^{-1}
\end{equation}

It is useful to also introduce the coset representative of $O(5,5)/O(5)\times O(5)$
in the spinorial representation ({\bf 16}) of $O(5,5)$, pertaining to vectors:
it is $U^{\alpha\dot\alpha}_{A\dot A}$,
where $\alpha\dot\alpha =1,\cdots ,16 \in O(5,5)$; $A,\dot A=1,\cdots ,4 \in SU(4) \sim O(5)$.
It appears in the supercovariant field-strength of vector fields, and in the supersymmetry
 transformation laws of fermions where  vectors are involved.
The corresponding kinetic matrix $\cN_{\alpha\dot\alpha\beta\dot\beta}$ is defined like
 for odd dimensional theories (no Gaillard--Zumino duality acts for one-form potentials
 in six dimensions):
$$
\cN_{\alpha\dot\alpha\beta\dot\beta} \equiv U_{\alpha\dot\alpha A\dot A}U_{\ \ \beta\dot\beta}^{A\dot A}
$$

The supercovariant field-strengths are:
\begin{eqnarray}
  \label{sucov6}
  \hat H^{ \Lambda + } &=& (dB^ \Lambda)^+  + (f^
\Lambda_{I +} (\gamma^I \otimes \bfone )_{\alpha\dot\alpha\beta\dot\beta}
F^{\alpha\dot\alpha} A^{\beta\dot\beta})^+\\
&+& f^\Lambda_{I +} \left( (\gamma^I)_{AB} \bar\psi ^A \Gamma _a \psi^B V^a
 + \bar\chi^{\dot A I} \Gamma_{ab}\psi_{\dot A} V^a V^b \right)\\
 \hat H^{ \Lambda - } &=& (dB^ \Lambda)^-  + (f^
\Lambda_{\dot I -} (\gamma^{\dot I} \otimes \bfone )_{\alpha\dot\alpha\beta\dot\beta}
F^{\alpha\dot\alpha} A^{\beta\dot\beta})^-\\
&+& f^\Lambda_{\dot I -} (\gamma^{\dot I})_{\dot A\dot B}
 (\bar\psi ^{\dot A} \Gamma _a \psi^{\dot B} V^a
 + \bar\chi^{ A\dot I} \Gamma_{ab}\psi_{ A} V^a V^b )\\
\hat F^{\alpha\dot\alpha}&=& dA^{\alpha\dot\alpha} + U^{\alpha\dot\alpha}_{A\dot A}(
\bar \psi^A \psi ^{\dot A} + \bar\chi^{\dot A I}\Gamma_a \psi^B (\gamma^I)^A_{\ B}V^a \nonumber\\
& + & \bar\chi^{ A \dot I}\gamma_a \psi^{\dot B} (\gamma^{\dot I})^{\dot A}_{\ \dot B} V^a)\\
\hat P_{IJ} &=& P_{IJ} - \bar \psi^A \chi_{\dot I}^{\ B} (\gamma_J)_{AB} -
\bar \psi^{\dot A} \chi_{ I}^{\ \dot B} (\gamma_{\dot J})_{\dot A\dot B}
\end{eqnarray}
The transformation laws of the chiral fermions are:
\begin{eqnarray}
  \label{trafer6}
  \delta\psi_A &=& D \epsilon _A + T_{I+\vert abc} \Delta^{abcd} (\gamma_I) ^{AB}
\epsilon_B V_d + T_{A\dot A\vert ab}\Delta^{abc}\epsilon^{\dot A}V_c + \cdots\\
\delta \chi^{I \dot A}&=& P^{I\dot I}_{,a} \Gamma^a\epsilon_{\dot B} (\gamma_{\dot I}) ^{\dot A \dot B}
+ T_{I+\vert abc} \Gamma^{abc}\epsilon_{\dot A} + T^{A\dot A}_{ \vert ab} \Gamma^{ab}\epsilon^B (\gamma^I)_{AB} + \cdots
\end{eqnarray}
where:
\begin{eqnarray}
  T_{I+} &=& f_{\Lambda I +}H^{+\Lambda}\\
T_{A\dot A} &=& U_{\alpha\dot\alpha A\dot A}F^{\alpha\dot\alpha}
\end{eqnarray}
The  central charges for the 3-- and 2-- forms are found  by integration
 of the corresponding 3-- and 2--forms $T_{+I},T_{A\dot A}$  on $S^3$ and $S^2$ respectively.
Defining:
\begin{eqnarray}
  \int_{S^3} H^\Lambda &=& g^\Lambda\\
\int_{S^3} \cG_\Lambda &=& e_\Lambda  \quad\quad
 (\cG_{\Lambda}^\mp \equiv {\partial \cL \over \partial
 H^{\Lambda \pm}}= \cN^\pm_{\Lambda\Sigma} H^{\Sigma\mp})\\
\int _{S^2} F^{\alpha\dot\alpha} & = & g^{\alpha\dot\alpha}
\end{eqnarray}
where $g$ and $e$ are magnetic and electric charges respectively, we find:
\begin{eqnarray}
  \label{z+}
  Z_{+ I} &=& \int f^\Lambda_{I+} (\cN_- - \cN_+ )_{\Lambda\Sigma} H^{\Lambda +}\nonumber\\
& &= \int \left( h_{\Lambda I+}  H^{\Lambda +}
 + f^\Lambda_{I+} \cG_\Lambda^+ +  h_{\Lambda I+}  H^{\Lambda -}
 + f^\Lambda_{I+} \cG_\Lambda^-\right)\nonumber\\
& & =  h_{\Lambda I+}  g^{\Lambda }
 + f^\Lambda_{I+} e_\Lambda
\end{eqnarray}
Note that in the second line of eq. (7.32) we have inserted the last
two terms which sum up to zero using the relations (7.18) and (7.30).
\par
In an analogous way one gets:
\begin{equation}
  \label{z-}
  Z_{- \dot I}=  h_{\Lambda\dot I -}  g^{\Lambda }
 + f^\Lambda_{\dot I-} e_\Lambda
\end{equation}
For the magnetic charge of the dressed vector fields we get:
\begin{equation}
  \label{zvec}
  Z_{A\dot A} = U_{\alpha\dot\alpha A\dot A}g^{\alpha\dot\alpha}
\end{equation}
The Maurer--Cartan equations give the following relations:
\begin{eqnarray}
  \label{mc6}
  \nabla Z_{J +} &=& P_{J\dot J} Z_{\dot J -}\\
\nabla Z_{A\dot A} &=& Z_{B\dot B}  (\gamma^I)_A^{\ \dot B}  (\gamma^{\dot J})_{\dot A}^{\  B}  P_{I \dot J}
\end{eqnarray}
The sum rules for the dyonic  charges associated to the 3--forms can be obtained
by using the same procedure as in the 4-dimensional case, provided we use
the (\ref{orto'6}) instead of (\ref{specdef}).
One gets:
\begin{equation}
  \label{sumrule6}
  Z_{+I} Z_{+I} + Z_{- \dot I} Z_{- \dot I} = (g,e) \cM(\cN_+,\cN_-) \pmatrix{g\cr e}\\
\end{equation}
where:
\begin{equation}
  \label{emme6}
  \cM(\cN_+,\cN_-) = \left(\matrix{
2({\cal N}_-^{-1} -{\cal N}_+^{-1})^{-1} & 
({\cal N}_- +{\cal N}_+)({\cal N}_- -{\cal N}_+)^{-1} \cr
- ({\cal N}_- -{\cal N}_+)^{-1}({\cal N}_- +{\cal N}_+) &
2({\cal N}_- -{\cal N}_+)^{-1}\cr}\right)
\end{equation}
and:
\begin{equation}
  Z_{+I} Z_{+I} - Z_{- \dot I} Z_{- \dot I} = 2ge.
\end{equation}
For the vector central charges one gets:
\begin{equation}
  Z_{A\dot A} Z^ {A\dot A} = g^{\alpha\dot\alpha}\cN_{\alpha\dot\alpha\beta\dot\beta} g^{\beta\dot\beta}.
\end{equation}
We observe that we have obtained five electric charges $e_\Lambda$ and five magnetic charges $g^\Lambda$.
However, as in the $D=6$, $N=1$ case, electric and magnetic charges are not independent. Indeed,
if we use the 3--form $H^x$ previously defined, we have the T--duality relations:
\begin{equation}
  \label{sagn}
  N_{xy} H^y = \eta_{xy} ^{\ \ \star}H^y
\end{equation}
where:
\begin{eqnarray}
  N_{xy} &=& L_{x I} L_{y I} + L_{x \dot I} L_{y \dot I} \\
 \eta_{xy} &=& L_{x I} L_{y I} - L_{x \dot I} L_{y \dot I}
\end{eqnarray}
and $L^x_I, L^x_{\dot I}$ are the representatives of $O(5,5)$
in the fundamental representation.
Therefore, as in the $D=6$, $N= (2,0)$ case:
\begin{equation}
  g^x = - e_x
\end{equation}
so that the distinction between electric and magnetic charge is
immaterial.
 From (\ref{sagn}) it follows  that $L_{xI}H^x$ has definite self-duality.
Indeed  we have:
\begin{equation}
  L_{x I} H^x = L_{xI} (N^{-1})^{ xy} \eta_{yz} ^{\ \ \ \star}H^z=
L^y_I \eta_{yz} ^{\ \ \ \star}H^z= L_{xI} ^{\ \ \ \star}H^x
\end{equation}
The analogous expression $L_{x\dot I} H^x$ is antiself-dual.
Finally, we observe that the matrix $N_{xy}$, contrary to what happens for
$\cN_\pm$,  transforms tensorially under $O(5,5)$.
}

The relation between the matrices $ {\cal N}_\pm$, transforming projectively, and the matrix
$N_{xy}$ defined above, transforming tensiorally under $O(5,5)$, turns out to be given by:
$$
\cM(\cN_+,\cN_-)
   \equiv C^T  N C
$$
where:
$$
  C=\frac{1}{\sqrt{2}}\left(\matrix{\bfone & \bfone \cr - \bfone & \bfone \cr}\right).
$$
\item{Let us now discuss the $N=2$, $D=8$ theory \cite{sase3}. \\
 Coset manifold:
\begin{equation}
G/H = {Sl(3,\IR) \over O(3) }   \times {Sl(2,\IR) \over U(1)}
\end{equation}
Field content:
\begin{equation}
( V^A_\mu, C_{\mu\nu\rho}, B^\Lambda_{\mu\nu}, A^{\Lambda\alpha}_\mu,
L^\Lambda_I , L^\alpha_i , \psi_A, \chi^{AI})
\end{equation}
where $ L^\Lambda_I$, ($\Lambda,I= 1,2,3$) is the coset representative
of $Sl(3,\IR)\over O(3)$ and  $\L^\alpha_i$ ($\alpha,i=1,2$) is the
representative of $Sl(2,\IR)\over O(2)$. $A=1,2$ is a $SU(2)$ index
of $H_{Aut}= SU(2)\times U(1)$ and $\psi_A,\chi^{AI}$ are left handed
spinors, the right handed parts being denoted by $\psi^A,\chi_A^I$. It is convenient
to decompose $\chi^{AI}$ into the ${3\over 2}$ and ${1\over 2}$ $SU(2)$ representations
according to:
\begin{equation}
  \chi^{IA} = {\buildrel\circ\over \chi}^{IA} +  (\sigma^I)^A_{\ B} \chi^B
\end{equation}
 The group theoretical assignments are displayed in Table \ref{tab8m}
\begin{table}[ht]
\caption{Transformation properties of fields in D=8, N=2}
\label{tab8m}
\begin{center}
\begin{tabular}{|c||c|c|c|c|c|c|c|c|c|c|}
\hline
 $D=8$, $N=2$ &$ V^a_\mu $& $C_{\mu\nu\rho}$ & $B^\Lambda_{\mu\nu}$
&$A^{\Lambda\alpha}_\mu$& $ L^\Lambda_{\ I} $&
$ L^\alpha _{\ i} $&  $ \psi^A_\mu $ & $ {\buildrel\circ\over \chi}^{I A}$ &
 $\chi^A$ & $R_H$ \\
\hline
\hline
 $Sl(3, \IR)$ & 1 & 1 & 3 & 3 & 3 & 1 & 1 & 1 &1 & - \\
\hline
$Sl(2, \IR)$ & 1 & 1 & 1 & 2 & 1 & 2 & 1 & 1 & 1 & - \\
\hline
$SU(2)$ & $1$ &  $1$ & $1$
&  $1$ &  $3$ &  $1$ &  $2$ &  $4$ & 2 & 5 \\
\hline
$U(1)$ & 0 & 0 & 0 & 0 & 0 & 1 &${ 1\over 2}$ & $-{ 1\over 2}$
 & ${ 3 \over 2}$ & 2 \\
\hline
\end{tabular}
\end{center}
\end{table}

We parametrize the coset representative of ${Sl(2,\IR) \over U(1)}$  in the following way:
\begin{equation}
L^\alpha_i = \pmatrix{L^1_+ & L^2_- \cr L^2_+ & L^1_- \cr}
\end{equation}
where:
\begin{eqnarray}
\eta_{\alpha\beta}L^\alpha_+ L^\beta_- &=& 1  \\
\bar L^\alpha_+ &=& L^\alpha_-
\end{eqnarray}
and we define:
\begin{equation}
f = L^1_+ +  L^2_+\quad ; \quad h=L^1_+ -  L^2_+
\end{equation}
The $Sl(2,\IR)$ group acts as a Gaillard--Zumino S--duality group on
the $Sl(2,\IR)$ vector $(H^{(4)\pm},\cG^{(4)\pm}) $ where
\begin{equation}
\cG_{abcd} =-{\rm i}/2 {\delta \cL \over \delta H^{abcd}}
\end{equation}
the kinetic $1\times 1$ ``matrix'' $\cN \equiv S $ being given by:
\begin{equation}
S= {L^1_+ -  L^2_+ \over L^1_+ +  L^2_+}.
\end{equation}
Furthermore, the same group $Sl(2,\IR)$ acts as a T--duality group on the doublet of
2--forms $F^\alpha_\Lambda$, ($\Lambda = 1,2,3 $) so that in the
supercovariant expression of $F^\alpha_\Lambda$ there appear
naturally $L^\alpha_\pm$, while in the supercovariant 4--form there
appear naturally the expressions $ f = L^1_+ +  L^2_+ $.
The supercovariant field strengths and vielbein  are:
\begin{eqnarray}
\hat H^{(4)} &=& dC + f \left(a_1  \bar\psi^A \Gamma_{ab} \psi^B \epsilon_{AB}
+ a_2 \bar\psi_A \Gamma_{abc} \chi^I_B (\sigma_I)^{AB}
+ h. c. \right) \nonumber\\
&+& (a_3 B^\Lambda \wedge F_\Lambda  + h. c. ) \\
\hat H^{(3)\Lambda } &=& dB^\Lambda + b_1
\epsilon^{\Lambda\Delta\Sigma} F^\alpha_\Delta A_{\alpha\Sigma}
\nonumber\\
&+& b_2(\bar\psi_A \Gamma_a \psi^B L^\Lambda_I (\sigma_I)^{\ A} _{B}V^a  +
b_3\bar\psi_A \Gamma_{ab} \chi^{IA} L^\Lambda_IV^a V^b + h. c. ) \\
\hat F^{\alpha}_\Lambda &=& dA^{\alpha}_\Lambda + \bigl(c_1L^\alpha_+ \bar
\psi^A \psi^B L_{\Lambda AB}  \nonumber\\
&+& c_2\bar\psi^A \Gamma_{a} \chi^{B}_I
L_{\Lambda J} (\sigma_K) _{AB} \epsilon^{IJK}V^a + h. c. \bigr)\\
\hat P^{IJ}&=&  P^{IJ} - (\bar {\buildrel\circ\over\chi}^{IA} \psi_B (\sigma^J)_A^{\ B}+ h.c.)\\
\hat P &=& P - (\bar\chi^A\psi_A + h.c.)
\end{eqnarray}
where $P^{IJ} = P^{IJ}_{,i} d\phi^i$ is the $Sl(3,\IR)/O(3)$ vielbein, symmetric and traceless in the
$O(3)$ indices $I,J$, while $P=P_{,S} dS$ is the complex 1--bein of $SU(1,1)/U(1) \sim Sl(2,\IR)/O(2)$.
\par
The transformation rules for the fermions are:
\begin{eqnarray}
\delta \psi _A &=& D \epsilon _A + d_1T^{(4) -} _{abcd}
\Delta ^{abcde} \epsilon^B V_e \epsilon _{AB} \nonumber\\
&+& d_2 T^{(3) } _{I abc} \Delta^{abcd} (\sigma^I) _A^{\
B}\epsilon_B V_d + d_3 T^{(2) } _{I - ab}
(\sigma^I)_{AB}
\Delta^{abc} \epsilon^B V_c + \cdots \\
\delta {\buildrel \circ \over \chi}^{IA} &=&
f_1P^{IJ} _{, i}\partial_a\phi^i \Gamma^a(\sigma_J) ^A_{\ B} \epsilon^B +
f_2T^{(3) }_{J abc} \Gamma^{abc}  ( \sigma^{IJ} -2
\delta^{IJ} ) ^A _{\ B} \epsilon^B \nonumber\\
&+&f_3T^{(2) } _{J + ab}   ( \sigma^{IJ} - 2
\delta^{IJ} ) ^{AB} \epsilon_B + \cdots \\
\delta \chi^A &=& g_1 P_{, S}\partial_a S \Gamma^a \epsilon^A + g_2T^{-(4)}_{abcd} \Gamma^{abcd} \epsilon^{AB}
\epsilon_B \nonumber\\
&+&g_3
T^{(2)}_{ ab -}\Gamma^{ab}
(\sigma^I )^{AB} \epsilon_B + \cdots
\end{eqnarray}
where:
\begin{eqnarray}
T^{(4)- }&=& \bar f^{-1} H^- \\
  T^{(3) } _I      &=& L_{\Lambda I} H^\Lambda \\
   T^{(2) } _{I \pm}  &=& L_{\alpha \pm }
   L^\Lambda_{ \ I} F^\alpha_{\Lambda}
\end{eqnarray}
Note that as in $D=4$ we have:
\begin{equation}
T^{(4) -}= \bar f^{-1} H^- = (\cN - \bar \cN )f H^- = h H^- - f \cG^-=
h H - f \cG = T^{(4)}
\end{equation}
  since:
  \begin{equation}
hH^+ - f \cG^+ = 0.
\end{equation}
Integrating on $S^{p+2}$ the ($p+2$)--forms $T^{(4)}, T^{(3)}_I, T^{(2)}_{I\pm}$
appearing in the fermions transformation
laws we obtain:
\begin{eqnarray}
Z^{(4)} &=& hg - fe \\
Z^{(3)}_I &=& L^\Lambda_I g^\Lambda\\
Z^{(2)}_{I \pm} &=& L_{\alpha \pm} L^\Lambda_I g^\alpha_\Lambda
\end{eqnarray}
 Note that the central charge associated to the 4--form is dyonic
 while those associated to the three and two forms are magnetic.
 The corresponding electric charges are retrieved as usual by
 integrating $\cN_{\Lambda\Sigma}H^\Sigma$ and
 $\cN_{\alpha\beta}^{\Lambda\Sigma} F^\beta_\Sigma $.
 \par
The Maurer--Cartan equations give rise to the following differential
relations among the charges:
\begin{eqnarray}
\nabla Z_{+ I} &=& Z_{+I} P_{++} + Z_{-I} P_{-+} \\
\nabla Z_I &=& Z_J P^J _{\ I} \\
\nabla Z &=& - \bar Z P
\end{eqnarray}
where $P_{++}$ and $P_{-+}$ are the components of the zweibein of
${Sl(2, \IR) \over O(2)}$ in a real notation.
Finally, from the explicit expression of the kinetic matrices:
\begin{eqnarray}
 \cN^{(4)} &=& h f^{-1} \\
 \cN_{\Lambda\Sigma} &=& L_{\Lambda I} L_{\Sigma I} \\
 \cN^{\Lambda\Sigma}_{\alpha\beta} &=& (L_{\alpha + } L_{\beta +}
 + L_{\alpha - } L_{\beta -} ) L^\Lambda_I L^\Sigma_I
\end{eqnarray}
we get the following sum rules:
\begin{eqnarray}
Z_I Z_I &=& g^\Lambda \cN_{\Lambda\Sigma} g^\Sigma \\
Z_{+ I} Z_{- I} &=& g^\alpha_\Lambda
\cN^{\Lambda\Sigma}_{\alpha\beta} g^\beta_\Sigma\\
\vert Z \vert ^2 &=& (g,e) \cM \pmatrix{ g \cr e}
\end{eqnarray}
\begin{equation}
\cM = \left( \matrix{ 1 & - Re \cN \cr 0 & 1 \cr}\right)
\left( \matrix{ Im \cN & 0 \cr 0 &Im \cN^{-1}\cr}\right)
\left( \matrix{ 1 & 0 \cr - Re \cN & 1 \cr}\right).
\end{equation}
}
\end{itemize}
\section*{Acknowledgments}
We acknowledge stimulating discussions with P. Fr\'e, C. Imbimbo and M. Trigiante.

\end{document}